\documentclass[11pt,a4paper]{article}
\pdfoutput=1
\usepackage{jheppub}

\usepackage[english]{babel}
\usepackage[utf8]{inputenc}
\usepackage{newunicodechar}
\usepackage{amsfonts,amsmath,amssymb,amsthm,bbold,bm,mathtools,slashed}
\usepackage{comment,enumerate,footnote,graphicx,subfloat,relsize}
\usepackage{array,tabularx,tabu,multirow,empheq}
\usepackage[usenames,dvipsnames]{xcolor}
\numberwithin{equation}{section}
\usepackage{float}
\usepackage[font=small,labelfont=bf]{caption}
\usepackage{cleveref}
\crefformat{pluralequation}{#2\black{eqs.~(}#1\black{)}#3}
\Crefformat{pluralequation}{#2\black{Equations~(}#1\black{)}#3}
\crefformat{pluralfigure}{#2\black{figs.~}#1#3}
\Crefformat{pluralfigure}{#2\black{Figures~}#1#3}
\usepackage{tikz}
\usetikzlibrary{arrows,shapes}
\usetikzlibrary{trees,patterns}
\usetikzlibrary{matrix,arrows} 				
\usetikzlibrary{positioning}				  
\usetikzlibrary{calc,through}				  
\usetikzlibrary{decorations.pathreplacing}  
\usepackage{pgffor}							

\usetikzlibrary{decorations.pathmorphing}	
\usetikzlibrary{decorations.markings}
\tikzset{
>=stealth', 
vector/.style={decorate, decoration={snake}, draw},
provector/.style={decorate, decoration={snake,amplitude=2.5pt}, draw},
antivector/.style={decorate, decoration={snake,amplitude=-2.5pt}, draw},
bigvector/.style={decorate, decoration={snake,amplitude=4pt}, draw},
fermion/.style={draw=black, postaction={decorate}, 
	decoration={markings,mark=at position .55 with {\arrow[draw=black]{>}}}},
fermionbar/.style={draw=black, postaction={decorate},
    decoration={markings,mark=at position .55 with {\arrow[draw=black]{<}}}},
fermionnoarrow/.style={draw=black},
doublefermion/.style={draw=black,double, postaction={decorate},
	decoration={markings,mark=at position .57 with {\arrow[draw=black]{>}}}},
doublefermionbar/.style={draw=black,double, postaction={decorate},
	decoration={markings,mark=at position .57 with {\arrow[draw=black]{<}}}},
doublefermionnoarrow/.style={draw=black,double},
gluon/.style={decorate, draw=black,
    decoration={coil,amplitude=4pt, segment length=5pt}},
scalar/.style={dashed,draw=black, postaction={decorate},
	decoration={markings,mark=at position .55 with {\arrow[draw=black]{>}}}},
scalarbar/.style={dashed,draw=black, postaction={decorate},
    decoration={markings,mark=at position .55 with {\arrow[draw=black]{<}}}},
scalarnoarrow/.style={dashed,draw=black},
momentum/.style={draw=black, postaction={decorate},
    decoration={markings,mark=at position 1 with {\arrow[draw=black]{>}}}},
antimomentum/.style={draw=black, postaction={decorate},
    decoration={markings,mark=at position 0.1 with {\arrow[draw=black]{<}}}}
}

\tikzstyle{block} = [draw, rectangle, minimum height=3em, minimum width=6em]
    
\newcommand{\nc}{\newcommand}
\nc{\pd}{\partial}
\nc{\bea}{\begin{eqnarray}}
\nc{\eea}{\end{eqnarray}}
\nc{\bal}{\begin{alignedat}}
\nc{\eal}{\end{alignedat}}
\nc{\beq}{\begin{equation}}
\nc{\eeq}{\end{equation}}
\nc{\bit}{\begin{itemize}}
\nc{\eit}{\end{itemize}}
\nc{\benu}{\begin{enumerate}}
\nc{\eenu}{\end{enumerate}}
\nc{\bdes}{\begin{description}}
\nc{\edes}{\end{description}}
\nc{\bma}{\begin{pmatrix}}
\nc{\ema}{\end{pmatrix}}

\newcommand{\black}[1]	{{\color{black} 	#1}}

\nc{\nn}{\nonumber}
\nc{\hc}{\text{h.c.}}
\nc{\cc}{\text{c.c.}}

\nc{\abs}[1]{\left| #1 \right|}
\def\[{\left[}
\def\]{\right]}
\def\({\left(}
\def\){\right)}
\def\<{\langle}
\def\>{\rangle}

\def\GeV{{\rm GeV}}

\def \cm{{\rm cm}}

\def\AnnP		{AnnP}
\def\AnnS		{AnnS}
\def\APhi		{{\cal A}_{\mathsmaller{T}}^{\mathsmaller{\Phi}}}
\def\AV			{\boldsymbol{{\cal A}}_{\mathsmaller{T}}^{\mathsmaller{V}}}
\def\ann		{{\rm ann}}
\def\aB			{\alpha_{\mathsmaller{B}}}
\def\aS			{\alpha_{\mathsmaller{S}}}
\def\aV			{\alpha_{\mathsmaller{V}}}
\def\aF			{\alpha_{\mathsmaller{\Phi}}}
\def\BSF		{{\rm \mathsmaller{BSF}}}
\def\BSFV		{BSF${}_{\mathsmaller{V}}$}
\def\BSFF		{BSF${}_{\mathsmaller{\Phi}}$}
\def\dec		{{\rm dec}}
\def\DM			{{\rm \mathsmaller{DM}}}
\def\DP			{{\rm \mathsmaller{DP}}}

\def\eff		{{\rm eff}}
\def\eq			{{\rm eq}}
\def\ellS		{\ell_{\mathsmaller{S}}}

\def\FOone		{\mathsmaller{\rm FO1}}

\def\GammaIon	{\Gamma_{\mathsmaller{\cal B}}^\ion}
\def\GammaDec	{\Gamma_{\mathsmaller{\cal B}}^\dec}
\def\GammaX		{\Gamma_{\mathsmaller{X}}}

\def\im			{\mathbb{i}}
\def\ion		{{\rm ion}}
\def\kappaB		{\kappa_{\mathsmaller{B}}}
\def\kappaS		{\kappa_{\mathsmaller{S}}}
\def\nX			{n_{\mathsmaller{X}}}
\def\nB			{n_{\mathsmaller{\cal B}}}
\def\mpl 		{m_{\rm \mathsmaller{Pl}}}
\def\mS			{m_{\mathsmaller{S}}}
\def\mF			{m_{\mathsmaller{\Phi}}}
\def\mX			{m_{\mathsmaller{X}}}
\def\mV			{m_{\mathsmaller{V}}}
\def\qF			{q_{\mathsmaller{\Phi}}}
\def\qX			{q_{\mathsmaller{X}}}

\def\PF			{P_{\mathsmaller{\Phi}}}
\def\PFvec		{{\bf P}_{\mathsmaller{\Phi}}}

\def\PVvec		{{\bf P}_{\mathsmaller{V}}}

\def\SM			{{\rm \mathsmaller{SM}}}
\def\twoPI		{\mathsmaller{\rm 2PI}}
\def\uni		{{\rm uni}}
\def\vrel		{v_{\rm rel}}
\def\VXX		{V_{\mathsmaller{XX}}}
\def\VXdaggerXdagger	{V_{\mathsmaller{X^{\dagger} X^{\dagger}}}}
\def\VXXdagger	{V_{\mathsmaller{XX^{\dagger}}}}
\def\xB			{x_{\mathsmaller{B}}}
\def\xS			{x_{\mathsmaller{S}}}
\def\XX			{{\mathsmaller{XX}}}
\def\XXdagger	{{\mathsmaller{XX^{\dagger}}}}

\def\zetaV		{\zeta_{\mathsmaller{V}}}
\def\zetaF		{\zeta_{\mathsmaller{\Phi}}}
\def\zetaB		{\zeta_{\mathsmaller{B}}}
\def\zetaS		{\zeta_{\mathsmaller{S}}}

\def\EB			{{\cal E}_{\mathsmaller{\cal B}}}
\def\YB			{Y_{\mathsmaller{\cal B}}}
\def\YX			{Y_{\mathsmaller{X}}}

\def\gstareffsqrt	{g_{*,\rm eff}^{1/2}}
\def\gstarS		{g_{*\mathsmaller{S}}}

\newcommand*\widefbox[1]{\fbox{\hspace{1em}#1\hspace{1em}}}
\colorlet{lightblue}{blue!30}
\colorlet{lightred}{red!30}
\preprint{Nikhef-2019-50}

\title{Dark matter bound state formation \\ via emission of a charged scalar}	

\author{Ruben Oncala and Kalliopi Petraki}

\affiliation{
	\href{https://www.nikhef.nl/en/}{\color{black} Nikhef}, 
	Science Park 105, 1098 XG Amsterdam, The Netherlands}

\bigskip

\affiliation{
	\href{http://www.lpthe.jussieu.fr/spip/index.php}{\color{black}
		Sorbonne Universit\'e, CNRS,  Laboratoire de Physique Th\'eorique et Hautes Energies (LPTHE)}, \\
	UMR 7589 CNRS \& Sorbonne Universit\'e, 
	4~Place Jussieu, F-75252, Paris, France}

\emailAdd{roncala@nikhef.nl}
\emailAdd{kpetraki@nikhef.nl}

\abstract{
The formation of stable or meta-stable bound states can dramatically affect the phenomenology of dark matter (DM). Although the capture into bound states via emission of a vector is known to be significant, the capture via scalar emission suffers from cancellations that render it important only within narrow parameter space. While this is true for neutral scalar mediators, here we show that bound-state formation via emission of a charged scalar can be extremely significant. To this end, we consider DM charged under a dark $U(1)$ force and coupled also to a light complex scalar that is charged under the same gauge symmetry. We compute the cross-sections for bound-state formation via emission of the charged scalar, and show that they can exceed those for capture via vector emission, as well as annihilation, by orders of magnitude. This holds even for very small values of the DM coupling to the charged scalar, and remains true in the limit of global symmetry.  We then compute the DM thermal freeze-out, and find that the capture into meta-stable bound states via emission of a charged scalar can cause a late period of significant DM depletion. Our results include analytical expressions in the Coulomb limit, and are readily generalisable to non-Abelian interactions. We expect them to have implications for Higgs-portal scenarios of multi-TeV WIMP DM, as well as scenarios that feature dark Higgses or (darkly-)charged inert scalars, including models of self-interacting DM.
}
\arxivnumber{1911.02605} 

\begin{document}
	\maketitle

\section{Introduction\label{Sec:Intro}}

Light scalar bosons that mediate a long-range force between dark matter (DM) particles appear in a variety of theories, including models of self-interacting DM and other hidden-sector constructions. Intriguingly, it has been recently shown that the 125~GeV Higgs boson can also mediate a long-range interaction between TeV-scale particles, that affects their annihilation rate and can bind them into bound states~\cite{Harz:2017dlj,Harz:2019rro}. This renders the dynamics of light scalar force carriers relevant as well to the phenomenology of DM consisting of Weakly interacting massive particles (WIMPs). The existence of bound levels is a generic feature of theories with light force mediators that has rich phenomenological implications.

The capture of unbound particles into bound states necessitates the dissipation of energy. This may occur radiatively, typically via emission of the mediator that is responsible for the long-range force. However, for a particle-antiparticle pair or a pair of identical particles, the radiative capture via emission of a scalar boson is rather suppressed due to cancellations in the amplitude that reflect in part the angular momentum selection rules of the process (cf.~\cref{sec:BSFviaScalarEmission_Amplitude})~\cite{Wise:2014jva,Petraki:2016cnz}. These cancellations concern the contributions to the radiative part of the amplitude that arise from the trilinear DM-DM-mediator coupling alone. The couplings of the scalar potential --- the self-couplings of the mediator, as well as the biquadratic couplings between DM and the mediator if DM is bosonic ---  also contribute to the radiative amplitude and may enhance the capture cross-sections~\cite{Oncala:2018bvl}. However for natural values of the parameters, the cross-sections remain mostly small.

In this work, we point out that the situation is markedly different if the emission of the scalar boson alters the potential between the interacting particles. This may occur if the scalar is charged under \emph{either a local or a global} symmetry. As we shall see, in this case, the leading-order contributions to the amplitude are proportional to the overlap of the initial-state and final-state wavefunctions, which now are not orthogonal since they are subject to different potentials. The large overlap between the incoming and outgoing states gives rise to strikingly large bound-state formation (BSF) cross-sections. This is akin to atomic transitions precipitated by ``sudden perturbations'', such as ionisation caused by a beta decay of the nucleus~\cite{Migdal:QualitativeQM}.

To demonstrate the phenomenological importance of the transitions we consider, we calculate the chemical decoupling of DM in the early universe taking into account the formation of particle-antiparticle bound states via charged-scalar emission, and their subsequent decay into radiation. The formation of metastable bound states in the early universe has been shown to deplete the DM abundance~\cite{vonHarling:2014kha}, with the effect being generally more pronounced if the bound states have sizeable binding energy. Then, they form and decay efficiently already at high temperatures, when the DM density is large~\cite{vonHarling:2014kha,Harz:2018csl,Harz:2019rro}. Here we find that, because of the largeness of the BSF cross-sections, shallow bound states can cause a second period of rapid DM depletion at low temperatures (of the order of their binding energy), much later than the traditional freeze-out. This alters the predicted couplings of DM to other species very significantly, thereby affecting its observable signatures.

For simplicity, in the present work, we carry out our computations in an Abelian model with scalar DM that is singly-charged under a dark $U(1)_D$ gauge force and is coupled to a doubly-charged light scalar via a trilinear coupling. We shall not assume that the light scalar obtains a vacuum expectation value (VEV). Even in models where it does, our computation remains essentially valid, provided that the VEV of the scalar is not much larger than its mass. Then, if the scalar mediator is light enough to be emitted during BSF, its mass and VEV must be smaller than all other relevant scales, and the symmetry is only mildly broken. Our results are also readily applicable to non-Abelian models, whose dynamics in the unbroken phase can be reduced to the Abelian case by an appropriate decomposition of the representations of the interacting particles~\cite{Kats:2009bv}.

The paper is organised as follows. 
In \cref{Sec:BSFviaScalarEmission}, we introduce the model, compute the cross-sections for BSF via emission of a charged scalar, provide analytical results in the Coulomb limit for capture into \emph{any} bound level, and discuss their features. We confront our results with partial-wave unitarity and discuss the resolution to its apparent violation where it occurs.  
In \cref{Sec:FreezeOut}, we consider the DM freeze-out in the presence of BSF via emission of a charged scalar, and show the effect on the DM relic density and predicted couplings. 
We conclude in \cref{Sec:Conclusion} with an outlook on the implications of our results.

\section{Bound-state formation via charged scalar emission \label{Sec:BSFviaScalarEmission}}

\subsection{The model \label{sec:BSFviaScalarEmission_Model}}

We assume that DM consists of a complex scalar field $X$ that couples to a dark Abelian gauge force $U(1)_{\mathsmaller{D}}$ with $V$ being the gauge boson, as well as to a light complex scalar $\Phi$ that is doubly charged under the same force. The interaction Lagrangian is
\begin{align}
{\cal L} =
&-\frac{1}{4} F_{\mu\nu} F^{\mu\nu}
+(D_\mu X)^\dagger (D^\mu X)
+(D_\mu \Phi)^\dagger (D^\mu \Phi)
-\mX^2 |X|^2 
-\mF^2 |\Phi|^2 
\nn \\
&-\frac{y \, \mX}{2}~\( X^2 \Phi^\dagger  + {X^{\dagger}}^2 \Phi \) 
- \frac{\lambda_{\mathsmaller{X}}}{4} |X|^4 
- \frac{\lambda_{\mathsmaller{\Phi}}}{4} |\Phi|^4 
- \lambda_{\mathsmaller{X\Phi}} |X|^2|\Phi|^2 ,
\label{eq:L}
\end{align}
with 
$F^{\mu\nu} \equiv \partial^\mu V^\nu -\partial^\nu V^\mu$ and 
$D_j^\mu \equiv \partial^\mu +\im q_j g V^{\mu}$, 
where the index $j$ denotes the particle of charge $q_j$. The charges are $\qX = 1$ and $\qF = 2$ for $X$ and $\Phi$ fields respectively. The quartic terms stabilize the scalar potential at large field values. It is possible that the dark sector couples also to the Standard Model (SM), via biquadratic couplings of the scalars to the Higgs, and/or kinetic mixing of $V$ with the Hypercharge gauge boson. Here we do not attempt a detailed phenomenological study of the model, but instead focus on computing the radiative capture into bound states via emission of a charged scalar, and simply showcasing its implications. We thus do not consider any couplings to the SM and do not derive any observational constraints.

We define the parameters that will appear in the non-relativistic potential,
\begin{align}
\aV \equiv \frac{g^2}{4\pi}
\qquad \text{and} \qquad
\aF \equiv \frac{y^2}{16\pi} \,.
\label{eq:alphas_definitions}
\end{align}
For convenience, we also define the total mass $M$ and the reduced mass $\mu$ of a pair of interacting particles; in our case $M=2\mX$ and $\mu = \mX/2$.

We emphasise here that we are interested in $\mF \ll \mX$. This hierarchy remains stable for momentum flows $Q < \mX$, which encompass the momentum transfer along the off-shell $\Phi$ bosons exchanged in the scattering and bound states (cf.~\cref{sec:BSFviaScalarEmission_Model_MomentaAndWF}). At $Q \gtrsim \mX$, $X$ loops generate corrections to the running mass of $\Phi$, $\delta (\mF^2) \propto \aF \mX^2$, that may far exceed its low-energy value.  Seen from a high-energy perspective, this amounts to a near cancellation between the high-energy value of the running $\Phi$ mass and the running contribution. While this may be fine-tuned, the value of the running $\Phi$ mass at high energies does not affect our computations. Moreover, \cref{eq:L} can be viewed as an effective theory valid below $\sim \mX$, that is potentially stabilised by additional physics at higher scales, such as supersymmetry.

\subsubsection{Non-relativistic potential \label{sec:BSFviaScalarEmission_Model_Potential}}

\begin{figure}[t!]
\centering
\begin{tikzpicture}[line width=1pt, scale=1]
\begin{scope}[shift={(0,1)}]
\begin{scope}[shift={(-2,0)}]
\node at (-1.3,1){$X$};
\node at (-1.3,0){$X^{\dagger}$};
\draw (-1,1) -- (1,1);\draw[fermion] 	(-1,1) -- (-0.4,1);\draw[fermion] 	(0.7,1) -- (1,1);
\draw (-1,0) -- (1,0);\draw[fermionbar] (-1,0) -- (-0.4,0);\draw[fermionbar](0.7,0) -- (1,0);
\node at (1.3,1){$X$};
\node at (1.3,0){$X^{\dagger}$};
\draw[fill=lightblue,shift={(0,0.5)}] (-0.5,-0.5) rectangle (0.5,0.5);
\node at (0,0.5){${\cal A}_{\mathsmaller{XX^{\dagger}}}^{\twoPI}$};
\end{scope}
\node at (0,0.5){$=$};
\begin{scope}[shift={(2,0)}]
\node at (-1.3,1){$X$};
\node at (-1.3,0){$X^{\dagger}$};
\draw[fermion] 		(-1,1) -- (0,1);
\draw[fermionbar] 	(-1,0) -- (0,0);	
\draw[vector]	(0,0) -- (0,1);
\node at (0.5,0.5){$V$};
\draw[fermion] 		(0,1) -- (1,1);
\draw[fermionbar] 	(0,0) -- (1,0);	
\node at (1.3,1){$X$};
\node at (1.3,0){$X^{\dagger}$};
\end{scope}
\node at (4,0.5){$+$};
\begin{scope}[shift={(6,0)}]
\node at (-1.3,1){$X$};
\node at (-1.3,0){$X^{\dagger}$};
\draw[fermion] 		(-1,1) -- (0,1);
\draw[fermionbar] 	(-1,0) -- (0,0);	
\draw[scalar]	(0,1) -- (0,0);
\node at (0.5,0.5){$\Phi$};
\draw[fermionbar] 	(0,1) -- (1,1);
\draw[fermion] 		(0,0) -- (1,0);	
\node at (1.3,1){$X^{\dagger}$};
\node at (1.3,0){$X$};
\end{scope}
\end{scope}
\begin{scope}[shift={(0,-1)}]
\begin{scope}[shift={(-2,0)}]
\node at (-1.3,1){$X$};
\node at (-1.3,0){$X$};
\draw (-1,1) -- (1,1);\draw[fermion] (-1,1) -- (-0.4,1);\draw[fermion] (0.7,1) -- (1,1);
\draw (-1,0) -- (1,0);\draw[fermion] (-1,0) -- (-0.4,0);\draw[fermion] (0.7,0) -- (1,0);
\node at (1.3,1){$X$};
\node at (1.3,0){$X$};
\draw[fill=lightred,shift={(0,0.5)}] (-0.5,-0.5) rectangle (0.5,0.5);
\node at (0,0.5){${\cal A}_{\mathsmaller{XX}}^{\twoPI}$};
\end{scope}
\node at (0,0.5){$=$};
\begin{scope}[shift={(2,0)}]
\node at (-1.3,1){$X$};
\node at (-1.3,0){$X$};
\draw[fermion]	(-1,1) -- (0,1);
\draw[fermion]	(-1,0) -- (0,0);	
\draw[vector]	(0,1) -- (0,0);
\node at (0.5,0.5){$V$};
\draw[fermion] 	(0,1) -- (1,1);
\draw[fermion] 	(0,0) -- (1,0);	
\node at (1.3,1){$X$};
\node at (1.3,0){$X$};
\end{scope}
\end{scope}
\end{tikzpicture}
\caption[]{\label{fig:2PI} 
The 2PI diagrams contributing to the non-relativistic potential for $XX^{\dagger}$ pairs (upper) and $XX$ or $X^{\dagger}X^{\dagger}$ pairs (lower). The arrows denote the flow of the $U(1)_D$ charge.}
\end{figure}
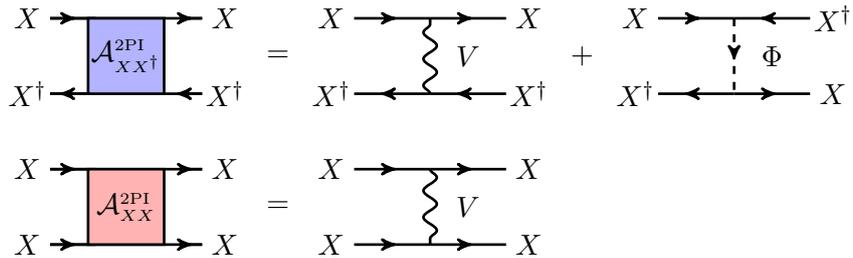

The long-range potential of $XX$, $X^{\dagger} X^{\dagger}$ and $X X^{\dagger}$ pairs is generated by the one-boson-exchange diagrams shown in \cref{fig:2PI}. Because the $\Phi$-exchange diagram for $XX^\dagger$ pairs is $u$-channel, the $\Phi$-generated potential depends on the angular momentum mode of the eigenstate; we clarify this subtlety in \cref{App:Potential}. Combining this with well-known results for vector-mediated and scalar-mediated potentials~\cite{Petraki:2015hla,Petraki:2016cnz,Oncala:2018bvl}, we obtain
\begin{subequations}
\label{eq:V}
\label[pluralequation]{eqs:V}
\begin{align}
\VXX (r) = \VXdaggerXdagger (r) &= +\frac{\aV}{r} \,,
\label{eq:V_XX}
\\
\VXXdagger (r) &= -\frac{\aV}{r} - (-1)^\ell \ \frac{\aF}{r} \, e^{-\mF r} \,,
\label{eq:V_XXdagger}
\end{align}
\end{subequations}
where $\aV$ and $\aF$ are defined in \cref{eq:alphas_definitions}. The potentials \eqref{eqs:V} distort the wavefunctions of pairs of unbound particles, a phenomenon known as the Sommerfeld effect~\cite{Sommerfeld:1931,Sakharov:1948yq}. For  $\aV + (-1)^\ell \aF > 0$, they also give rise to $XX^{\dagger}$ bound states.\footnote{In fact, the condition for the existence of $XX^\dagger$ bound levels is somewhat more relaxed since the repulsive contribution to the potential \eqref{eq:V_XXdagger} arising from $\Phi$ exchange for $\ell$ odd is of finite range, while the attractive term is of infinite range.}

\subsubsection{Radiative capture processes \label{sec:BSFviaScalarEmission_Model_CaptureProcesses}} 

The capture of unbound particles into bound states can occur radiatively, via emission of a vector or scalar boson, according to the processes
\begin{align}
X+X^{\dagger} &\to {\cal B} (XX^{\dagger}) +V , 
\label{eq:BSF_VectorEmission}
\end{align}
and
\begin{subequations}
\label{eq:BSF_ScalarEmission}
\label[pluralequation]{eqs:BSF_ScalarEmission}
\begin{align}
X+X &\to {\cal B} (XX^{\dagger}) +\Phi , 
\label{eq:BSF_XXToBandPhi} \\
X^{\dagger}+X^{\dagger} &\to {\cal B} (XX^{\dagger}) +\Phi^\dagger .
\label{eq:BSF_XbarXbarToBandPhidagger} 
\end{align}
\end{subequations}
We shall refer to the processes \eqref{eq:BSF_VectorEmission} and \eqref{eq:BSF_ScalarEmission} as \BSFV~and \BSFF~respectively. The leading order Feynman diagrams are shown in \cref{fig:BSF_VectorEmission,fig:BSF_ScalarEmission}. 

\BSFV~has been computed in  \cite{Petraki:2015hla,Petraki:2016cnz} (see \cite{Harz:2018csl} for non-Abelian generalisations), and a number of papers have considered its effects on the DM relic density~\cite{vonHarling:2014kha,Baldes:2017gzw,Baldes:2017gzu,Harz:2018csl,Cirelli:2018iax,Harz:2019rro,Fukuda:2018ufg} and indirect signals~\cite{Pospelov:2008jd,Pearce:2015zca,An:2016gad,Cirelli:2016rnw,Baldes:2017gzw,Baldes:2017gzu,Cirelli:2018iax}. Here, the coupling of DM to the light scalar gives rise to an additional contribution to the \BSFV \ amplitude at leading order, shown in \cref{fig:BSF_VectorEmission}. We review and adapt the computation of \BSFV \ to the present model in \cref{App:BSFV}.

In the rest of this section, we focus on the \BSFF \ cross-sections.

\begin{figure}[t!]
\centering
\begin{tikzpicture}[line width=1pt, scale=1]
\begin{scope}[shift={(0,2)}]
\node at (-4.3,1){$X$};
\node at (-4.3,0){$X^{\dagger}$};
\draw[fermion]		(-2.1,1) -- (-2,1);
\draw[fermionbar]	(-2.1,0) -- (-2,0);
\draw (-4,1) -- (0,1);
\draw (-4,0) -- (0,0);
%
\draw[fill=lightblue,shift={(-3,0.5)}] (-0.5,-0.5) rectangle (0.5,0.5);
\draw[fill=lightblue,shift={(-1,0.5)}] (-0.5,-0.5) rectangle (0.5,0.5);
\node at (-3,0.5){${\cal A}_{\mathsmaller{XX^{\dagger}}}^{\twoPI}$};
\node at (-1,0.5){${\cal A}_{\mathsmaller{XX^{\dagger}}}^{\twoPI}$};
\node at (-2,0.5){$\cdots$};
\draw[vector]	(0,1) -- (0.5,1.8);
\node at (0.7,2){$V$};
\node at (4.3,1){$X$};
\node at (4.3,0){$X^{\dagger}$};
\draw[fermion]		(2,1) -- (2.1,1);
\draw[fermionbar]	(2,0) -- (2.1,0);
\draw (0,1) -- (4,1);
\draw (0,0) -- (4,0);
%
\draw[fill=lightblue,shift={(3,0.5)}] (-0.5,-0.5) rectangle (0.5,0.5);
\draw[fill=lightblue,shift={(1,0.5)}] (-0.5,-0.5) rectangle (0.5,0.5);
\node at (3,0.5){${\cal A}_{\mathsmaller{XX^{\dagger}}}^{\twoPI}$};
\node at (1,0.5){${\cal A}_{\mathsmaller{XX^{\dagger}}}^{\twoPI}$};
\node at (2,0.5){$\cdots$};
\draw[fill=none,gray,line width=1.5pt] (2.1,0.5) ellipse (1.8 and 0.75);
\node at (4.6,0.5){${\cal B}$};
\end{scope}
\node at (0,1.5){$+$};
\begin{scope}[shift={(0,0)}]
\node at (-4.3,1){$X$};
\node at (-4.3,0){$X^{\dagger}$};
\draw[fermion]		(-2.1,1) -- (-2,1);
\draw[fermionbar]	(-2.1,0) -- (-2,0);
\draw (-4,1) -- (0,1);
\draw (-4,0) -- (0,0);
%
\draw[fill=lightblue,shift={(-3,0.5)}] (-0.5,-0.5) rectangle (0.5,0.5);
\draw[fill=lightblue,shift={(-1,0.5)}] (-0.5,-0.5) rectangle (0.5,0.5);
\node at (-3,0.5){${\cal A}_{\mathsmaller{XX^{\dagger}}}^{\twoPI}$};
\node at (-1,0.5){${\cal A}_{\mathsmaller{XX^{\dagger}}}^{\twoPI}$};
\node at (-2,0.5){$\cdots$};
\draw[vector]	(0,0) -- (0.5,-0.8);
\node at (0.7,-1){$V$};
\node at (4.3,1){$X$};
\node at (4.3,0){$X^{\dagger}$};
\draw[fermion]		(2,1) -- (2.1,1);
\draw[fermionbar]	(2,0) -- (2.1,0);
\draw (0,1) -- (4,1);
\draw (0,0) -- (4,0);
%
\draw[fill=lightblue,shift={(3,0.5)}] (-0.5,-0.5) rectangle (0.5,0.5);
\draw[fill=lightblue,shift={(1,0.5)}] (-0.5,-0.5) rectangle (0.5,0.5);
\node at (3,0.5){${\cal A}_{\mathsmaller{XX^{\dagger}}}^{\twoPI}$};
\node at (1,0.5){${\cal A}_{\mathsmaller{XX^{\dagger}}}^{\twoPI}$};
\node at (2,0.5){$\cdots$};
\draw[fill=none,gray,line width=1.5pt] (2.1,0.5) ellipse (1.8 and 0.75);
\node at (4.6,0.5){${\cal B}$};
\end{scope}
\node at (0,-1.5){$+$};
\begin{scope}[shift={(0,-3)}]
\node at (-4.3,1){$X$};
\node at (-4.3,0){$X^{\dagger}$};
\draw[fermion]		(-2.1,1) -- (-2,1);
\draw[fermionbar]	(-2.1,0) -- (-2,0);
\draw (-4,1) -- (0,1);
\draw (-4,0) -- (0,0);
%
\draw[fill=lightblue,shift={(-3,0.5)}] (-0.5,-0.5) rectangle (0.5,0.5);
\draw[fill=lightblue,shift={(-1,0.5)}] (-0.5,-0.5) rectangle (0.5,0.5);
\node at (-3,0.5){${\cal A}_{\mathsmaller{XX^{\dagger}}}^{\twoPI}$};
\node at (-1,0.5){${\cal A}_{\mathsmaller{XX^{\dagger}}}^{\twoPI}$};
\node at (-2,0.5){$\cdots$};
\draw[scalar]	(0,1) 	-- (0,0.3);
\draw[scalarnoarrow]	(0,0.3) -- (0,0);
\draw[vector]	(0,0.4) -- (0.7,-0.45);
\node at (0.9,-0.65){$V$};
\node at (4.3,1){$X^{\dagger}$};
\node at (4.3,0){$X$};
\draw[fermionbar]	(2,1) -- (2.1,1);
\draw[fermion]		(2,0) -- (2.1,0);
\draw (0,1) -- (4,1);
\draw (0,0) -- (4,0);
%
\draw[fill=lightblue,shift={(3,0.5)}] (-0.5,-0.5) rectangle (0.5,0.5);
\draw[fill=lightblue,shift={(1,0.5)}] (-0.5,-0.5) rectangle (0.5,0.5);
\node at (3,0.5){${\cal A}_{\mathsmaller{XX^{\dagger}}}^{\twoPI}$};
\node at (1,0.5){${\cal A}_{\mathsmaller{XX^{\dagger}}}^{\twoPI}$};
\node at (2,0.5){$\cdots$};
\draw[fill=none,gray,line width=1.5pt] (2.1,0.5) ellipse (1.8 and 0.75);
\node at (4.6,0.5){${\cal B}$};
\end{scope}
\end{tikzpicture}
\caption[]{
\label{fig:BSF_VectorEmission} 
The capture into bound states via emission of a vector boson (\BSFV), $X+ X^{\dagger} \to {\cal B} (XX^{\dagger}) + V$. While the bottom diagram appears to be naively of higher order, the momentum exchange along the two $\Phi$ propagators scales with the couplings, and renders this diagram of the same order as the two upper diagrams. We refer to \cref{App:BSFV} for the computation.}
	
\bigskip
	
\begin{tikzpicture}[line width=1pt, scale=1]
\begin{scope}[shift={(0,1)}]
\node at (-4.3,1){$X$};
\node at (-4.3,0){$X$};
\draw[fermion]	(-2.1,1) -- (-2,1);
\draw[fermion]	(-2.1,0) -- (-2,0);
\draw (-4,1) -- (0,1);
\draw (-4,0) -- (0,0);
%
\draw[fill=lightred,shift={(-3,0.5)}] (-0.5,-0.5) rectangle (0.5,0.5);
\draw[fill=lightred,shift={(-1,0.5)}] (-0.5,-0.5) rectangle (0.5,0.5);
\node at (-3,0.5){${\cal A}_{\mathsmaller{XX}}^{\twoPI}$};
\node at (-1,0.5){${\cal A}_{\mathsmaller{XX}}^{\twoPI}$};
\node at (-2,0.5){$\cdots$};
\draw[scalar]	(0,0) -- (0.6,-0.75);
\node at (0.8,-0.8){$\Phi$};
\node at (4.3,1){$X$};
\node at (4.3,0){$X^{\dagger}$};
\draw[fermion]		(2,1) -- (2.1,1);
\draw[fermionbar]	(2,0) -- (2.1,0);
\draw (0,1) -- (4,1);
\draw (0,0) -- (4,0);
%
\draw[fill=lightblue,shift={(3,0.5)}] (-0.5,-0.5) rectangle (0.5,0.5);
\draw[fill=lightblue,shift={(1,0.5)}] (-0.5,-0.5) rectangle (0.5,0.5);
\node at (3,0.5){${\cal A}_{\mathsmaller{XX^{\dagger}}}^{\twoPI}$};
\node at (1,0.5){${\cal A}_{\mathsmaller{XX^{\dagger}}}^{\twoPI}$};
\node at (2,0.5){$\cdots$};
\draw[fill=none,gray,line width=1.5pt] (2.1,0.5) ellipse (1.8 and 0.75);
\node at (4.6,0.5){${\cal B}$};
\end{scope}
\node at (0,-0.25){$+$};
\begin{scope}[shift={(0,-1.75)}]
\node at (-4.3,1){$X$};
\node at (-4.3,0){$X$};
\draw[fermion]	(-2.1,1) -- (-2,1);
\draw[fermion]	(-2.1,0) -- (-2,0);
\draw (-4,0) -- (-3.5,1);\draw (-3.5,1) -- (0,1);
\draw (-4,1) -- (-3.5,0);\draw (-3.5,0) -- (0,0);
%
\draw[fill=lightred,shift={(-3,0.5)}] (-0.5,-0.5) rectangle (0.5,0.5);
\draw[fill=lightred,shift={(-1,0.5)}] (-0.5,-0.5) rectangle (0.5,0.5);
\node at (-3,0.5){${\cal A}_{\mathsmaller{XX}}^{\twoPI}$};
\node at (-1,0.5){${\cal A}_{\mathsmaller{XX}}^{\twoPI}$};
\node at (-2,0.5){$\cdots$};
\draw[scalar]	(0,0) -- (0.6,-0.75);
\node at (0.8,-0.8){$\Phi$};
\node at (4.3,1){$X$};
\node at (4.3,0){$X^{\dagger}$};
\draw[fermion]		(2,1) -- (2.1,1);
\draw[fermionbar]	(2,0) -- (2.1,0);
\draw (0,1) -- (4,1);
\draw (0,0) -- (4,0);
%
\draw[fill=lightblue,shift={(3,0.5)}] (-0.5,-0.5) rectangle (0.5,0.5);
\draw[fill=lightblue,shift={(1,0.5)}] (-0.5,-0.5) rectangle (0.5,0.5);
\node at (3,0.5){${\cal A}_{\mathsmaller{XX^{\dagger}}}^{\twoPI}$};
\node at (1,0.5){${\cal A}_{\mathsmaller{XX^{\dagger}}}^{\twoPI}$};
\node at (2,0.5){$\cdots$};
\draw[fill=none,gray,line width=1.5pt] (2.1,0.5) ellipse (1.8 and 0.75);
\node at (4.6,0.5){${\cal B}$};
\end{scope}
\end{tikzpicture}
\caption[]{
\label{fig:BSF_ScalarEmission} 
The capture into bound states via emission of a charged scalar (\BSFF), $X+ X \to {\cal B} (XX^{\dagger}) + \Phi$. There is also the conjugate process, $X^{\dagger}+X^{\dagger} \to {\cal B} (XX^{\dagger}) + \Phi^\dagger$. The arrows denote the flow of the $U(1)_D$ charge. 
Note that in this case, the diagrams in which the incoming $XX$ particles emit off-shell $V$ and $\Phi$ that fuse to produce the final state $\Phi$ are of higher order, and we do not consider them here.} 
\end{figure}

\subsubsection{Momentum decomposition and wavefunctions \label{sec:BSFviaScalarEmission_Model_MomentaAndWF}}

We focus on the processes \eqref{eq:BSF_ScalarEmission}. For simplicity, in the following we neglect the mass of $\Phi$, except in the phase-space integration. This will allow us to obtain analytical expressions for the \BSFF~cross-sections, and gain important insight. Taking fully into account the mass of $\Phi$ (and potentially also a non-zero mass for $V$) requires computing the wavefunctions numerically, as done in \cite{Petraki:2016cnz} for a Yukawa potential and \cite{Harz:2017dlj,Harz:2019rro} for mixed Coulomb and Yukawa potentials.
In the Coulomb approximation, the $XX,~X^{\dagger}X^{\dagger}$ scattering states and the $XX^{\dagger}$ bound states are governed respectively by the potentials
\begin{align}
V_{\mathsmaller{S}} = -\aS/r \qquad \text{and} \qquad V_{\mathsmaller{B}} = -\aB/r , 
\label{eq:VSandVB}
\end{align}
with 
\begin{align}
\aS=-\aV, 
\qquad 
\aB = \aV + (-1)^\ell \aF .
\label{eq:alphaSalphaB}
\end{align}

The momentum assignments for the particles participating in \BSFF~are shown in \cref{fig:BSF_Scalar_RadiativePart}. In order to separate the motion of the center-of-momentum from the relative motion in the scattering and bound states, we decompose the momenta as follows~\cite{Petraki:2015hla}
\begin{subequations}
\label{eq:MomentumTrans}	
\label[pluralequation]{eqs:MomentumTrans}	
\begin{align}
k_1^{(\prime)} &\equiv K/2 + k^{(\prime)},& 	
k_2^{(\prime)} &\equiv K/2 - k^{(\prime)},
\label{eq:MomentumTrans_k1k2} \\
p_{\mathsmaller{X}} &\equiv P/2 + p,& 		
p_{\mathsmaller{X^\dagger}} &\equiv P/2 - p . 
\label{eq:MomentumTrans_p1p2}
\end{align}
\end{subequations}

\paragraph*{Scattering states.} In \cref{eq:MomentumTrans_k1k2}, the unprimed momenta correspond to infinite separation of $XX$, while the primed momenta denote the corresponding values in the $XX$ wavepacket, which is distorted by the long-range interaction; this is the well-known Sommerfeld effect~\cite{Sommerfeld:1931,Sakharov:1948yq}. 
It is easy to see from \cref{eq:MomentumTrans_k1k2} that in the non-relativistic regime, ${\bf k} = \mu {\bf v}_{\rm rel}$, with ${\bf v}_{\rm rel}$ being the relative velocity of the incoming $XX$ pair.
The non-relativistic on-shell relations for $k_1^0$ and $k_2^0$ imply that the total energy of the scattering state is $K^0 \simeq M + {\cal E}_{\bf k} + {\bf K^2}/(2M)$, with ${\cal E}_{\bf k} = {\bf k^2}/(2\mu) = \mu \vrel^2/2$. 
The scattering states are described by the wavefunctions $\phi_{\bf k}^\XX ({\bf r})$ in position space and $\tilde{\phi}_{\bf k}^\XX ({\bf k'})$ in momentum space; ${\bf k}$ is the expectation value of ${\bf k}'$. The wavefunctions $\phi_{\bf k}^\XX ({\bf r})$ obey the Schr\"odinger equation with the potential \eqref{eq:V_XX} and energy eigenvalue ${\cal E}_{\bf k}$.

\paragraph*{Bound states.}
They are described by the wavefunctions $\psi_{n\ell m}^\XXdagger ({\bf r})$ and $\tilde{\psi}_{n\ell m}^\XXdagger ({\bf p})$ in position and momentum space respectively, where $n\ell m$ are the standard principal and angular momentum quantum numbers in a central potential that determine the expectation value of ${\bf p}$. 
The energy of the bound states is $P^0 \simeq M - |{\cal E}_{n}| + {\bf P^2}/(2M)$, where the binding energies can be parametrised as ${\cal E}_{n} = -\kappaB^2/(2 n^2 \mu)$, with $\kappaB \equiv \mu \aB$ being the Bohr momentum of the system. Note that $p_1^0$ and $p_2^0$ do \emph{not} obey on-shell relations individually. 	
The wavefunctions $\psi_{n\ell m}^\XXdagger ({\bf r})$ obey the Schr\"odinger equation with the potential \eqref{eq:V_XXdagger} and energy eigenvalue ${\cal E}_{n}$.

\paragraph{Hierarchy of scales.}

The emergence of non-perturbative phenomena -- the Sommerfeld effect and the existence of bound states -- is largely due to the different scales involved in the $XX$ and $XX^\dagger$ scattering. For the scattering states and the bound states, 
\begin{subequations}
\label{eq:HierarchiesOfScales}
\label[pluralequation]{eqs:HierarchiesOfScales}
\begin{gather}
\mu \vrel^2 / 2 \ll \mu \vrel
\ll \mu \lesssim M ,
\\
\mu \aB^2 /(2n^2)
\ll \mu \aB/n 
\ll \mu  \lesssim M ,
\end{gather}
\end{subequations}
or equivalently,  
\begin{subequations}
\label{eq:HierarchiesOfScales2}
\label[pluralequation]{eqs:HierarchiesOfScales2}
\begin{gather}
{\cal E}_{\bf k} \ll |{\bf k}| \sim |{\bf k}'| \ll K^0
\\
|{\cal E}_{n}| \ll \kappaB/n \sim |{\bf p}| \ll P^0
\end{gather}  
\end{subequations}
In our computations, we make approximations based on these hierarchies.

\paragraph{Energy-momentum conservation.}

Taking into account the above relations, the conservation of energy and momentum, $K = P + \PF$, implies that $\Phi$ takes away the kinetic energy of the relative motion in the scattering state and the binding energy of the bound state~\cite{Petraki:2015hla,Petraki:2016cnz},
\beq
\sqrt{ \abs{\PFvec}^2 + \mF^2 }  = \omega 
\equiv {\cal E}_{\bf k} - {\cal E}_{n} 
= \frac{{\bf k}^2 +\kappaB^2/n^2}{2\mu} = \frac{\mu}{2} (\aB^2/n^2 + \vrel^2) ,
\label{eq:omega}
\eeq 
where we neglected the recoil of the bound state, as per~\eqref{eqs:HierarchiesOfScales}.  \Cref{eq:omega} can be recast as $\abs{\PFvec} = \omega \, s_{ps}^{1/2}$, with the phase-space suppression factor being 
\begin{align}
s_{ps} \equiv 1-\mF^2/\omega^2 .
\label{eq:PhaseSpaceSuppr}
\end{align}

\paragraph{Parametrisation.}
Throughout, we shall thus assume and use the well-known analytical solutions for the energy eigenstates and eigenvalues in a Coulomb potential (see e.g.~ref.~\cite{Messiah:1962}), which we review in \cref{App:OverlapIntegral}. We discuss the range of validity of the Coulomb approximation in \cref{sec:FreezeOut_CoulombApprox}, in the context of the phenomenological application of \BSFF \ on the DM freeze-out, that we present in \cref{Sec:FreezeOut}. In the Coulomb regime, we need the following two variables to parametrise the cross-sections in a minimal fashion~\cite{Petraki:2015hla},
\begin{align}
\zetaS \equiv \aS / \vrel, \qquad \zetaB \equiv \aB/\vrel. 
\label[pluralequation]{eqs:zetaSzetaB}
\end{align}
Taking the couplings \eqref{eq:alphaSalphaB} into account, $\zetaS$ and $\zetaB$ can be re-expressed in terms of
\begin{align}
\zetaV \equiv \aV / \vrel, \qquad \zetaF \equiv \aF/\vrel. 
\label{eqs:zetaVzetaF}
\end{align}
Outside the Coulomb regime, the wavefunctions can be computed as in ref.~\cite{Petraki:2016cnz} (see~\cite{Harz:2017dlj,Harz:2019rro} for results in a mixed Coulomb and Yukawa potential).

\subsection{Amplitude \label{sec:BSFviaScalarEmission_Amplitude}}
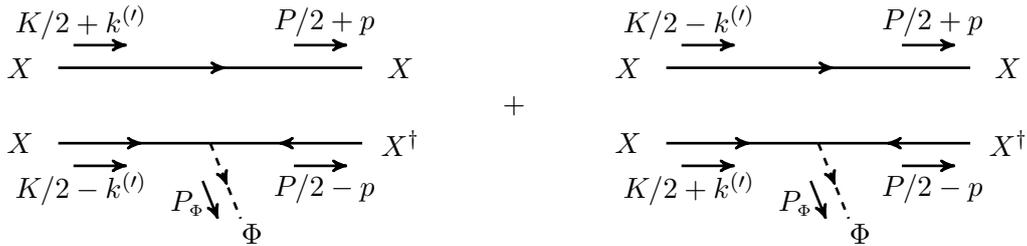
\begin{figure}[t!]
\centering
\begin{tikzpicture}[line width=1pt, scale=1]
\begin{scope}[shift={(-4,0)}]
\node at (-2.5,1){$X$};
\node at (-2.5,0){$X$};
\node at (2.5,1){$X$};
\node at (2.5,0){$X^{\dagger}$};
\node at (-1.7,1.6){$K/2+k^{(\prime)}$};		\node at (1.5,1.6){$P/2+p$};
\draw[momentum] 	(-1.8,1.3) 	-- (-1.1, 1.3);	\draw[momentum] 	(1.1, 1.3) 	-- (1.8, 1.3);
\draw[fermion] 		(-2,  1) 	-- ( 2,   1);
\draw[fermion] 		(-2,  0) 	-- ( 0,   0);	\draw[fermionbar] 	( 0,  0) 	-- ( 2,  0);
\draw[momentum] 	(-1.8,-0.3) -- (-1.1, -0.3);\draw[momentum] 	(1.1,-0.3) 	-- (1.8,-0.3);
\node at (-1.7,-0.6){$K/2-k^{(\prime)}$};		\node at (1.5,-0.6){$P/2-p$};
\node at (0.55,-1.2){$\Phi$};
\draw[scalar]	(0,0) -- (0.4,-1);
\draw[momentum]	(-0.1,-0.5) -- (0.1,-1);
\node at (-0.3,-0.8){$P_{\mathsmaller{\Phi}}$};
\end{scope}
\node at (0,0.5){$+$};
\begin{scope}[shift={(4,0)}]
\node at (-2.5,1){$X$};
\node at (-2.5,0){$X$};
\node at (2.5,1){$X$};
\node at (2.5,0){$X^{\dagger}$};
\node at (-1.7,1.6){$K/2-k^{(\prime)}$};		\node at (1.5,1.6){$P/2+p$};
\draw[momentum] 	(-1.8,1.3) 	-- (-1.1, 1.3);	\draw[momentum] 	(1.1, 1.3) 	-- (1.8, 1.3);
\draw[fermion] 		(-2,  1) 	-- ( 2,   1);
\draw[fermion] 		(-2,  0) 	-- ( 0,   0);	\draw[fermionbar] 	( 0,  0) 	-- ( 2,  0);
\draw[momentum] 	(-1.8,-0.3) -- (-1.1, -0.3);\draw[momentum] 	(1.1,-0.3) 	-- (1.8,-0.3);
\node at (-1.7,-0.6){$K/2+k^{(\prime)}$};		\node at (1.5,-0.6){$P/2-p$};
\node at (0.55,-1.2){$\Phi$};
\draw[scalar]	(0,0) -- (0.4,-1);
\draw[momentum]	(-0.1,-0.5) -- (0.1,-1);
\node at (-0.3,-0.8){$P_{\mathsmaller{\Phi}}$};
\end{scope}
\end{tikzpicture}
\caption[]{
\label{fig:BSF_Scalar_RadiativePart} 
The radiative part of transitions via emission of a charged scalar, $XX \to XX^{\dagger} + \Phi$. The arrows on the field lines denote the flow of the $U(1)_D$ charge.} 
\end{figure}

The amplitude for \BSFF~is~\cite{Petraki:2015hla}
\beq
\im {\cal M}_{{\bf k} \to n\ell m}^{\mathsmaller{\Phi}} 
\simeq 
\int 
\frac{d^3{\bf k'}}{(2\pi)^3}
\frac{d^3{\bf p}}{(2\pi)^3} 
\  \tilde{\phi}_{\bf k}^\XX ({\bf k'})
\ \im \APhi ({\bf k',p}) 
\ \frac{[\tilde{\psi}_{n\ell m}^\XXdagger ({\bf p})]^*}{\sqrt{2\mu}} ,
\label{eq:BSF_ScalarEmission_M_def}
\eeq
where the factor $1/\sqrt{2\mu}$ has arisen in switching from the relativistic to the non-relativistic normalisation for the fields participating in the bound state~\cite{Petraki:2015hla}.  $\APhi$ is the (amputated) amplitude of the radiative part of the process, 
\begin{align}
X (K/2 + k') + X (K/2 - k') \to X (P/2 + p) + X^{\dagger} (P/2 - p) + \Phi (\PF),
\end{align}
where the parentheses denote the momenta of each field; $K$ and $P$ are the total 4-momenta of the $XX$ scattering state and the $XX^{\dagger}$ bound state respectively. The leading order diagrams with the precise momentum assignments are shown in \cref{fig:BSF_Scalar_RadiativePart}. Because these diagrams are not fully connected, the virtuality of the $X,X^{\dagger}$ fields has to be integrated out as described in \cite{Petraki:2015hla} (see~\cite{Harz:2018csl} for a more recent summary). Adapting the result of \cite{Petraki:2015hla}, we find the leading order contributions to be
\begin{align}
\im \APhi ({\bf k',p}) 
&\simeq -\im 2y \, M\mu \, (2\pi)^3 
\[\delta^3 ( {\bf k'} - {\bf p} + \PFvec/2)  +  \delta^3 ({\bf k'} + {\bf p} - \PFvec/2)\] .
\label{eq:BSF_ScalarEmission_AT}
\end{align}
Combining \cref{eq:BSF_ScalarEmission_M_def,eq:BSF_ScalarEmission_AT}, Fourier transforming the wavefunctions, and taking into account that $\psi_{n\ell m} ({\bf -r}) = (-1)^\ell \psi_{n\ell m} ({\bf r})$, we obtain
\begin{align}
\im {\cal M}_{{\bf k} \to n\ell m}^{\mathsmaller{\Phi}} 
\simeq - \im y \, M\sqrt{2\mu} \int d^3{\bf r}
\ \phi_{\bf k}^\XX ({\bf r}) 
\ [\psi_{n\ell m}^\XXdagger ({\bf r})]^*
\[ e^{+\im {\PFvec \cdot {\bf r} /2}} +(-1)^{\ell} \, e^{-\im {\PFvec \cdot {\bf r} /2}} \] .
\label{eq:BSF_ScalarEmission_M}
\end{align}

Amplitudes of this type can be computed by expanding in powers of $\PFvec \cdot {\bf r}/2$~\cite{Petraki:2016cnz}. Indeed, the bound-state wavefunction is exponentially suppressed at $r \gtrsim n/\kappaB$. Moreover, the scattering state wavefunction oscillates at $r>1/k$. Thus the integrand is significant roughly only for $r\lesssim 1/\sqrt{\kappaB^2/n^2+{\bf k}^2}$. Taking \cref{eq:omega} into account, this implies $\PFvec \cdot {\bf r}/2 \lesssim \sqrt{\kappaB^2/n^2+{\bf k}^2}/(4\mu) = \sqrt{\aB^2/n^2 + \vrel^2}/4 \ll 1$.
If the scattering and the bound states were subject to the same potential, the zeroth order term in this expansion would vanish due to the orthogonality of the wavefunctions.\footnote{\label{foot:Cancellation}
For two particles $X_1, X_2$ coupled to a light \emph{neutral} scalar $\varphi$ via 
$\delta {\cal L} = - y_j m_j X_j^\dagger X_j \varphi$, 
the amplitude for the formation of $X_1X_2$ bound states via $\varphi$ emission is~\cite{Petraki:2015hla,Petraki:2016cnz}
\begin{align}
\im {\cal M}_{{\bf k} \to n\ell m} \simeq - \im \, M\sqrt{2\mu} \int d^3{\bf r}
\ \phi_{\bf k} ({\bf r}) \, \psi_{n\ell m}^* ({\bf r})
\[ y_1 e^{-\im \eta_2 \PFvec \cdot {\bf r}} + y_2 e^{\im \eta_1 \PFvec \cdot {\bf r}} \], 
\nn
\end{align}
with  $\eta_{1,2} \equiv m_{1,2}/(m_1+m_2)$. In the $\PFvec \cdot {\bf r}$ expansion, the zeroth order terms vanish due to the orthogonality of the wavefunctions, and for $y_1=y_2$ and $m_1=m_2$ also the first order terms cancel with each other. Then, the dominant contributions arise from the $(\PFvec \cdot {\bf r})^2$ terms (plus corrections of the same order that have been omitted in the above expression). Note that the second cancellation indicates the angular momentum selection rule $\Delta \ell = {\rm even}$.  Thus, for a particle-antiparticle pair or a pair of identical particles, the capture cross-section is suppressed and becomes phenomenologically important mostly for large couplings.}

The essential point of our calculation is that because $\Phi$ carries away charge, the scattering and bound state wavefunctions are governed by different potentials and thus are not orthogonal.\footnote{
While this is inevitable in an Abelian theory, in non-Abelian theories it is possible for a pair of particles to emit a charged boson without changing their combined representation. For example, because ${\rm adj} \otimes {\rm adj} \supset {\rm adj}$, two particles each transforming in the adjoint representation of a group, can begin from a combined adjoint configuration, emit an adjoint boson and end up again in an adjoint combined state.}
%
Therefore, to lowest order, \cref{eq:BSF_ScalarEmission_M} becomes
\begin{align}
\im {\cal M}_{{\bf k} \to n\ell m}^{\mathsmaller{\Phi}} 
&\simeq - \im \, \delta_{\ell,\rm even} \, 2y M\sqrt{2\mu} 
\int d^3{\bf r}
\ \phi_{\bf k}^\XX ({\bf r}) 
\ [\psi_{n\ell m}^\XXdagger ({\bf r})]^*  .
\label{eq:BSF_ScalarEmission_M_LO}
\end{align}

Since the scattering state consists of a pair of identical bosons $XX$, the wavefunction is related to that of two distinguishable particles (DP), $\phi_{{\bf k}}^\DP ({\bf r})$, as follows
\begin{align}
\phi_{\bf k}^\XX ({\bf r}) 
= \frac{\phi_{\bf k}^\DP ({\bf r}) +\phi_{\bf k}^\DP ({\bf-r}) }{\sqrt{2}} 
= \frac{\phi_{\bf k}^\DP ({\bf r}) +\phi_{\bf-k}^\DP ({\bf r}) }{\sqrt{2}}  
= \sqrt{2} \sum_{\ellS = \rm even } \phi_{{\bf k},\ellS}^\DP ({\bf r}) ,
\label{eq:Wavefunctions_Symmetrization}
\end{align}
where $\phi_{{\bf k},\ellS}^\DP ({\bf r})$ denotes the $\ellS$ angular mode of $\phi_{{\bf k}}^\DP ({\bf r})$. We now define the overlap integral of wavefunctions of distinguishable particles\footnote{
In the notation of refs.~\cite{Petraki:2015hla,Petraki:2016cnz}, this is the overlap integral ${\cal I}_{{\bf k}, n\ell m} ({\bf b})$, evaluated at ${\bf b=0}$ and up to the overall constant $\kappaB^{3/2}$, here introduced to make ${\cal R}_{{\bf k},n\ell m}$ dimensionless.}
\begin{align}
{\cal R}_{{\bf k}, n\ell m} \equiv 
\kappaB^{3/2} \, \int d^3 r \ [\psi_{n\ell m}^\DP ({\bf r})]^* \, \phi_{\bf k}^\DP ({\bf r}) ,
\label{eq:Rcal_def}
\end{align}
for scattering and bound states that are subject to the potentials \eqref{eq:VSandVB}. We calculate ${\cal R}_{{\bf k}, n\ell m}$ analytically in \cref{App:OverlapIntegral} (without specifying the couplings $\aS$ and $\aB$). The angular integration in \cref{eq:Rcal_def} imposes the selection rule
\begin{align}
\ellS = \ell .
\label{eq:SelectionRule}
\end{align}
Then, collecting \cref{eq:BSF_ScalarEmission_M_LO,eq:Wavefunctions_Symmetrization,eq:Rcal_def,eq:SelectionRule}, we find
\begin{align}
{\cal M}_{{\bf k} \to n\ell m}^{\mathsmaller{\Phi}} 
&\simeq - \frac{4y}{\aB^{3/2}} \frac{M}{\mu} 
{\cal R}_{{\bf k}, n\ell m}~\delta_{\ell,\rm even} .
\label{eq:BSF_ScalarEmission_M_final}
\end{align}
We will use \cref{eq:BSF_ScalarEmission_M_final} to compute the \BSFF \ cross-section in \cref{sec:BSFviaScalarEmission_CrossSection}. Before doing so, some clarifications are important.

For $\ell=$~odd, we must keep the first order terms in the $(\PFvec \cdot {\bf r})$ expansion of the integrand of \cref{eq:BSF_ScalarEmission_M}. Then the amplitude becomes non-vanishing for $\ell=$~odd, and is proportional to the overlap integral $\int d^3{\bf r} \( \PFvec \cdot {\bf r}\) \phi_{\bf k}^\XX ({\bf r}) \ [\psi_{n\ell m}^\XXdagger ({\bf r})]^*$, which imposes the selection rule $|\ell-\ellS|=1$. As per \cref{eq:Wavefunctions_Symmetrization}, the scattering state wavefunction contains only modes $\ellS=$~even. Thus, this contribution survives, and yields a cross-section that is larger than BSF via emission of a neutral scalar (cf.~\cref{foot:Cancellation}) and of the equivalent order as BSF via emission of a vector boson.\footnote{
For fermionic DM, we find that the $(\PFvec \cdot {\bf r})^0$ contributions would survive for $\ell+s+1=$~even, with $s=$0~or~1 being the total spin. However, the $XX$ wavefunctions contain only $\ellS+s+1=$~odd modes. Given the $\ellS=\ell$ selection rule \eqref{eq:SelectionRule}, these contributions cancel. The $(\PFvec \cdot {\bf r})^1$ contributions have the same fate: they would survive for $\ell+s+1=$~odd, however the selection rule now becomes $|\ell-\ellS|=1$. Since $\ellS+s+1=$~odd, these contributions cancel as well.}

Nevertheless, the cancellation of the zeroth order terms in $(\PFvec \cdot {\bf r})$ for odd $\ell$ is a particularity of the model we are considering here, rather than a generic feature of BSF via charged scalar emission. For example, in a (coannihilation) scenario where the two incoming particles transform under different representations of the underlying symmetry (i.e.~in the $U(1)$ case, they have different charges), there in no generic cancellation between the contributing diagrams. Thus, to remain focused on the main point of our computation, we shall consider only the lowest order contributions given by \cref{eq:BSF_ScalarEmission_M_LO}.

The result \eqref{eq:BSF_ScalarEmission_M_final} should also make evident that for \BSFF, the diagrams in which the final-state $\Phi$ is produced from the fusion of off-shell $V$ and $\Phi$ emitted by the incoming $XX$ pair, are subleading. Diagrams where the radiated boson is emitted from an off-shell propagator exchanged between the two interacting particles, are known to give leading-order contributions to \BSFV~\cite{Asadi:2016ybp}. However, as shown here, the diagrams of \cref{fig:BSF_Scalar_RadiativePart} for \BSFF \ yield lower order contributions than the corresponding diagrams for \BSFV, where the vector-emission vertices introduce a momentum suppression in the wavefuntion overlap integral and thus in the amplitude (see \cref{App:BSFV} for more details). Thus, with respect to the diagrams of \cref{fig:BSF_Scalar_RadiativePart}, the $\Phi$ emission from $V\Phi$ fusion must be subleading; in fact, it turns out to be even of higher order than the corresponding diagrams in \BSFV, due to the different Lorentz structure of the vertices involved.

\subsection{Cross-Section \label{sec:BSFviaScalarEmission_CrossSection}}

The cross-section times relative velocity for the \BSFF~processes \eqref{eq:BSF_ScalarEmission} is
\beq
\sigma_{n\ell m}^{\mathsmaller{\Phi}} \vrel  
= \frac{|\PFvec|}{2^6 \pi^2 M^2\mu} \, \int d\Omega \, 
\left| {\cal M}_{{\bf k} \to n\ell m}^{\mathsmaller{\Phi}} \right|^2 ,
\label{eq:sigma2to2_general}
\eeq
where the momentum of the emitted scalar $\PFvec$ is given by \cref{eq:omega}.

Collecting \cref{eq:sigma2to2_general,eq:omega,eq:BSF_ScalarEmission_M_final,eq:RcalSquared}, we find for the capture cross-section,
\begin{align}
\sigma_{n\ell}^{\mathsmaller{\Phi}} \vrel \equiv
\sum_{m=-\ell}^\ell \sigma_{n\ell m}^{\mathsmaller{\Phi}} \vrel ,
\end{align}
the following
\begin{empheq}[box=\widefbox]{align}
\sigma_{n\ell}^{\mathsmaller{\Phi}} \vrel
&\simeq
s_{ps}^{1/2} \, \frac{\pi}{\mu^2} 
\frac{\aF}{\aB}\(1-\frac{\aS}{\aB}\)^2
~\delta_{\ell,\rm even}
~\frac{2^{4\ell+9} \, n^2}{2\ell+1} \frac{(n+\ell)!}{(n-\ell-1)!} \[\frac{\ell!}{(2\ell)!}\]^2
\nn \\  
&\times
S_\ell (\zetaS)
\[ \frac{(\zetaB^2/n^2)^{\ell+3}}{(1+\zetaB^2/n^2)^{2\ell+3}} \]
e^{-4\zetaS \, {\rm arccot} (\zetaB/n)}
\label{eq:sigmav_BSF_ScalarEmission}
\\  
&\times
\left|
{}_2F_1 \( 
1+\ell-n; ~ 
1+\ell +\im \zetaS; ~
2\ell+2; ~
\frac{4 \im \zetaB/n}{(1+\im \zetaB/n)^2}
\)
\right|^2 , 
\nn 
\end{empheq}
where $S_\ell (\zetaS)$ is the Sommerfeld factor for $\ell$-wave processes~\cite{Cassel:2009wt}, 
\beq
S_\ell(\zetaS) =
\frac{2\pi\zetaS}{1-e^{-2\pi\zetaS}}~\prod_{j=1}^{\ell} \(1+\frac{\zetaS^2}{j^2}\) ,
\label{eq:S_ell}
\eeq
and in our model
\begin{subequations}
\label{eq:aSaB}
\label[pluralequation]{eqs:aSaB}
\begin{align}
\aS&=-\aV,			&	\zetaS&\equiv \aS/\vrel = -\zetaV ,
\label{eq:aSzetaS}
\\
\aB &=+\aV + (-1)^{\ell} \aF,	&	\zetaB&\equiv \aB/\vrel = +\zetaV +  (-1)^{\ell} \zetaF ,
\label{eq:aBzetaB}
\end{align}
\end{subequations}
with $\aV$ and $\aF$ defined in \cref{eq:alphas_definitions}. ${}_2F_1$ is the (ordinary) hypergeometric function, and $s_{ps}$ is the phase-space suppression factor defined in \cref{eq:PhaseSpaceSuppr}.

\begin{figure}[t!]
\centering	
\includegraphics[width=0.95\textwidth]{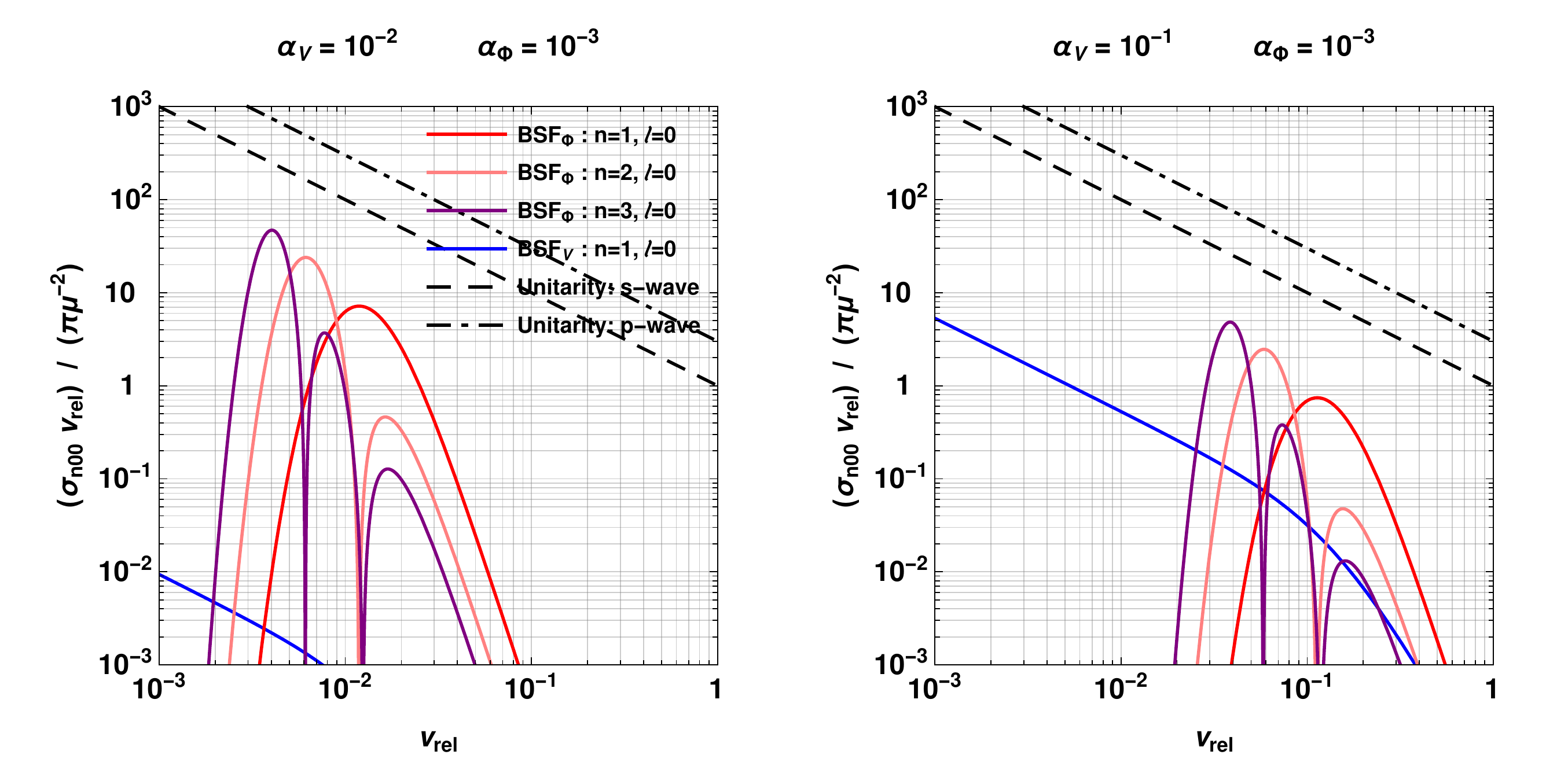}
\caption[]{\label{fig:CrossSections_BSF_ellZeroStates}
The velocity-weighted cross-section for capture into zero-angular momentum $XX^{\dagger}$ bound states, $n00$: $XX \to {\cal B}_{n00} (XX^{\dagger}) + \Phi$ for $n=1,2,3$, and  $XX^{\dagger} \to {\cal B}_{100} (XX^{\dagger}) + V$. For $V$ emission, the capture into $n>1$ states is subdominant to capture into $n=1$~\cite{Petraki:2016cnz}, and we do not show them here. Also shown are the $s$- and $p$-wave unitarity limits on inelastic cross-sections; for capture into $n00$,  \BSFF~is $s$-wave, while \BSFV~is $p$-wave.  All cross-sections have been normalised to $\pi/\mu^2$, with $\mu$ being the reduced mass of the interacting particles, and we have neglected any phase-space suppression due to the mass of $\Phi$.}

\bigskip

\includegraphics[width=0.95\textwidth]{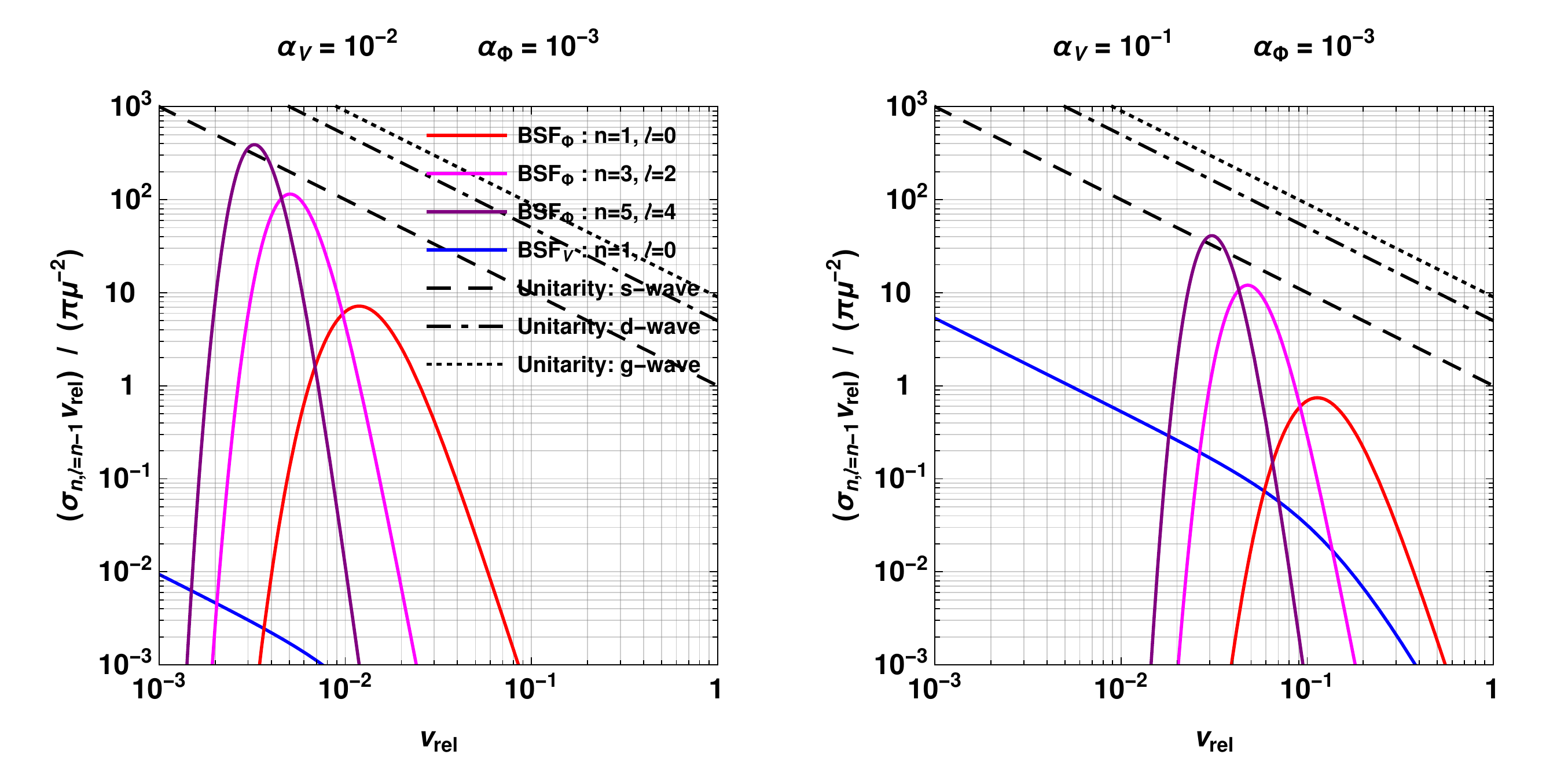}
\caption{\label{fig:CrossSections_BSF_ellEQUALSnminus1}
Same as in \cref{fig:CrossSections_BSF_ellZeroStates}, for capture into bound levels with $\ell=n-1$, which have the highest multiplicity for a given $n$. We also show the respective unitarity limits for each process.}
\end{figure}

The cross-sections \eqref{eq:sigmav_BSF_ScalarEmission} is the main result of this section. They are readily generalisable to unbroken perturbative non-Abelian theories: as in ref.~\cite{Harz:2018csl}, the appropriate colour factors arising in the amplitudes \eqref{eq:BSF_ScalarEmission_M_final} upon projection of the initial and final states onto states of definite colour must be included, $\aS$ and $\aB$ have to be chosen according to the initial and final colour representations~\cite{Kats:2009bv}, and the (anti)symmetrisation of the wavefunctions in the case of identical particles must be taken into account. Considering the couplings~\eqref{eqs:aSaB}, we illustrate \cref{eq:sigmav_BSF_ScalarEmission} in  \cref{fig:CrossSections_BSF_ellZeroStates,fig:CrossSections_BSF_ellEQUALSnminus1}, and compare \BSFF \ with \BSFV~\cite{Petraki:2015hla,Petraki:2016cnz}.

A few remarks are in order.
\begin{itemize}
\item
The computed contribution \eqref{eq:sigmav_BSF_ScalarEmission} to the BSF cross-section vanishes if $\aB=\aS$, as expected from the orthogonality of the wavefunctions.
\item
The hypergeometric function in \cref{eq:sigmav_BSF_ScalarEmission} is a finite polynomial in its last argument (which can be also cast as $1+e^{\im 4{\rm arccot}(\zetaB/n)}$) because its first argument is a non-positive integer, $1+\ell-n \leqslant 0$. For $\ell = n-1$, this factor reduces to 1. For an arbitrary $\ell$, it tends to 1 both at large and at small velocities ($\zetaB,|\zetaS| \ll 1$ and $\zetaB,|\zetaS| \gg 1$). At intermediate velocities, it gives rise to cancellations, as seen for example in \cref{fig:CrossSections_BSF_ellZeroStates}. 
\item
At large velocities, $\vrel \gg \aB/n = (\aV+\aF)/n$ (i.e.~$\zetaB/n \ll 1$), the overlap of the scattering and bound state wavefunctions is small, as seen from the term inside the square brackets in the second line of \cref{eq:sigmav_BSF_ScalarEmission}. The \BSFF~cross-sections are suppressed by $\zetaB^{2(\ell+3)} \ll 1$. 
\item
At low velocities, $\vrel \lesssim |\aS| = \aV$ (i.e.~$|\zetaS| \gtrsim 1$), the \BSFF~cross-sections become suppressed due to the repulsion in the scattering state, by the Sommerfeld factor $S_\ell (\zetaS)$, with $\zetaS = -\zetaV <0$. The suppression becomes exponential at very low velocities $\zetaS \ll - 1$.
\item
In between, the \BSFF~cross-sections become very significant. While the velocity at which they peak depends on $n$, $\ell$ and the ratio $\aS/\aB$, it can be roughly approximated by $\vrel \sim \aB/n = (\aV+\aF)/n$, as seen in \cref{fig:CrossSections_BSF_ellZeroStates,fig:CrossSections_BSF_ellEQUALSnminus1}.
\item
For $\zetaB/n = \aB/(n\vrel) \gtrsim 1$, the factor in \cref{eq:sigmav_BSF_ScalarEmission} next to $S_\ell(\zetaS)$ yields the characteristic behaviour $\vrel^{2\ell}$ of $\ell$-wave processes without Sommerfeld. Combined with $S_\ell(\zetaS)$, we see that the velocity suppression of higher partial waves disappears, and all partial waves exhibit the velocity dependence of $S_0$.
\end{itemize}

Let us also now examine two different limits of \cref{eq:sigmav_BSF_ScalarEmission}.

\paragraph{Limit $\boldsymbol{\aV \to 0}$.}
In the limit of vanishing gauge coupling, the symmetry that is responsible for the dynamics we are considering becomes essentially global. This limit can be also effectively attained if the gauge boson has a very large mass $\mV \gg \mu \aV$ such that it mediates a contact type interaction. In this case,  $V$ decouples from the  low-energy effective theory, leaving a remnant unbroken global symmetry. In this regime, $\aB\to\aF,~\zetaB \to \zetaF$ and $\aS \to 0,~\zetaS \to 0$. Then, \cref{eq:sigmav_BSF_ScalarEmission} becomes
\begin{align}
\sigma_{n\ell}^{\mathsmaller{\Phi}} \vrel
&\simeq
s_{ps}^{1/2}
~\frac{\pi}{\mu^2} 
~\delta_{\ell,\rm even}
~\frac{2^{4\ell+9} n^2}{2\ell+1} \frac{(n+\ell)!}{(n-\ell-1)!} \[\frac{\ell!}{(2\ell)!}\]^2
\nn \\ &\times 
\[ \frac{(\zetaF^2/n^2)^{\ell+3}}{(1+\zetaF^2/n^2)^{2\ell+3}} \]
\times
\left|
{}_2F_1 \( 
1+\ell-n; ~ 
1+\ell; ~
2\ell+2; ~
\frac{4 \im \zetaF/n}{(1+\im \zetaF/n)^2}
\)
\right|^2 .
\label{eq:sigmav_BSF_ScalarEmission_gaugeZero}
\end{align}
Since there is now no repulsion in the scattering state, this cross-section is not exponentially suppressed at very low velocities. Nevertheless, because there is also no long-range attraction in the incoming state, the cross-section scales as $\sigma_{n\ell}^{\mathsmaller{\Phi}} \vrel \propto \vrel^{2\ell}$ at $\vrel \ll \aB/n$.

\paragraph{Limit $\boldsymbol{\aV \gg \aF}$.}\label{sec:} 
In this regime, $\aB \to \aV,~\zetaB\to\zetaV$. As always $\aS = -\aV,~\zetaS=-\zetaV$. Then, \cref{eq:sigmav_BSF_ScalarEmission} becomes
\begin{align}
\sigma_{n\ell}^{\mathsmaller{\Phi}} \vrel
&\simeq
s_{ps}^{1/2}
~\frac{\pi (\aF/\aV)}{\mu^2} 
~\delta_{\ell,\rm even}
~\frac{2^{4\ell+11} n^2}{2\ell+1} \frac{(n+\ell)!}{(n-\ell-1)!} \[\frac{\ell!}{(2\ell)!}\]^2
~S_\ell (-\zetaV)
\[ \frac{(\zetaV^2/n^2)^{\ell+3}}{(1+\zetaV^2/n^2)^{2\ell+3}} \]
\nn \\ &\times 
e^{4\zetaV \, {\rm arccot} (\zetaV/n)}
\left|
{}_2F_1 \( 
1+\ell-n; ~ 
1+\ell - \im \zetaV; ~
2\ell+2; ~
\frac{4 \im \zetaV/n}{(1+\im \zetaV/n)^2}
\)
\right|^2 .
\label{eq:sigmav_BSF_ScalarEmission_gaugeLarge}
\end{align}
Despite the very small $\aF$, this can exceed the \BSFV~cross-section for a significant velocity range, as seen in \cref{fig:CrossSections_BSF_ellZeroStates,fig:CrossSections_BSF_ellEQUALSnminus1}.

\subsubsection*{Capture via scattering}

While in this work we focus on radiative BSF, it is possible that the dissipation of energy necessary for the capture into bound states occurs via scattering on other particles through exchange of an off-shell mediator, if the mediator couples also to other light degrees of freedom. 
Although of higher order, such processes can be extremely efficient inside a relativistic thermal bath, where the density of the light particles is very high, as was recently shown in \cite{Binder:2019erp} and previously suggested in~\cite{Kim:2016zyy,Biondini:2017ufr,Biondini:2018xor}. 

Reference~\cite{Binder:2019erp} found that the rate of capture via scattering factorises into the radiative cross-section (albeit without any phase-space suppression due to the mass of the emitted scalar), times a part that includes the kinematics and dynamics of the bath particles. The cross-sections \eqref{eq:sigmav_BSF_ScalarEmission} can then be recast to calculate BSF via off-shell exchange of a charged scalar. A corollary of this and the largeness of the radiative cross-sections \eqref{eq:sigmav_BSF_ScalarEmission} is that the corresponding bath scattering processes must also be very significant in the early universe in the presence of light relativistic particles coupled to the charged scalar. In the case of an unbroken or mildly broken gauge symmetry, the gauge bosons and charged scalars already provide the relativistic bath necessary for such scattering processes to occur, $XX +V \to {\cal B}(XX^\dagger) + \Phi$ and $XX + \Phi^\dagger \to {\cal B}(XX^\dagger) + V$. 

In contrast to the radiative capture, BSF via bath scattering is not kinematically blocked if the mediator mass is larger than the energy available to be dissipated [cf.~\cref{eq:omega}]. This implies that bound-state effects are not only enhanced, but also relevant to a broader parameter space.

\subsection{Partial-wave unitarity \label{sec:BSFviaScalarEmission_Unitarity}}

The unitarity of the $S$ matrix implies an upper bound on the partial wave inelastic cross-section~\cite{Griest:1989wd}
\begin{align}
\sigma_{\rm inel}^{(J)} \vrel^{} \leqslant 
\sigma_{\uni}^{(J)} \vrel^{} = \frac{\pi (2J+1)}{\mu^2 \vrel} ,
\label{eq:UnitarityLimit}
\end{align}
where $J$ denotes the partial wave of the scattering state wavefunction that participates in the process. For \BSFF, this is the same as that of the bound state formed, thus we consider the ratio
\begin{align}
\sigma_{n\ell}^{\mathsmaller{\Phi}}  / \sigma_\uni^{(\ell)}
= \aF \times
\tilde{f}_{n\ell} (\zetaS,\zetaB) 
\, \delta_{\ell,\rm even} ,
\label{eq:sigmaRatio_BSFoverUNI}
\end{align}
with
\begin{align}
\tilde{f}_{n\ell}  (\zetaS,\zetaB) &\equiv
\frac{2^{4\ell+9} \, n^2~(n+\ell)!}{(n-\ell-1)!} 
\[\frac{\ell!}{(2\ell+1)!}\]^2
\(1-\frac{\zetaS}{\zetaB}\)^2
\frac{S_\ell (\zetaS)}{\zetaB}
\[ \frac{(\zetaB^2/n^2)^{\ell+3}}{(1+\zetaB^2/n^2)^{2\ell+3}} \]
\nn \\  &\times
e^{-4\zetaS \, {\rm arccot} (\zetaB/n)}
\left|
{}_2F_1 \( 
1+\ell-n; ~ 
1+\ell +\im \zetaS; ~
2\ell+2; ~
\frac{4 \im \zetaB/n}{(1+\im \zetaB/n)^2}
\)
\right|^2 ,
\label{eq:sigmaRatio_BSFoverUNI_f}
\end{align}
where we have neglected the phase-space suppression factor $s_{ps}^{1/2}$. As $\vrel$ varies, $\zetaS$ and $\zetaB$ scan a range of values, but the ratio $r\equiv \zetaS / \zetaB$ of course remains constant (neglecting the possible running of the coupling, which may render $\aS$ mildly dependent on $\vrel$). Unitarity must be respected for all $\vrel$. We thus define
\begin{align}
f_{n\ell;r} (\zetaB) \equiv \tilde{f}_{n\ell} (r \, \zetaB, \zetaB) .
\label{eq:sigmaRatio_BSFoverUNI_ftilde}
\end{align}
Then \cref{eq:sigmaRatio_BSFoverUNI} implies that unitarity is respected provided that $\aF$ is sufficiently small, 
\begin{equation}
\label{eq:alpha_Phi_max}
\aF < 1/\max[f_{nl;r}(\zetaB)] ,
\end{equation}
where $f_{nl;r}$ is maximized with respect to $\zetaB$. (Note that $\zetaB>0$ always for bound states to exist.)

\begin{figure}[t!]
\centering	
\includegraphics[height=0.43\textwidth]{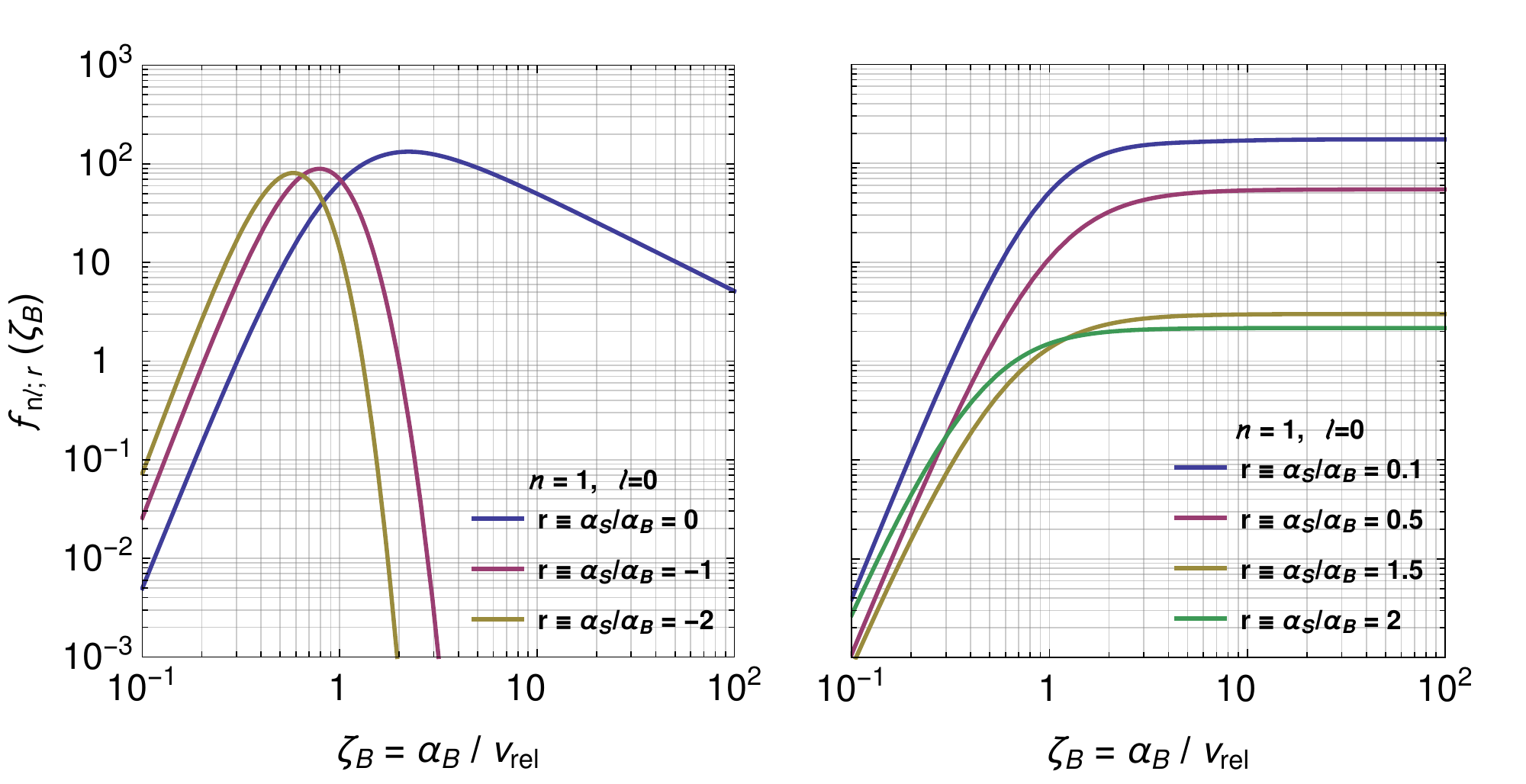}
\caption[]{
\label{fig:Unitarity_sigmaRatio_BSFoverUNI}  
$f_{n\ell;r} \equiv \aF^{-1} ( \sigma_{n\ell}^{\mathsmaller{\Phi}} / \sigma_{\uni}^{(\ell)} )$ vs $\zetaB \equiv \aB/\vrel$, for  $n=1,\ell=0$ and various values of $r\equiv \aS/\aB$. The coupling $\aF$ must be sufficiently small, $\aF < \max (f_{n\ell;r})$, such that unitarity is respected for all velocities. This is possible because $f_{n\ell;r}$ is bounded from above.}

\bigskip

\includegraphics[height=0.43\textwidth]{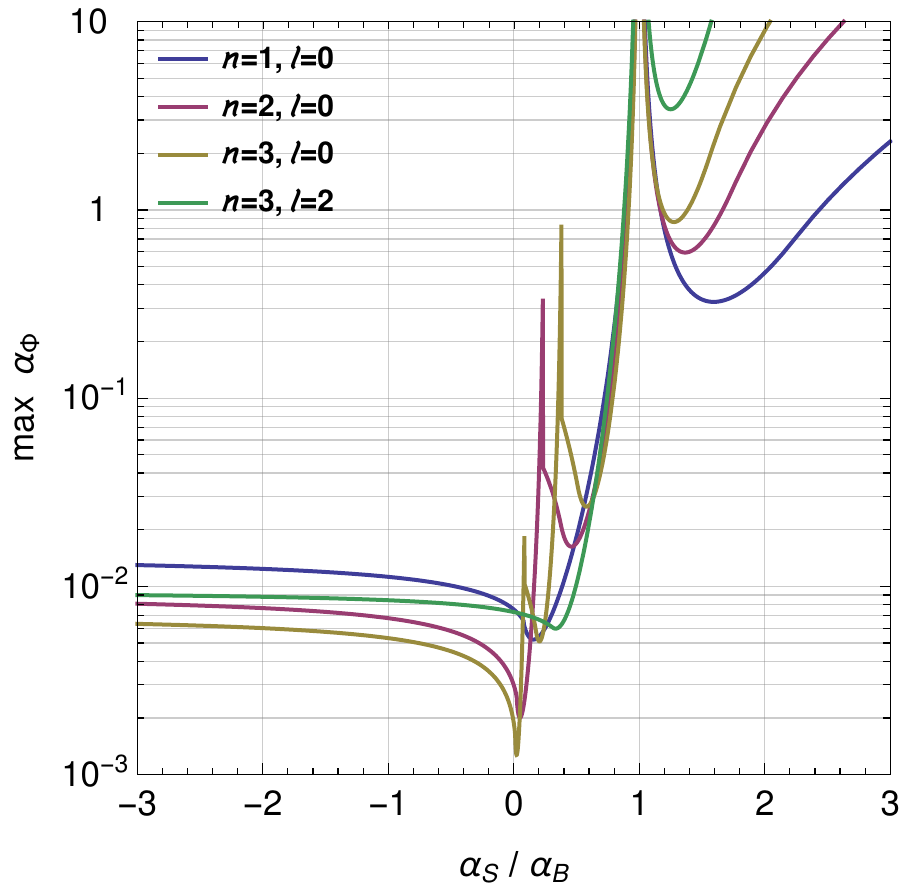}
\caption[]{\label{fig:Unitarity_alphaPhiMax}  
The maximum value of $\aF$ vs.~$r\equiv \aS/\aB$, for which the \BSFF~cross-sections~\eqref{eq:sigmav_BSF_ScalarEmission} remain below the unitarity limit for all velocities. In the model considered in this work $r = -\aV/(\aV+\aF)$, thus $-1 \leqslant r \leqslant 0$.}
\end{figure}

In \cref{fig:Unitarity_sigmaRatio_BSFoverUNI}, we present $f_{n\ell;r} (\zetaB)$ vs.~$\zetaB$ for $n=1,\ell=0$ and various values of $r$.  In the present model, $r=-\aV/(\aV+\aF) \in [-1,0]$. However, in any model where transitions of the type considered here occur, the BSF amplitudes will be proportional to the overlap integrals \eqref{eq:Rcal_def}, and the cross-sections will be similar to \cref{eq:sigmav_BSF_ScalarEmission}, up to a possible numerical factor. Thus, to get a broader insight into the implications of unitarity, in \cref{fig:Unitarity_sigmaRatio_BSFoverUNI} we consider a wider range of $r$ values. As seen, $f_{n\ell;r}$ is bounded from above; this remains true for all $n,\ell$. It is then indeed possible to find a maximum value for $\aF$, below which our calculation is consistent with unitarity, but above which it evidently fails. We determine this numerically and present it in \cref{fig:Unitarity_alphaPhiMax}. Notably, for $\aS/\aB <0$, our computation fails already at rather small values of $\aF$. This is a consequence of the very large overlap between the initial and final states. The high peak of the \BSFF~cross-sections at $\vrel \sim \aB/n$, explained in \cref{sec:BSFviaScalarEmission_CrossSection}, results in a rather stringent upper bound on $\aF$.

What is the underlying reason for this apparent violation of unitarity, and how can unitarity be restored in the computation of the \BSFF~cross-sections? At such low values of $\aF$ it is unlikely that higher order corrections to the perturbative  part of the amplitude $\APhi$ [cf.~\cref{eq:BSF_ScalarEmission_AT}] may have any significant effect on the cross-section. Moreover, it has been pointed out that the breakdown of unitarity in perturbative calculations at low energies suggests that the two-particle interactions at infinity must be resummed~\cite[section~5]{Baldes:2017gzw}. In our computation, this has been done at leading order, by the resummation of the one-boson exchange diagrams of \cref{fig:2PI} that give rise to the potentials~\eqref{eqs:V}. However, by the optical theorem, all the inelastic processes to which the two interacting particles may participate also contribute to the self-energy of this two-particle state. Such contributions are typically neglected because they are of higher order than the one-boson exchange diagrams, and give rise to shorter-range (or contact) potentials that may have only limited impact on the large-distance behaviour of the wavefunctions. Still, the fact that the \BSFF~cross-sections can become so large suggests that their contribution to the two-particle self-energy may be significant, thus it must be resummed. The effect of this resummation will likely be significant mostly for incoming momenta around the peak of the \BSFF~cross-sections,  $k_{\rm peak}$. While the corrected \BSFF~cross-sections should be consistent with unitarity, we expect them to remain very significant for $k \sim k_{\rm peak}$, and essentially unaffected for $k \gg k_{\rm peak}$ and $k \ll k_{\rm peak}$. Therefore, we still expect significant phenomenological implications. We leave this computation for future work.

\section{Freeze-out of thermal-relic dark matter \label{Sec:FreezeOut}}

To showcase the phenomenological applications of the above, we consider the effect of \BSFF \ on the density of thermal-relic DM. Below, we list the pertinent cross-sections and rates, and present the Boltzmann equations that govern the evolution of the unbound and bound DM particle densities. We then describe how freeze-out is modified due to \BSFF, and compute the couplings that reproduce the observed DM density. For simplicity, we assume that the DM particles and the radiation to which they couple are at the same temperature as the SM plasma, and use the standard time parameter
\begin{align}
x=\mX/T . 
\label{eq:x=mX/T}
\end{align} 
The generalisation to different dark sector and SM temperatures is straightforward, see e.g.~\cite{Baldes:2017gzu}. 

\subsection{Interaction rates}

\subsubsection{Annihilation  \label{sec:FreezeOut_ANN}}
\begin{subfigures}
\label{fig:Annihilation}
\label[pluralfigure]{figs:Annihilation}
\begin{figure}[t!]
\centering
\begin{tikzpicture}[line width=1pt, scale=0.9]
\begin{scope}[shift={(0,2)}]
\node at (-8,0.5){$XX^\dagger \to VV:$};
\begin{scope}[shift={(-4,0)}]
\node at (-1.3,1){$X$};
\node at (-1.3,0){$X^{\dagger}$};
\draw[fermion]	(-1,1) -- (0,1);
\draw[fermion]	(0,1) -- (0,0);
\draw[fermion]	(0,0) -- (-1,0);	
\draw[vector] 	(0,1) -- (1,1);
\draw[vector] 	(0,0) -- (1,0);	
\node at (1.3,1){$V$};
\node at (1.3,0){$V$};
\end{scope}
\node at (-2,0.5) {$+$};
\begin{scope}[shift={(0,0)}]
\node at (-1.3,1){$X$};
\node at (-1.3,0){$X^{\dagger}$};
\draw[fermion]	(-1,1) -- (0,1);
\draw[fermion]	(0,1) -- (0,0);
\draw[fermion]	(0,0) -- (-1,0);	
\draw[vector] 	(0,1) -- (1,0);
\draw[vector] 	(0,0) -- (1,1);	
\node at (1.3,1){$V$};
\node at (1.3,0){$V$};
\end{scope}
\node at (2,0.5) {$+$};
\begin{scope}[shift={(3.5,0)}]
\node at (-0.8,1){$X$};
\node at (-0.8,0){$X^{\dagger}$};
\draw[fermion]	(-0.5,1) -- (0,0.5);
\draw[fermion]	(0,0.5) -- (-0.5,0);	
\draw[vector] 	(0,0.5) -- (0.5,1);
\draw[vector] 	(0,0.5) -- (0.5,0);	
\node at (0.8,1){$V$};
\node at (0.8,0){$V$};
\end{scope}
\end{scope}
\begin{scope}[shift={(0,0)}]
\node at (-8,0.5){$XX^\dagger \to \Phi\Phi^\dagger:$};
\begin{scope}[shift={(-4,0)}]
\node at (-1.3,1){$X$};
\node at (-1.3,0){$X^{\dagger}$};
\draw[fermion]	(-1,1) -- (0,1);
\draw[fermionbar]	(0,1) -- (0,0);
\draw[fermion]	(0,0) -- (-1,0);	
\draw[scalar] 	(0,1) -- (1,1);
\draw[scalarbar] 	(0,0) -- (1,0);	
\node at (1.3,1){$\Phi$};
\node at (1.4,0){$\Phi^\dagger$};
\end{scope}
\node at (-2,0.5) {$+$};
\begin{scope}[shift={(0,0)}]
\node at (-1.3,1){$X$};
\node at (-1.3,0){$X^{\dagger}$};
\draw[fermion]	(-1,1) -- (-0.4,0.5);
\draw[fermion]	(-0.4,0.5) -- (-1,0);	
\draw[vector] 	(-0.4,0.5) -- (0.4,0.5);
\draw[scalar] 	(0.4,0.5) -- (1,1);
\draw[scalarbar] 	(0.4,0.5) -- (1,0);	
\node at (1.3,1){$\Phi$};
\node at (1.4,0){$\Phi^\dagger$};
\end{scope}
\node at (2,0.5) {$+$};
\begin{scope}[shift={(3.5,0)}]
\node at (-0.8,1){$X$};
\node at (-0.8,0){$X^{\dagger}$};
\draw[fermion]	(-0.5,1) -- (0,0.5);
\draw[fermion]	(0,0.5) -- (-0.5,0);	
\draw[scalar] 	(0,0.5) -- (0.5,1);
\draw[scalarbar](0,0.5) -- (0.5,0);	
\node at (0.8,1){$\Phi$};
\node at (0.9,0){$\Phi^\dagger$};
\end{scope}
\end{scope}
\begin{scope}[shift={(0,-2)}]
\node at (-8,0.5){$XX \to V\Phi:$};
\begin{scope}[shift={(-4,0)}]
\node at (-1.3,1){$X$};
\node at (-1.3,0){$X$};
\draw[fermion]	(-1,1) -- (0,1);
\draw[fermion]	( 0,1) -- (0,0);
\draw[fermion]	(-1,0) -- (0,0);
\draw[vector] 	(0,1) -- (1,1);		
\draw[scalar] 	(0,0) -- (1,0);
\node at (1.3,1){$V$};
\node at (1.3,0){$\Phi$};
\end{scope}
\node at (-2,0.5) {$+$};
\begin{scope}[shift={(0,0)}]
\node at (-1.3,1){$X$};
\node at (-1.3,0){$X$};
\draw[fermion]	(-1,0) -- (-0.5,0.5);\draw[fermionnoarrow]	(-0.5,0.5) -- (0,1);
\draw[fermion]	( 0,1) -- (0,0);
\draw[fermion]	(-1,1) -- (-0.5,0.5);\draw[fermionnoarrow]	(-0.5,0.5) -- (0,0);	
\draw[vector] 	(0,1) -- (1,1);		
\draw[scalar] 	(0,0) -- (1,0);
\node at (1.3,1){$V$};
\node at (1.3,0){$\Phi$};
\end{scope}
\node at (2,0.5) {$+$};
\begin{scope}[shift={(4,0)}]
\node at (-1.3,1){$X$};
\node at (-1.3,0){$X$};
\draw[fermion]	(-1,1) -- (-0.4,0.5);
\draw[fermion]	(-1,0) -- (-0.4,0.5);	
\draw[scalar] 	(-0.4,0.5) -- (0.4,0.5);
\draw[vector] 	(0.4,0.5) -- (1,1);	
\draw[scalar] 	(0.4,0.5) -- (1,0);
\node at (1.3,1){$V$};
\node at (1.3,0){$\Phi$};
\end{scope}
\end{scope}
\end{tikzpicture}
\caption[]{\label{fig:Ann_Tree} 
The tree level diagrams contributing to the annihilation of $XX^{\dagger}$, $XX$ and $X^{\dagger} X^{\dagger}$ pairs. Note that for the $XX^{\dagger} \to VV$ and  $XX \to \Phi V$ annihilation processes, there are both $t$- and $u$-channel diagrams. In contrast, there is no $u$-channel diagram for $XX^{\dagger} \to \Phi\Phi^\dagger$. The arrows denote the flow of the $U(1)_D$ charge.}
\bigskip\bigskip
\begin{tikzpicture}[line width=1pt, scale=0.9]
\begin{scope}[shift={(-4,0)}]
\node at (-4.3,1){$X$};
\node at (-4.3,0){$X^{\dagger}$};
\draw[fermion]		(-4,1) -- (-3.4,1);
\draw[fermionbar]	(-4,0) -- (-3.4,0);
\draw (-4,1) -- (0,1);
\draw (-4,0) -- (0,0);
\draw (0,1) -- (1,0);\draw[fermion] 	(0,1) -- (0.5,0.5);
\draw (0,0) -- (1,1);\draw[fermionbar] 	(0,0) -- (0.5,0.5);
%
\draw[fill=lightblue,shift={(-3,0.5)}] (-0.5,-0.5) rectangle (0.5,0.5);
\draw[fill=lightblue,shift={(-1,0.5)}] (-0.5,-0.5) rectangle (0.5,0.5);
\node at (-3,0.5){${\cal A}_{\mathsmaller{XX^{\dagger}}}^{\twoPI}$};
\node at (-1,0.5){${\cal A}_{\mathsmaller{XX^{\dagger}}}^{\twoPI}$};
\node at (-2,0.5){$\cdots$};
\draw[fill=black] (0.5,0.5) circle  (0.25);
\end{scope}
\begin{scope}[shift={(4,0)}]
\node at (-4.3,1){$X$};
\node at (-4.3,0){$X$};
\draw[fermion]	(-4,1) -- (-3.4,1);
\draw[fermion]	(-4,0) -- (-3.4,0);
\draw (-4,1) -- (0,1);
\draw (-4,0) -- (0,0);
\draw (0,1) -- (1,0);\draw[fermion]	(0,1) -- (0.5,0.5);
\draw (0,0) -- (1,1);\draw[fermion]	(0,0) -- (0.5,0.5);
%
\draw[fill=lightred,shift={(-3,0.5)}] (-0.5,-0.5) rectangle (0.5,0.5);
\draw[fill=lightred,shift={(-1,0.5)}] (-0.5,-0.5) rectangle (0.5,0.5);
\node at (-3,0.5){${\cal A}_{\mathsmaller{XX}}^{\twoPI}$};
\node at (-1,0.5){${\cal A}_{\mathsmaller{XX}}^{\twoPI}$};
\node at (-2,0.5){$\cdots$};
\draw[fill=black] (0.5,0.5) circle  (0.25);
\end{scope}
\end{tikzpicture}
\caption[]{\label{fig:Ann_SE}
The long-range interaction affects the rate of the annihilation processes, and necessitates the resummation of the 2PI interactions at infinity. The 2PI kernels are shown in \cref{fig:2PI}. The black blob stands for the tree-level annihilation diagrams of \cref{fig:Ann_Tree}.	 
}
\end{figure}
\end{subfigures}

The tree-level annihilation channels for $XX$, $X^{\dagger} X^{\dagger}$ and $XX^{\dagger}$ pairs are shown in \cref{fig:Ann_Tree}. The annihilation processes are affected by the Sommerfeld effect as depicted in \cref{fig:Ann_SE}. We consider only $s$-wave contributions, at leading order in each coupling and zeroth order in $\vrel$. The full velocity-weighted cross-sections are
\begin{subequations}
\label{eq:sigma_ANN}
\label[pluralequation]{eqs:sigma_ANN}
\begin{align}
\sigma_{\mathsmaller{XX^{\dagger}\to VV}} \,\vrel 
&\simeq 
\frac{2\pi \aV^2}{\mX^2} \ S_0 (\zetaV+\zetaF) ,
\label{eq:sigmav_XXdaggerToVV}
\\
\sigma_{\mathsmaller{XX^{\dagger}\to \Phi \Phi^\dagger}} \,\vrel 
&\simeq 
\frac{2\pi [\aF-\lambda_{\mathsmaller{X\Phi}}/(8\pi)]^2}{\mX^2} \ S_{0} (\zetaV+\zetaF) ,
\label{eq:sigmav_XXdaggerToPhiPhidagger}
\\
\sigma_{\mathsmaller{XX \to \Phi V}} \,\vrel =
\sigma_{\mathsmaller{X^{\dagger}X^{\dagger} \to \Phi^\dagger V}} \,\vrel 
&\simeq 0 ,
\label{eq:sigmav_XXToPhiV}
\end{align}
\end{subequations}
where we recall that $\zetaV \equiv \aV / \vrel$ and $\zetaF = \aF / \vrel$, and $S_0(\zetaS) \equiv 2\pi \zetaS/(1-e^{-2\pi \zetaS})$ is the $s$-wave Sommerfeld factor [cf.~\cref{eqs:zetaVzetaF,eq:S_ell}]. Note that the cross-section \eqref{eq:sigmav_XXdaggerToVV} for $XX^\dagger \to VV$ is twice as large as the spin-averaged cross-section for the annihilation of a fermionic particle-antiparticle pair into two Abelian vector bosons~\cite{vonHarling:2014kha,Petraki:2016cnz}.  For the $XX^{\dagger}\to \Phi \Phi^\dagger$ annihilation, the $s$-channel diagram (annihilation via off-shell $V$) is $p$-wave and we have neglected it in \cref{eq:sigmav_XXdaggerToPhiPhidagger}. For simplicity, in the following we shall also ignore the $\lambda_{\mathsmaller{X\Phi}}$ contribution. This coupling does not affect \BSFF, and ignoring it will allow us to compare more easily the strength of the processes that arise from the essential couplings of the model, $\aF$ and $\aV$.

Thus, the total velocity-weighted annihilation cross-section we will consider is
\beq
\sigma_\ann \vrel 
\simeq 
\frac{2\pi (\aV^2 + \aF^2)}{\mX^2} \ S_0 (\zetaV+\zetaF) ,
\label{eq:sigmaAnn_Tot}
\eeq
with its thermal average being
\begin{align}
\<\sigma_{\ann} \vrel \> &= \frac{x^{3/2}}{2\sqrt{\pi}}
\int_0^\infty d\vrel \, \vrel^2 \, (\sigma_\ann \vrel) \, e^{- x\vrel^2/4} .
\label{eq:sigmaAnn_Averaged}
\end{align}

\subsubsection{Bound-state formation, ionisation and decay  \label{sec:FreezeOut_BoundStates}}

\subsubsection*{Formation}

As already discussed, $XX^{\dagger}$ bound states can form via emission of a $V$ or a $\Phi$ boson, according to the processes \eqref{eq:BSF_VectorEmission} and \eqref{eq:BSF_ScalarEmission}, with the Feynman diagrams shown in \cref{fig:BSF_VectorEmission,fig:BSF_ScalarEmission}. For simplicity, we shall consider the capture into the ground state only, $n=1,\ell=m=0$, for both \BSFV~and \BSFF. The larger binding energy and decay rate of the ground state render the ionisation processes unimportant earlier on, and imply that the capture into the ground state has a higher efficiency in depleting DM than the other BSF processes.  Moreover, for \BSFV, the capture into the ground state is the dominant contribution~\cite[fig.~2]{Petraki:2016cnz}. On the other hand, for \BSFF, the rate of capture into excited states may exceed that of capture into the ground state in some velocity range, as seen in \cref{fig:CrossSections_BSF_ellZeroStates,fig:CrossSections_BSF_ellEQUALSnminus1}.  While we do expect that the capture into excited states plays an important role, here we only aim at showcasing the effect of BSF via emission of a charged scalar. We leave more detailed phenomenological studies for future work.

The effect of \BSFV~on the DM relic density was shown in \cite{vonHarling:2014kha}, in a setup where $V$ was the sole mediator; the corresponding cross-sections have been computed in~\cite{Petraki:2015hla,Petraki:2016cnz,Harz:2018csl}. In \cref{App:BSFV}, we review the computation and adapt it to the present model.

Taking the above into account, the BSF cross-sections we will consider are
\begin{subequations}
\label{eq:sigmav_BSF_100}
\label[pluralequation]{eqs:sigmav_BSF_100}
\begin{align}
\sigma_{100}^{\mathsmaller{V}} \vrel  
&= \frac{2^9\pi}{3\mX^2} 
~\aV (\aV+\aF)~ \(1+\frac{2\aF}{\aV+\aF}\)^2 \times
\nn \\ 
&\times
S_0 (\zetaV-\zetaF) 
\, [1+(\zetaV-\zetaF)^2] 
\, \frac{(\zetaV+\zetaF)^4}{[1+(\zetaV+\zetaF)^2]^3}
\, e^{-4 (\zetaV - \zetaF) {\rm arccot} (\zetaV + \zetaF)} ,
\label{eq:sigmav_100_VectorEmission}
\\
\sigma_{100}^{\mathsmaller{\Phi}} \vrel 
&\simeq
\frac{2^{11} \pi}{\mX^2} 
~\frac{\aF(2\aV+\aF)^2}{(\aV+\aF)^3}~
S_{0} (-\zetaV)
\, \[ \frac{(\zetaV+\zetaF)^2}{1+(\zetaV+\zetaF)^2} \]^3
\, e^{4\zetaV \, {\rm arccot} (\zetaV+\zetaF)} ,
\label{eq:sigmav_100_ScalarEmission}
\end{align}
\end{subequations}
where $S_0(\zetaS) \equiv 2\pi \zetaS / (1-e^{-2\pi \zetaS})$ is the $s$-wave Sommerfeld factor [cf.~\cref{eq:S_ell}], and we neglect any phase-space suppression due to the mass of the $\Phi$ (cf.~\cref{sec:FreezeOut_CoulombApprox}). The total BSF cross-section is then
\beq
\sigma_{\BSF} \vrel 
= \sigma_{100}^{\mathsmaller{\Phi}} \vrel 
+ \sigma_{100}^{\mathsmaller{V}} \vrel .
\label{eq:sigmaBSF_Tot}
\eeq
The thermally averaged BSF cross-section is
\begin{align}
\< \sigma_{\BSF} \vrel \> = \frac{x^{3/2}}{2\sqrt{\pi}}
\int_0^\infty d\vrel \, \vrel^2 (\sigma_{\BSF} \vrel) e^{-x \vrel^2/4}
\(1+\frac{1}{e^{x [\vrel^2 + (\aV+\aF)^2] / 4 }-1}\) ,
\label{eq:sigmaBSF_Averaged}
\end{align}
where the last factor accounts for the Bose enhancement due to the low-energy boson ($V$ or $\Phi$) emitted in the capture process; including the Bose enhancement is necessary in order to ensure detailed balance at large temperatures~\cite{vonHarling:2014kha}. 

\subsubsection*{Ionization}\label{sec:ionization} 

The ionisation rate of the bound states can be found either by using the Milne relation between the capture and ionization cross-sections (see~\cite[appendix~D]{Harz:2018csl} for the proof), or more directly, by invoking detailed balance,
\begin{align}
\GammaIon
= \< \sigma_{\BSF} \vrel \>  \times   (\nX^\eq)^2 / \nB^\eq 
= \< \sigma_{\BSF} \vrel \>  \times s (\YX^\eq)^2 / \YB^\eq ,
 \label{eq:GammaIon}
\end{align}
where $s$ is the entropy density of the universe, and the equilibrium yields of the unbound particles and the bound states, $\YX^\eq  \equiv \nX^\eq / s$ and $\YB^\eq  \equiv \nB^\eq /s$,  are given in \cref{sec:FreezeOut_BoltzEq} below.  Using these densities, we obtain  
\begin{align}
\GammaIon \simeq \< \sigma_{\BSF} \vrel \>  
\(
\frac{\mX T}{4\pi} 
\)^{3/2}
\ e^{-|\EB|/T} ,
\label{eq:GammaIon_T}
\end{align}
where $\EB = {\cal E}_{10}$ is the binding energy of the ground state. 

\subsubsection*{Decay into radiation} 

The dominant decays of the ground state are
\begin{align}
{\cal B}_{100} (XX^{\dagger})  &\to VV , \  \Phi\Phi^\dagger ,
\label{eq:DecayModes}
\end{align}
with total rate
\begin{align}
\GammaDec \simeq |\psi_{100}^{\XXdagger} (0)|^2 (\sigma_{\rm ann} \vrel)_0^{\rm pert}
\label{eq:DecayRate_def} 
\end{align}
where $(\sigma_{\rm ann} \vrel)_0^{\rm pert}$ is the perturbative $s$-wave velocity-weighted annihilation cross-section (to zeroth order in $\vrel$), which is contained in \cref{eq:sigmaAnn_Tot}.  Then,
\begin{align}
\GammaDec &\simeq
\frac{\mX^3(\aV+\aF)^3}{2^3\pi} \(
\frac{2\pi \aV^2}{\mX^2} + \frac{2\pi \aF^2}{\mX^2} 
\)  
= \frac{\mX}{4}  (\aV+\aF)^3  (\aV^2 + \aF^2) .
\label{eq:DecayRate} 
\end{align}

\subsection{Boltzmann equations and effective depletion rate \label{sec:FreezeOut_BoltzEq}}

Let $\YX = \nX / s$ and $\YB = \nB / s$ be the yields of the unbound $X$ particles and the bound states respectively. The Boltzmann equations that govern the evolution of the densities are~\cite{vonHarling:2014kha}\footnote{We use the Planck mass $\mpl = 1.22 \times 10^{19}~\GeV$.}
\begin{subequations}
\label{eq:BoltzmannEqs}
\label[pluralequation]{eqs:BoltzmannEqs}
\begin{align}
\frac{d\YX}{dx} = 
&- \sqrt{\frac{\pi}{45}} \frac{\mpl\,\mX\,\gstareffsqrt}{x^2} 
\left\{ 
\<\sigma_\ann \vrel\> \[\YX^2 - (\YX^{\eq})^2\]  
+ \<\sigma_{\BSF} \, \vrel\> \[\YX^2 - \frac{\YB}{\YB^{\eq}} (\YX^{\eq})^2 \] 
\right\} ,
\label{eq:BoltzmannEqs_Free} 
\\
\frac{d\YB}{dx} = 
&+ \sqrt{\frac{\pi}{45}} \frac{\mpl\,\mX\,g_{*,\eff}^{1/2}}{x^2}
\<\sigma_{\BSF} \, \vrel\> \[\YX^2 - \frac{ \YB}{\YB^{\eq}}(\YX^{\eq})^2 \]
- \sqrt{\frac{45}{4\pi^3}} 
\frac{\mpl}{\mX^2}
\frac{\gstareffsqrt}{\gstarS}
~x \, \GammaDec \(\YB - \YB^{\eq} \) ,
\label{eq:BoltzmannEqs_Bound} 
\end{align}
\end{subequations}
where $\< \sigma_\ann \vrel \>$, $\< \sigma_\BSF \vrel \>$ and $\GammaDec$ have been given in \cref{sec:FreezeOut_ANN,sec:FreezeOut_BoundStates}. In \cref{eqs:BoltzmannEqs}, 
\beq
\gstareffsqrt \equiv \frac{\gstarS}{\sqrt{g_*}}
\(1 - \frac{x}{3g_{*\mathsmaller{S}}} \frac{dg_{*\mathsmaller{S}}}{dx} \) \,,
\label{eq:gstareff}
\eeq
with $g_*$ and $\gstarS$ being the energy and entropy relativistic degrees of freedom.  We will take $g_* = \gstarS = g_*^{\SM} + 4$ to account for the SM plus the $V$ and $\Phi$ degrees of freedom, during the DM freeze-out. We recall that the entropy density of the universe is $s=(2\pi^2/45) \gstarS T^3$. 
In the non-relativistic regime, the equilibrium yields $\YX^\eq$ and $\YB^\eq$ are
\begin{subequations}
\label{eq:YieldsEq}
\label[pluralequation]{eqs:YieldsEq}
\begin{align}
\YX^\eq &\simeq \frac{90}{(2\pi)^{7/2}} \ \frac{1}{\gstarS} \ x^{3/2} \ e^{-x} \,, 
\label{eq:YX_eq} 
\\
\YB^\eq &\simeq \frac{90}{(2\pi)^{7/2}} \ \frac{1}{\gstarS} 
\ (2x)^{3/2} 
\ e^{-2x [1-(\aV+\aF)^2/8]}  \,.
\label{eq:YB_eq}  
\end{align}
\end{subequations}

The relic density of the $X,X^{\dagger}$ particles is
\beq
\Omega_{\mathsmaller{X}} = 2\mX \YX s_0 /\rho_c ,
\label{eq:Omega}
\eeq
where $s_0\simeq 2840~\cm^{-3}$ and $\rho_c \simeq 3.67 \times 10^{-47} \GeV^4$ are the entropy and critical energy density of the universe today~\cite{Aghanim:2018eyx}. We require that $\Omega_{\mathsmaller{X}} = \Omega_{\DM}$, where the observed DM density is $\Omega_{\DM} \simeq 0.264$~\cite{Aghanim:2018eyx}.

We note that in the minimal setup considered here, the $\Phi$ particles are stable, being the lightest degrees of freedom charged under $U(1)_{\mathsmaller{D}}$. As such, they contribute to the DM density. However, due to their lightness, even a small $\aV$ suffices to ensure that they annihilate into vector bosons efficiently, via the $t/u$-channel process $\Phi \Phi^\dagger \to VV$, thereby reaching a cosmologically negligible relic abundance. Moreover, the radiation density due to $V$ evades the CMB and BBN constraints provided that the dark sector temperature is somewhat lower than that of the SM plasma at the corresponding times. If $\Phi$ acquires a VEV and/or the dark sector couples to the SM, then there are more possibilities for the cosmological abundance of $\Phi$ and $V$, as well as constraints. A complete phenomenological study is beyond the scope of this work.

\subsubsection*{Effective depletion cross-section \label{sec:FreezeOut_sigmav_effective}}

Instead of the system of coupled \cref{eqs:BoltzmannEqs}, the effect of bound states on the relic density can be described by a single Boltzmann equation for the DM particles and an \emph{effective} annihilation cross-section that includes BSF weighted by the fraction of bound states that decay rather than being ionised. We define first the effective BSF cross-section
\begin{align}
\<\sigma_\BSF \vrel\>_\eff \equiv  \<\sigma_{\BSF} \vrel\> \times
\( \frac{\GammaDec}{\GammaDec + \GammaIon} \) .
\label{eq:sigmaBSF_Eff}
\end{align}
The effective DM depletion cross-section is
\begin{align}
\<\sigma \vrel\>_\eff = \<\sigma_\ann \vrel\> +  \<\sigma_\BSF \vrel\>_\eff .
\label{eq:sigmaEff}
\end{align}
We may compute the $X$ relic density by solving the Boltzmann equation~\cite{Liew:2016hqo}
\begin{align}
\frac{d\YX}{dx} = 
&- \sqrt{\frac{\pi}{45}} \frac{\mpl\,\mX\,\gstareffsqrt}{x^2}
~\<\sigma \vrel\>_\eff \[\YX^2 - (\YX^{\eq})^2\] .
\label{eq:BoltzmannEqs_Eff} 
\end{align}

\begin{figure}[t!]
\centering	
{}\hfill
\includegraphics[width=0.45\textwidth]{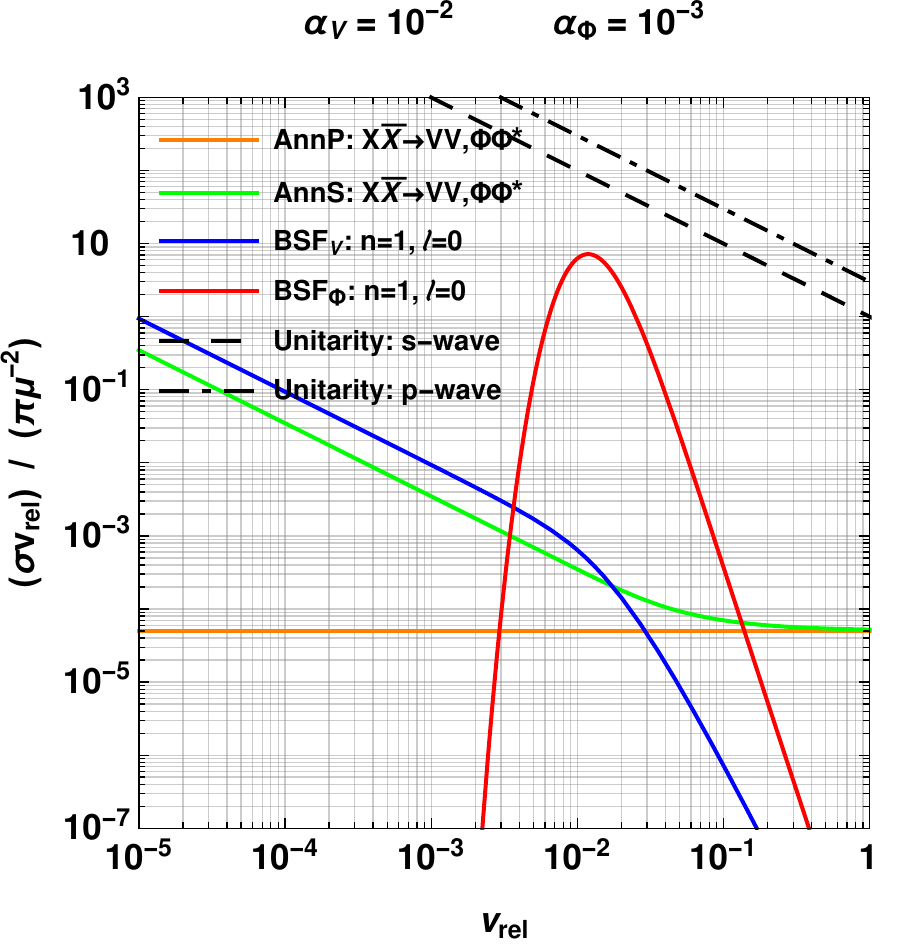}
\hfill
\includegraphics[width=0.45\textwidth]{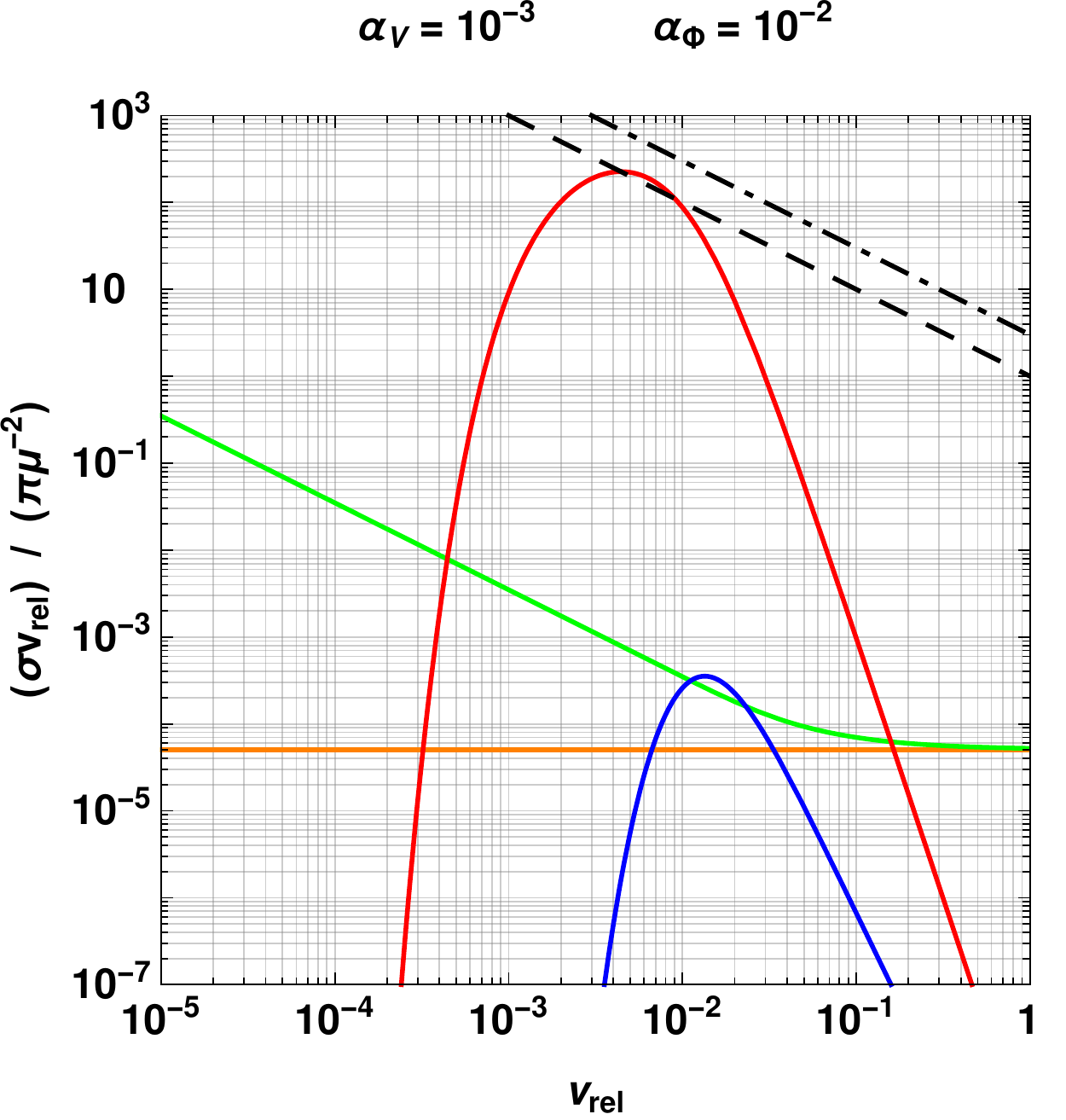}
\hfill{}
\caption{\label{fig:CrossSectionsFO}
The BSF and annihilation velocity-weighted cross-sections that we consider in the computation of the DM freeze-out, vs. $\vrel$. All cross-sections have been normalised to $\pi/\mu^2$, with $\mu=\mX/2$ being the reduced mass of the interacting particles. Also shown, the unitarity limits on $s$-wave and $p$-wave inelastic processes, which have to be respected by \BSFF~and \BSFV, respectively, for capture into the ground state.}
	
\bigskip 
	
{}\hfill
\includegraphics[width=0.45\textwidth]{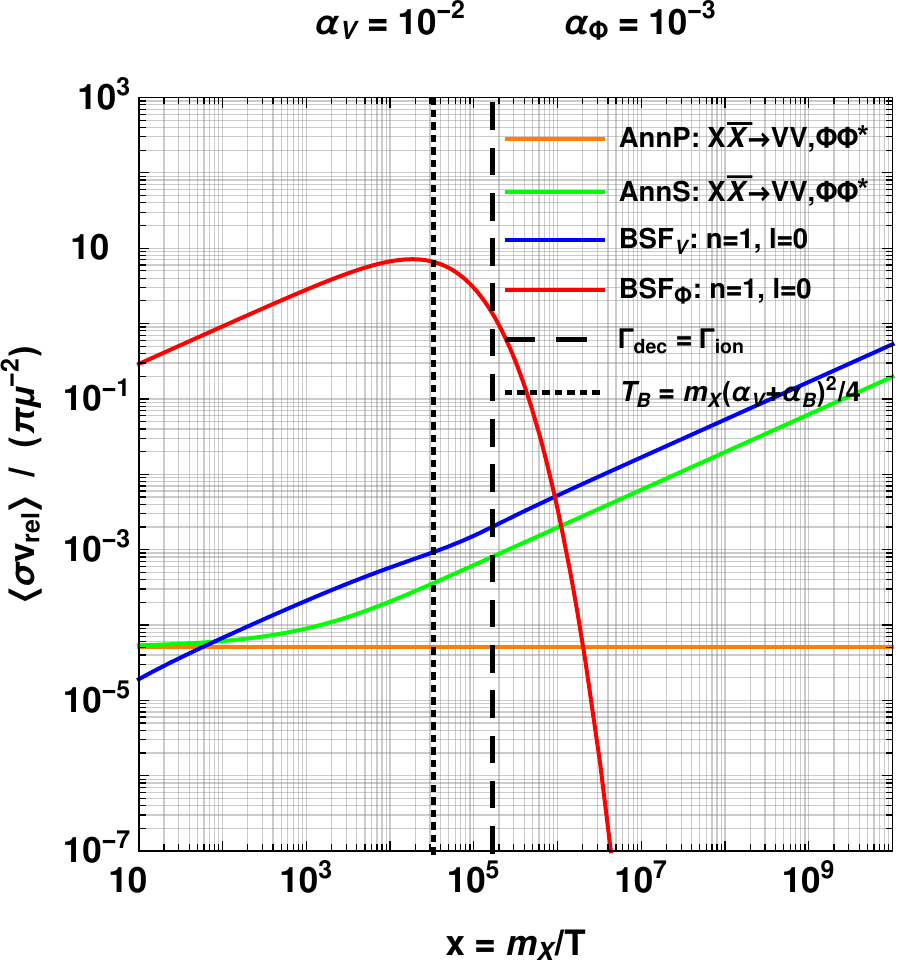}
\hfill
\includegraphics[width=0.45\textwidth]{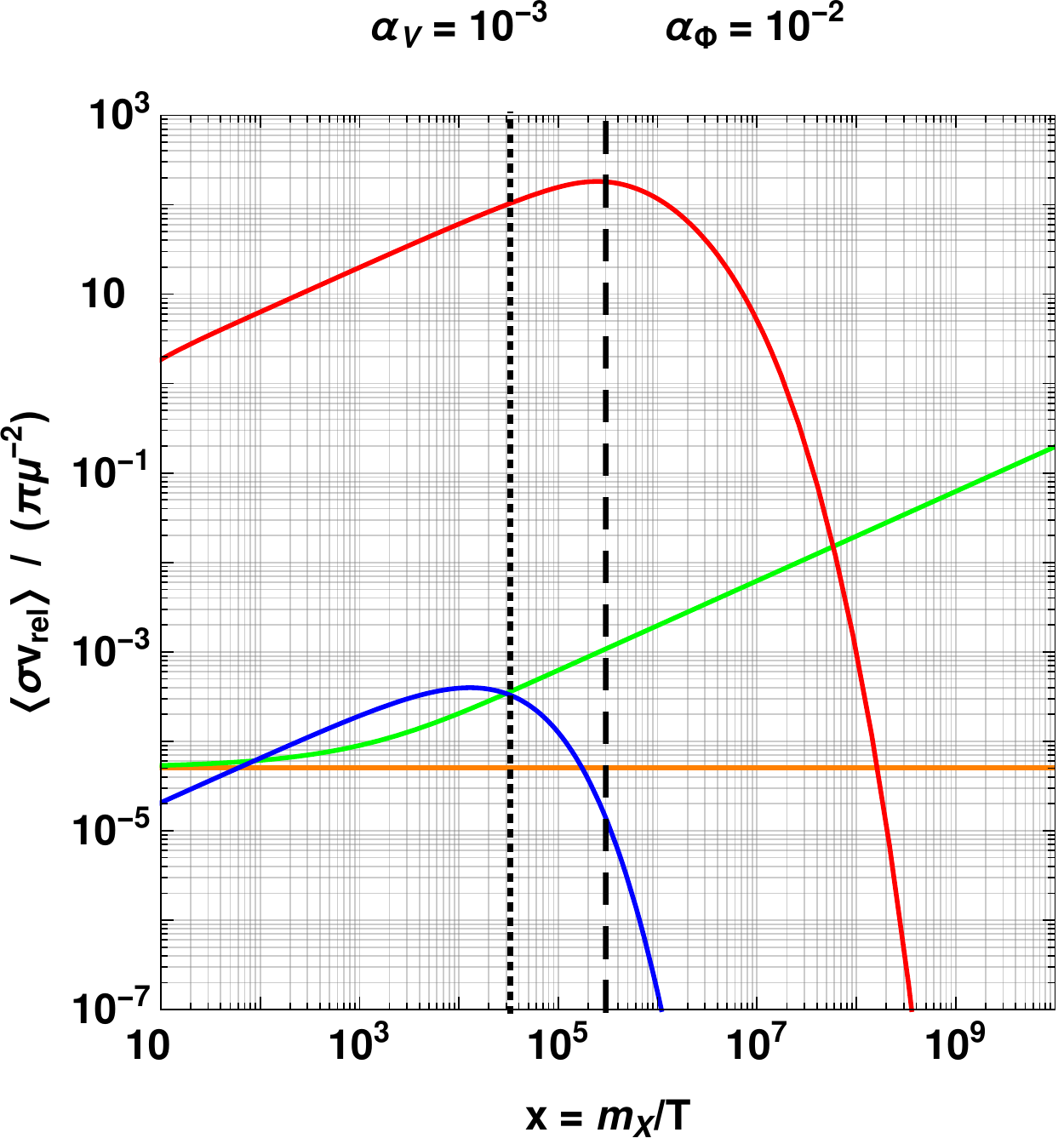}
\hfill{}
\caption{\label{fig:CrossSectionsFO_Averaged}
The thermally averaged BSF and annihilation velocity-weighted cross-sections that we consider in the computation of the DM freeze-out, vs. $x \equiv \mX/T$. All cross-sections have been normalised to $\pi/\mu^2$, with $\mu=\mX/2$ being the reduced mass of the interacting particles. We also mark two important mileposts: (i) the time when the temperature equals the binding energy $T_B = \mX (\aV+\aF)^2/4$, at and below which the equilibrium occupation number of the bound states becomes very significant, and (ii) the end of ionisation equilibrium, below which the DM depletion rate saturates to the BSF rate and thus becomes sensitive to the BSF cross-section.
}
\end{figure}

In \cref{fig:CrossSectionsFO,fig:CrossSectionsFO_Averaged}, we show the DM annihilation and BSF cross-sections, and their thermal averages. The \BSFF~cross-sections can exceed both the \BSFV \ and annihilation cross-sections by orders of magnitude, even for very small values of the couplings. However, the effect on the DM density depends on the interplay among the bound-state formation, ionisation and decay processes. To anticipate the result, it is useful to discern between two phases during the DM chemical decoupling.

\begin{enumerate}[(a)]
\item
While the temperature is large enough to ensure that $\GammaIon \gg \GammaDec$, the system is in a state of \emph{ionization equilibrium}~\cite{Binder:2018znk}, where the effective DM depletion rate due to bound states is essentially independent of the BSF cross-section. Indeed, combining  \cref{eq:GammaIon_T,eq:DecayRate_def,eq:sigmaBSF_Eff} under the aforementioned condition, we obtain 
\begin{align}
\<\sigma_{\BSF} \vrel\>_{\eff} \simeq \<\sigma_{\ann} \vrel\>_0^{\rm pert}
\times 8\sqrt{\pi} \( \frac{|\EB|}{T} \)^{3/2} e^{+|\EB|/T} .
\label{eq:sigmaBSF_eff_IonEquil}
\end{align}
Note that \cref{eq:sigmaBSF_eff_IonEquil} does not rely on the specific couplings or interactions of the model considered here, but is rather general. It is easy to check that \eqref{eq:sigmaBSF_eff_IonEquil} is small in comparison to the annihilation cross-section, unless or until the temperature approaches or drops below the binding energy. In a $U(1)$ model where DM couples only to the gauge boson, the ionization equilibrium ends at a temperature somewhat higher than the binding energy, thus \eqref{eq:sigmaBSF_eff_IonEquil} remains mostly small and most of the BSF effect on the DM density arises after that the end of ionisation equilibrium~\cite{vonHarling:2014kha}. However, in the present model, the largeness of the \BSFF \ cross-section sustains ionization equilibrium down to temperatures below the binding energy (cf.~\cref{fig:CrossSectionsFO_Averaged}), thereby rendering the DM depletion significant during this phase. Clearly, while \cref{eq:sigmaBSF_eff_IonEquil} is independent of $\<\sigma_{\BSF} \vrel\>$, the duration of ionisation equilibrium depends on it.

\item
The ionisation equilibrium ends when the ionisation rate drops below the decay rate, $\GammaIon \lesssim \GammaDec$. Then, the DM depletion rate approaches rapidly the actual BSF rate, and is therefore sensitive to the exact BSF cross-section, 
$\<\sigma_{\BSF} \vrel\>_{\eff} \simeq \<\sigma_{\BSF} \vrel\>$. 
\end{enumerate}

\begin{figure}[t!]
\centering
	
{}\hfill
\includegraphics[width=0.45\textwidth]{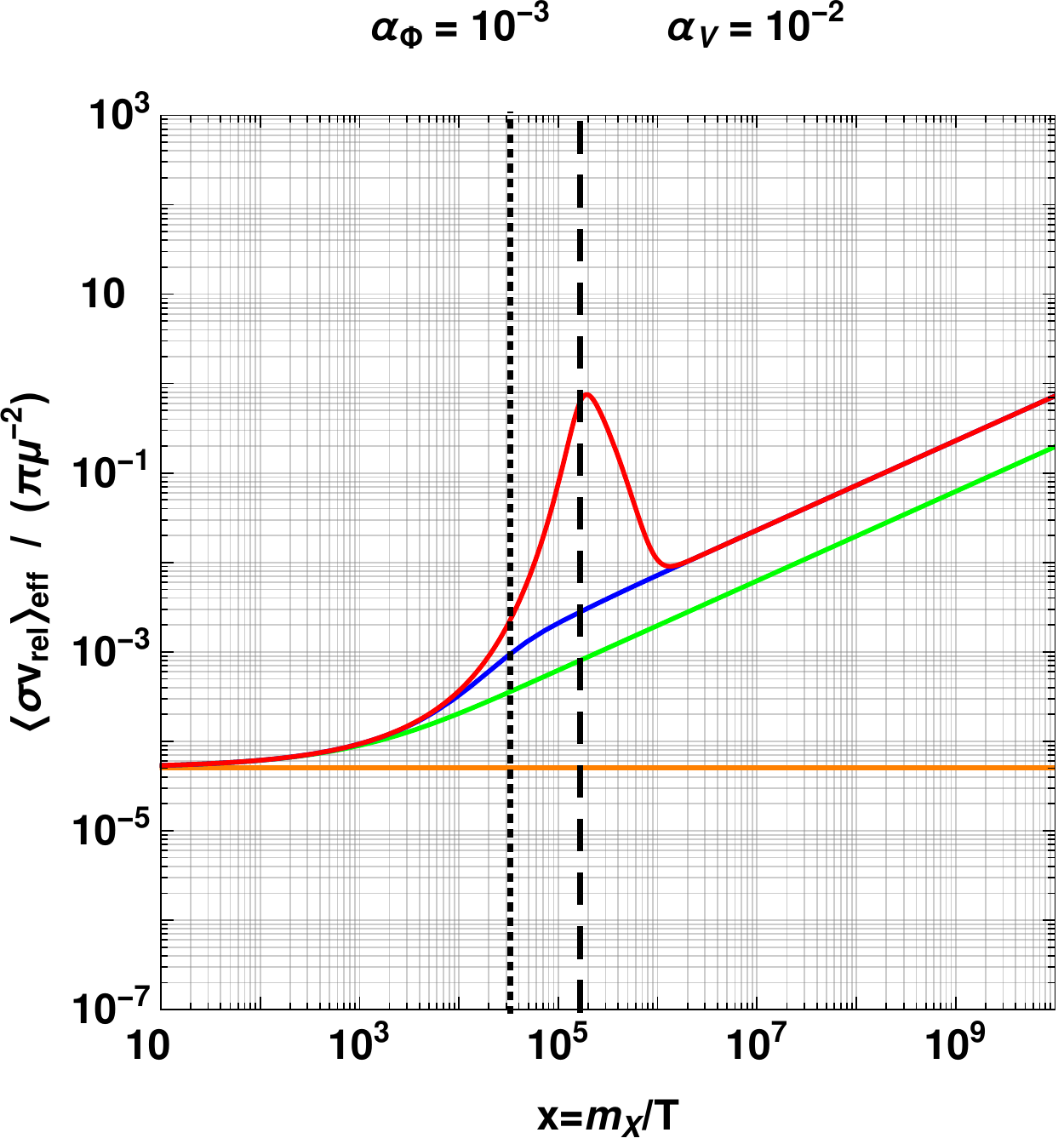}
\hfill
\includegraphics[width=0.45\textwidth]{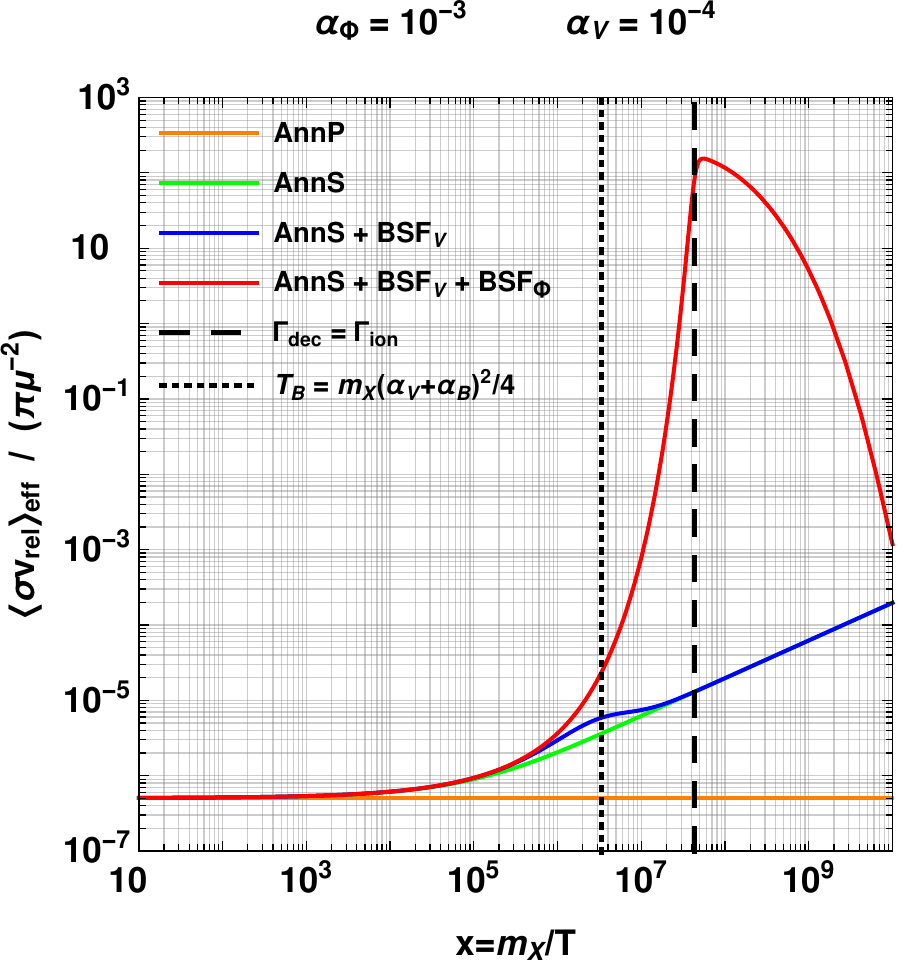}
\hfill{}
\caption{\label{fig:CrossSectionsFO_Effective_x}
The thermally-averaged  velocity-weighted effective cross-section as a function $x = \mX/T$, for fixed $\aF = 10^{-3}$ and two different values of $\aV$.}

\bigskip

\includegraphics[width=0.45\textwidth]{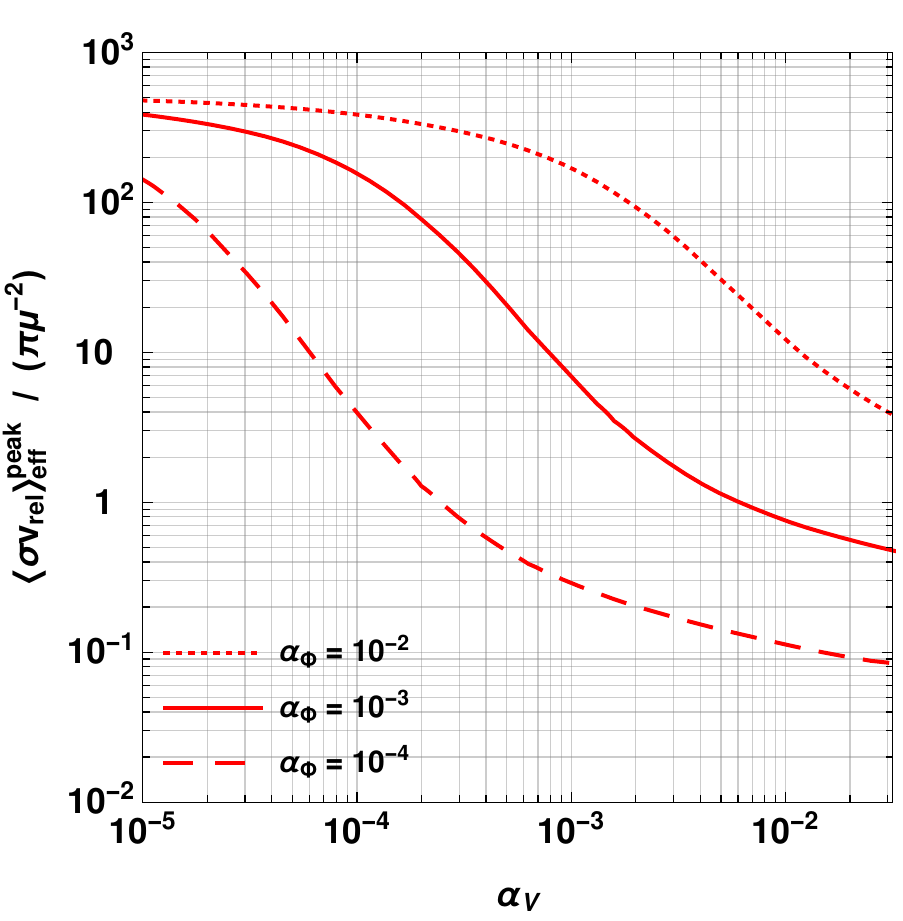}
\caption{\label{fig:CrossSectionsFO_Effective_alphaV}
The value of the thermally averaged velocity-weighted effective cross-section at its peak as a function of $\aV$, for different values of $\aF$. Note that the time at which the peak occurs, $x=x_{\rm peak}$, depends on $\aV$ and $\aF$, and is chosen accordingly. We observe that, by varying $\aV$,  $\<\sigma \vrel\>_{\eff}^{\rm peak}$ rises at $\aV \lesssim \aF$, and becomes most significant at the limit of global symmetry, $\aV \to 0$.}	
\end{figure}

Taking into account the above considerations, and in order to gain insight on whether \BSFF~may affect the DM density, in \cref{fig:CrossSectionsFO_Effective_alphaV,fig:CrossSectionsFO_Effective_x} we present the effective DM depletion cross-section for the following four cases:
\begin{description}
\item[\AnnP:] Perturbative annihilation only (diagrams shown in \cref{fig:Ann_Tree}).
\item[\AnnS:] Annihilation including the Sommerfeld effect due to both $V$ and $\Phi$ exchange [cf.~\cref{eq:sigmaAnn_Tot}].
\item[\AnnS~+~\BSFV:] Annihilation with Sommerfeld effect due to $V$ and $\Phi$ exchange, plus \BSFV \  [cf.~\cref{eq:sigmav_100_VectorEmission}] with the ionization of the bound states [cf.~\cref{eq:GammaIon}] caused by $V$ only.
\item[\AnnS~+~\BSFV~+~\BSFF:] 
Annihilation with Sommerfeld effect due to $V$ and $\Phi$ exchange, plus \BSFV \ and \BSFF \ [cf.~\cref{eq:sigmaBSF_Tot}], with the ionization of the bound states [cf.~\cref{eq:GammaIon}] caused by $V$ or $\Phi$.
\end{description}
In \cref{fig:CrossSectionsFO_Effective_x} we show the evolution of $\<\sigma \vrel\>_{\eff}$ as the temperature drops. We choose a small value for the DM coupling to the charged scalar, $\aF =10^{-3}$, to be well within the range that is consistent with unitarity (cf.~\cref{sec:BSFviaScalarEmission_Unitarity}). We observe that the \BSFV \ and \BSFF \ contributions to the effective cross-section begin to rise at $T \sim |\EB|$, as implied by \cref{eq:sigmaBSF_eff_IonEquil}. Later on, the ionisation equilibrium ends, and $\<\sigma_{\BSF} \vrel\>_{\eff}$ saturates to $\<\sigma_{\BSF} \vrel\>$; because this occurs at $T < |\EB|$, when $\sqrt{\<\vrel^2\>} < \aB = \aV+\aF < \aV$,  the effective BSF cross-section $\<\sigma_{\BSF} \vrel\>_{\eff}$ decreases beyond this point, due to the repulsive potential in the $XX$ and $X^\dagger X^\dagger$ scattering states.  The largeness of $\<\sigma_{\BSF} \vrel\>$ and consequently of $\<\sigma_{\BSF} \vrel\>_{\eff}$ around its peak, suggest that the DM depletion processes may recouple, even at a very low temperature, as we shall see in \cref{sec:FreezeOut_BoltzEq_Solutions}.

Comparing the two plots of \cref{fig:CrossSectionsFO_Effective_x}, corresponding to $\aV=10^{-2}$ and $\aV=10^{-4}$, we also note that the rise and the peak of \BSFF \ become more pronounced for smaller $\aV$ values, which allow for the effective \BSFF \ cross-sections to grow larger before the suppression due to the repulsion in the scattering state settles in. To investigate this further, in \cref{fig:CrossSectionsFO_Effective_alphaV} we show the dependence of $\<\sigma \vrel\>_{\eff}$ evaluated at its peak, on $\aV$. We see that $\<\sigma \vrel\>_{\eff}^{\rm peak}$ rises at $\aV \lesssim \aF$, and becomes most significant at $\aV \to 0$, i.e.~in the limit where the symmetry becomes global.

\subsection{Solutions of the Boltzmann equations \label{sec:FreezeOut_BoltzEq_Solutions}}

We now solve the Boltzmann \cref{eq:BoltzmannEqs_Eff}, discuss the qualitative features of the solutions and present numerical results. We focus mostly on small values of $\aF$, roughly $\aF \lesssim 10^{-2}$, in order to be consistent with unitarity (cf.~\cref{sec:BSFviaScalarEmission_Unitarity}), and mark any parameter space where the \BSFF \ cross-section violates it.

\subsubsection{Freeze-out and recoupling of DM depletion processes}

\begin{figure}[t!]
\centering
{}\hfill
\includegraphics[width=0.45\textwidth]{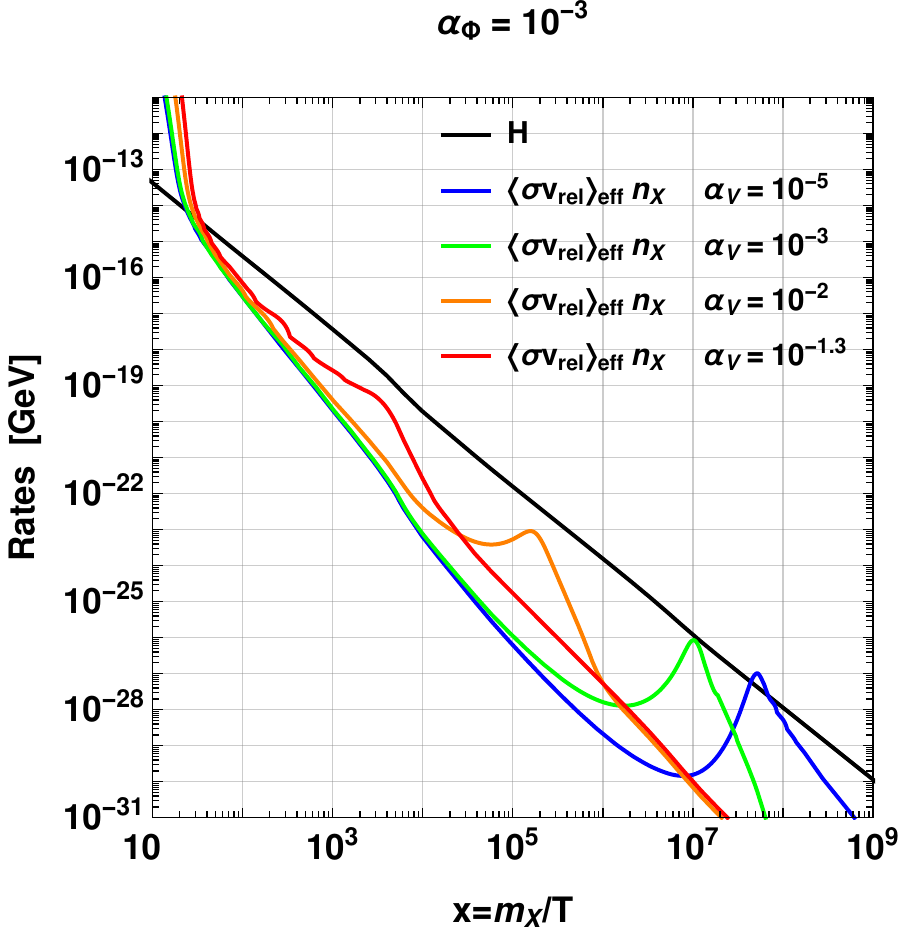}
\hfill
\includegraphics[width=0.425\textwidth]{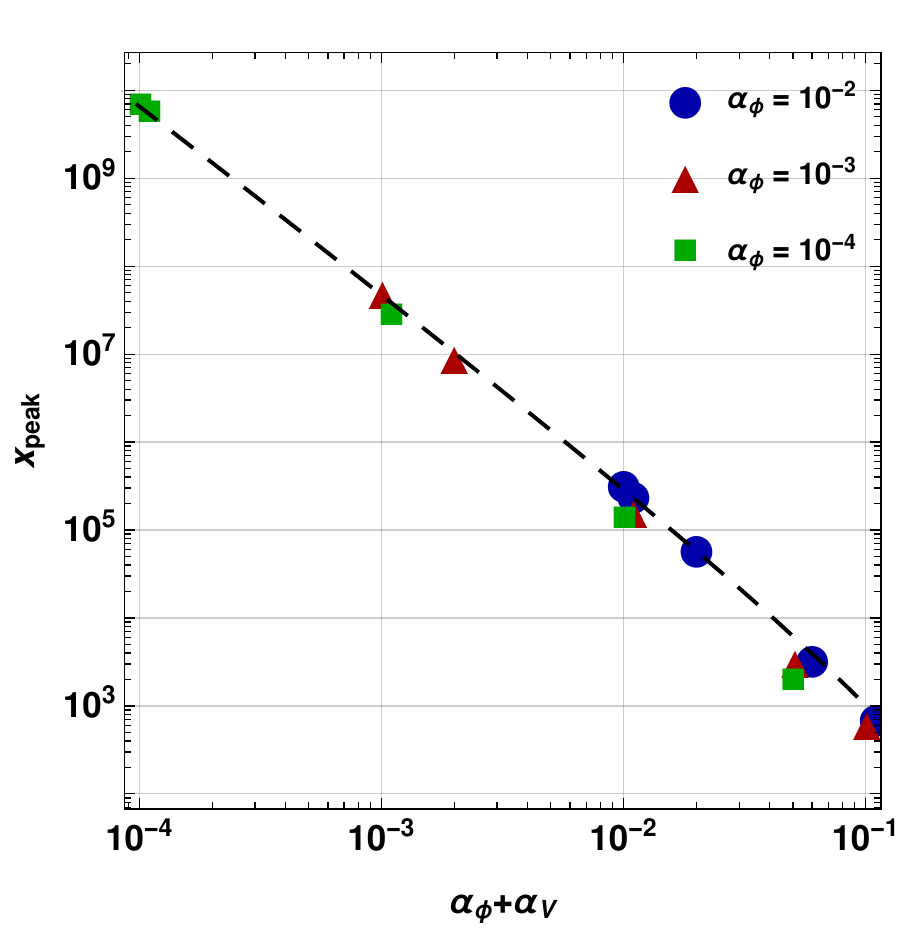}
\hfill{}
\caption{\label{fig:Rates}
\emph{Left:} The DM depletion rate $\GammaX = \nX \<\sigma \vrel\>_{\eff}$ compared to the Hubble parameter $H$, for $\aF = 10^{-3}$ and different values of $\aV$. We have used $\mX=10^3~\GeV$. 
\emph{Right:} $x_{\rm peak}$ is the value of $x \equiv \mX/T$ at which $\<\sigma \vrel\>_{\eff}$ peaks. The symbols correspond to the values determined numerically, while the line is the semi-analytical prediction \cref{eq:xpeak}.}
\end{figure}

\begin{figure}[t!]
\centering
{}\hfill
\includegraphics[width=0.40\textwidth]{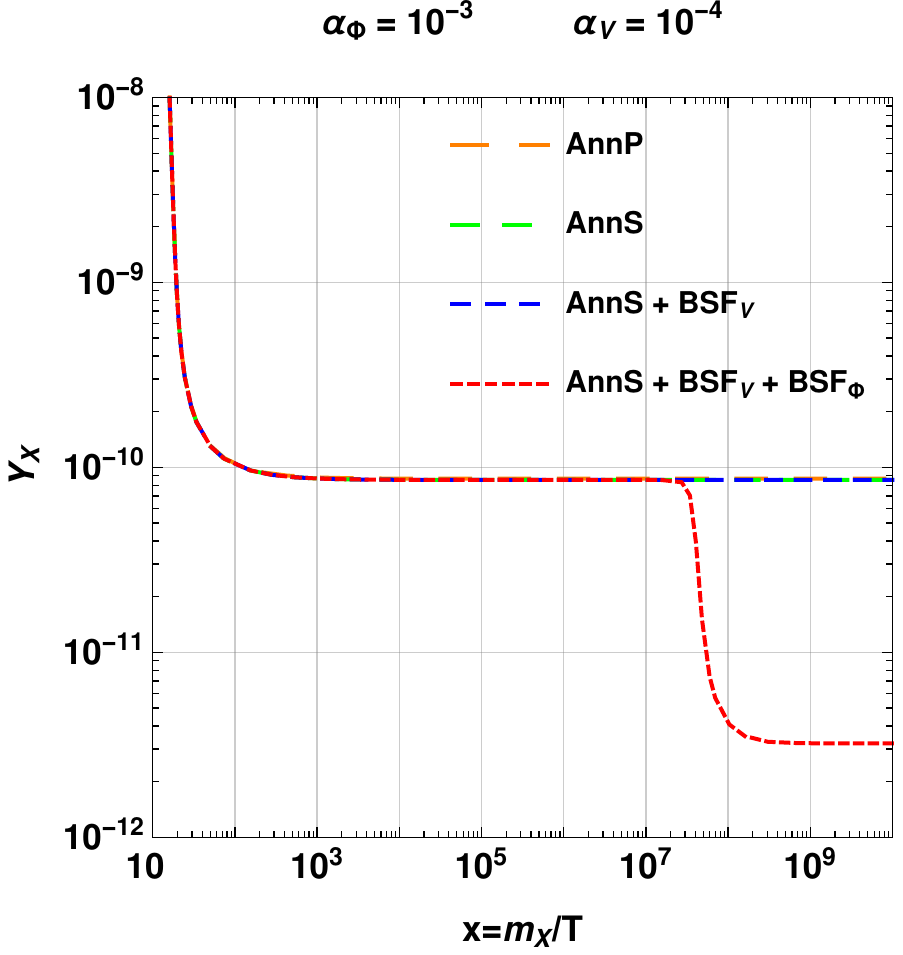}
\hfill
\includegraphics[width=0.40\textwidth]{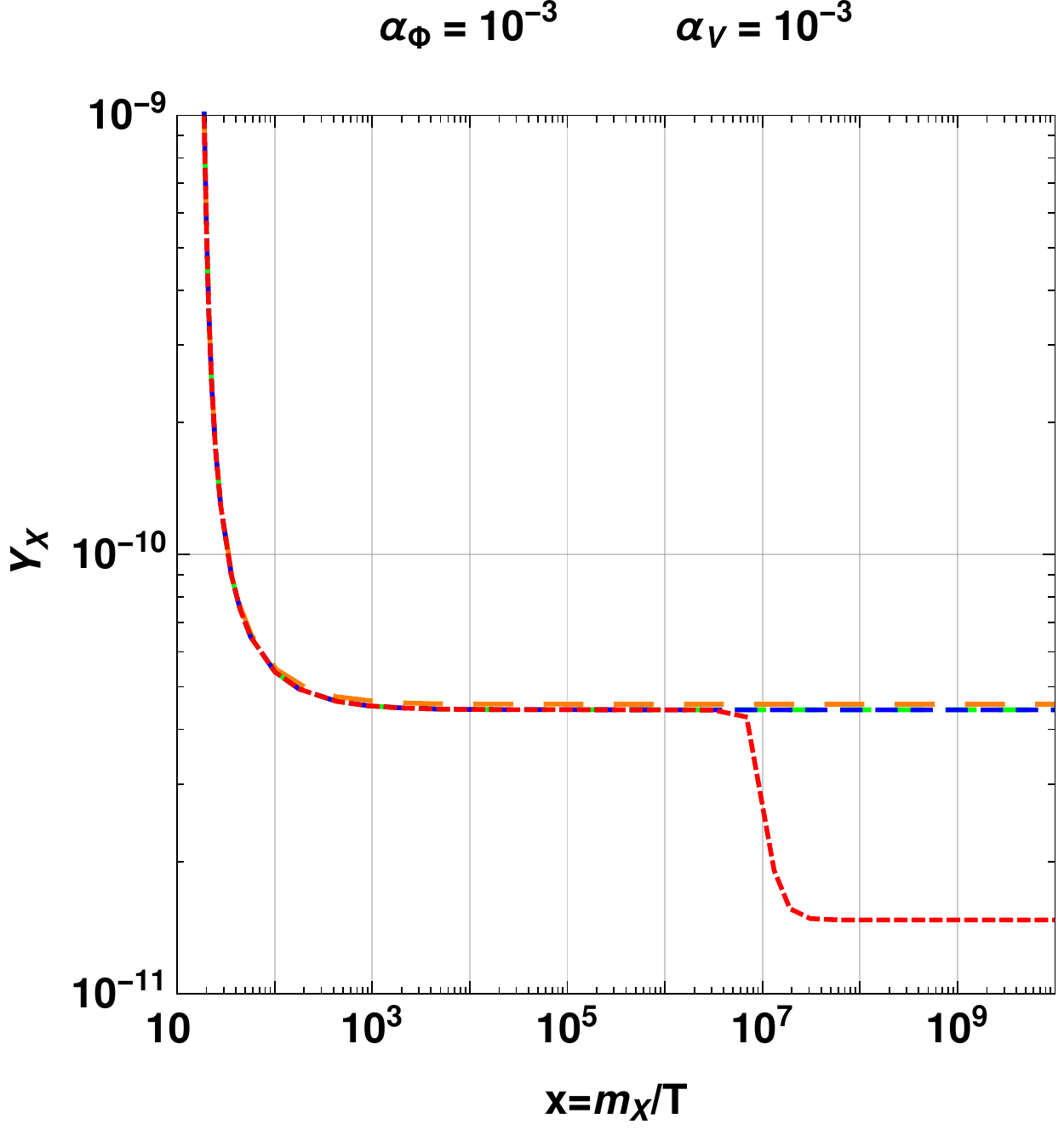}
\hfill{}
	
\bigskip
	
{}\hfill
\includegraphics[width=0.40\textwidth]{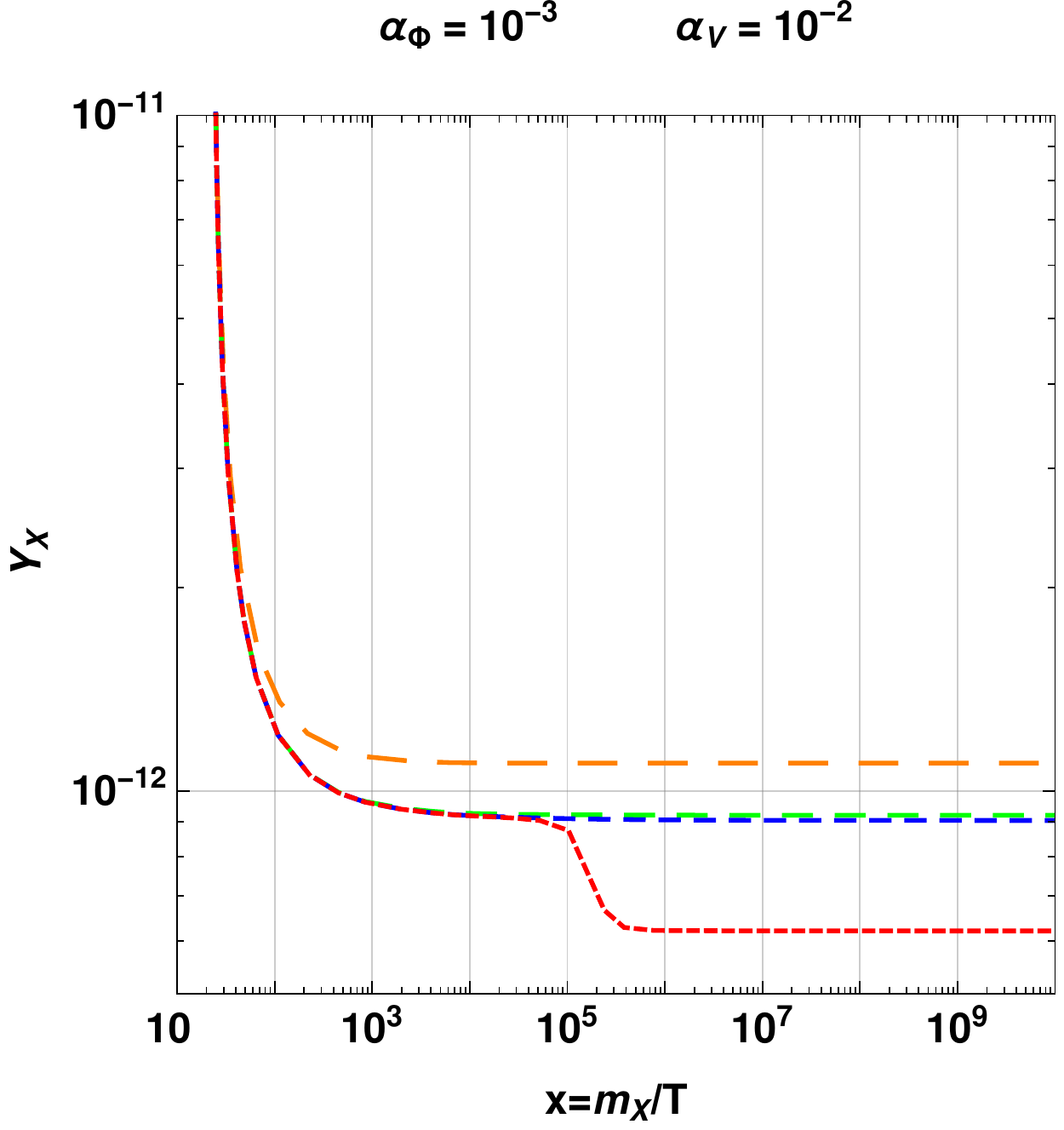}
\hfill
\includegraphics[width=0.40\textwidth]{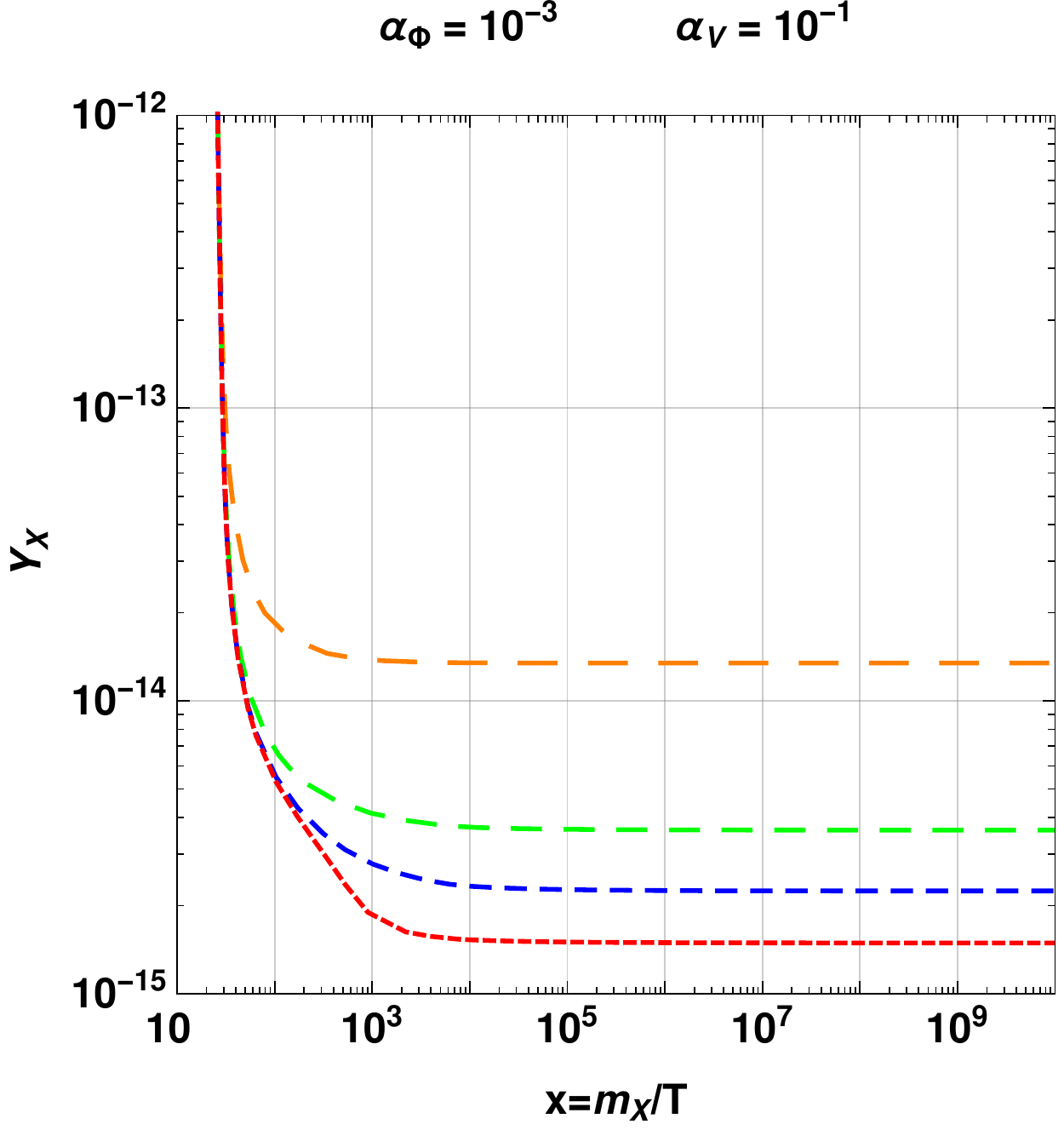}
\hfill{}
\caption{\label{fig:Abundances}
The evolution of the DM yield $\YX \equiv \nX/s$ vs. $x\equiv \mX/T$ for the four cases defined in \cref{sec:FreezeOut_sigmav_effective}. We have fixed $\mX=10^3$~GeV.}
\end{figure}

The Boltzmann \cref{eq:BoltzmannEqs_Eff} describes the balance between the DM depletion processes and the expansion of the universe. Motivated by the sharp increase of $\<\sigma \vrel\>_{\eff}$ at low temperatures seen in \cref{fig:CrossSectionsFO_Effective_x}, we compare the $X$ depletion rate $\GammaX \equiv \nX \<\sigma \vrel\>_{\eff}$ with the expansion rate of the universe $H = \sqrt{4\pi^3 g_*/45} \, T^2/\mpl$ in the left plot of \cref{fig:Rates}. Then in \cref{fig:Abundances}, we show the evolution of the DM density for different sets of parameters, in the four cases defined in \cref{sec:FreezeOut_sigmav_effective}. The DM chemical decoupling is marked by two important events.

\paragraph{First freeze-out.} As is standard, for $x \lesssim x_{\FOone} \approx 30$, the DM depletion and creation processes are in equilibrium, i.e.~$\GammaX \gtrsim H$. Beyond this point, the $X,X^\dagger$ densities depart from their equilibrium values, their exponential drop is stalled, and they begin to freeze-out. However, because the annihilation and BSF cross-sections increase with decreasing temperature, the depletion of DM continues to be important until somewhat later, and may lead to the reduction of the DM density by a factor of a few. The Sommerfeld enhancement of the annihilation processes is important for $\vrel \lesssim 10(\aV + \aF)$, which upon thermal averaging implies $x \gtrsim 10^{-2} (\aV+\aF)^{-2}$. Thus if $\aV + \aF \gtrsim 10^{-2}$, the Sommerfeld enhancement becomes significant already at $x \sim 10^{2}$, i.e.~soon after freeze-out, while the DM density is still quite large. For BSF, a somewhat larger coupling, $\aV + \aF \gtrsim 0.03$, is required. In this range of couplings, the DM chemical decoupling is prolonged beyond freeze-out, as clearly seen in the bottom right plot of \cref{fig:Abundances}.

\paragraph{Recoupling of DM depletion and second freeze-out.}

If $\aV+\aF \lesssim 10^{-2}$, the ionisation processes impede the DM depletion via BSF until quite late, when the DM density is rather low. However, the largeness of the \BSFF \ cross-section may compensate for the smallness of the DM density, and result in the recoupling of the DM depletion processes around the time when $\<\sigma \vrel\>_{\eff}$ peaks, at $T \lesssim |\EB|$.\footnote{Recoupling of the DM depletion at late times may occur also due to the strong velocity dependence of Sommerfeld-enhanced cross-sections at (or close to) parametric resonance points~\cite{vandenAarssen:2012ag,Binder:2017lkj}. The recoupling observed here is not due to resonant features, and applies to broader parameter space.}
We may estimate if and when this occurs as follows. The $X,X^\dagger$ yield after the first freeze-out is estimated by the standard result,  
$\mX \YX^\FOone \<\sigma_{\ann} \vrel\>^{\FOone} \sim 8 \times 10^{-19}~\GeV^{-1}$ (see e.g.~\cite{Gondolo:1990dk}), where we assumed that the direct annihilation dominates the DM depletion rate at that time; this is indeed true for the range of $\aV,\aF$ where the recoupling may occur.  The DM depletion recouples if $\YX^\FOone \<\sigma \vrel\>_{\eff} \gtrsim H/s$, which implies 
$\<\sigma \vrel\>_{\eff}  / \<\sigma_{\ann} \vrel\>^{\FOone} \gtrsim  0.4 \, x / \sqrt{g_*}$. 
If this occurs, it does so shortly before DM exits the ionisation equilibrium, i.e.~while \cref{eq:sigmaBSF_eff_IonEquil} is still approximately valid. (As already mentioned, at later times \BSFF \ decreases exponentially due to the repulsion in the scattering state.) Thus, the recoupling condition becomes 
$\sqrt{\pi/S_{\ann}^\FOone}(\aV+\aF)^3 \, x^{3/2} \, e^{x (\aV+\aF)^2/4} \gtrsim  0.4 \, x / \sqrt{g_*}$, where $S_{\ann}^\FOone \sim {\cal O}(1)$ stands for the thermally averaged Sommerfeld factor of the annihilation processes around the time of the first freeze-out. For large $\aV+\aF$, this condition yields a time close to the first freeze-out, and corresponds to the case when the DM chemical decoupling is simply delayed due to the BSF processes, as discussed above (cf.~bottom right plot in \cref{fig:Abundances}). However, for smaller couplings, we obtain the following estimate for the time of recoupling and approximately the peak of $\<\sigma \vrel\>_{\eff}$,
\begin{align}
x_{\rm peak} \sim \frac{8 \, \ln (\aV+\aF)^{-1}}{(\aV+\aF)^2} ,
\label{eq:xpeak}
\end{align}
where we kept the leading order logarithmic correction in $x_{\rm peak}$.
Note that $x_{\rm peak}$ is independent of the DM mass $\mX$. In the right plot of \cref{fig:Rates}, we compare the semi-analytical prediction \eqref{eq:xpeak} with values of $x_{\rm peak}$ determined numerically, and we find them in very good agreement. For $\aV+\aF \gtrsim 10^{-2}$, \eqref{eq:xpeak} occurs much after the first chemical decoupling. In the top and the bottom left plots of \cref{fig:Abundances}, this manifests as a second plateau of the DM yield at large $x$. Clearly, the recoupling of the depletion processes at low temperatures results in very significant decrease of the DM abundance. This impels the re-determination of the couplings that give rise to the observed DM density.

\begin{figure}[t!]
\centering
{}\hfill
\includegraphics[width=0.45\textwidth]{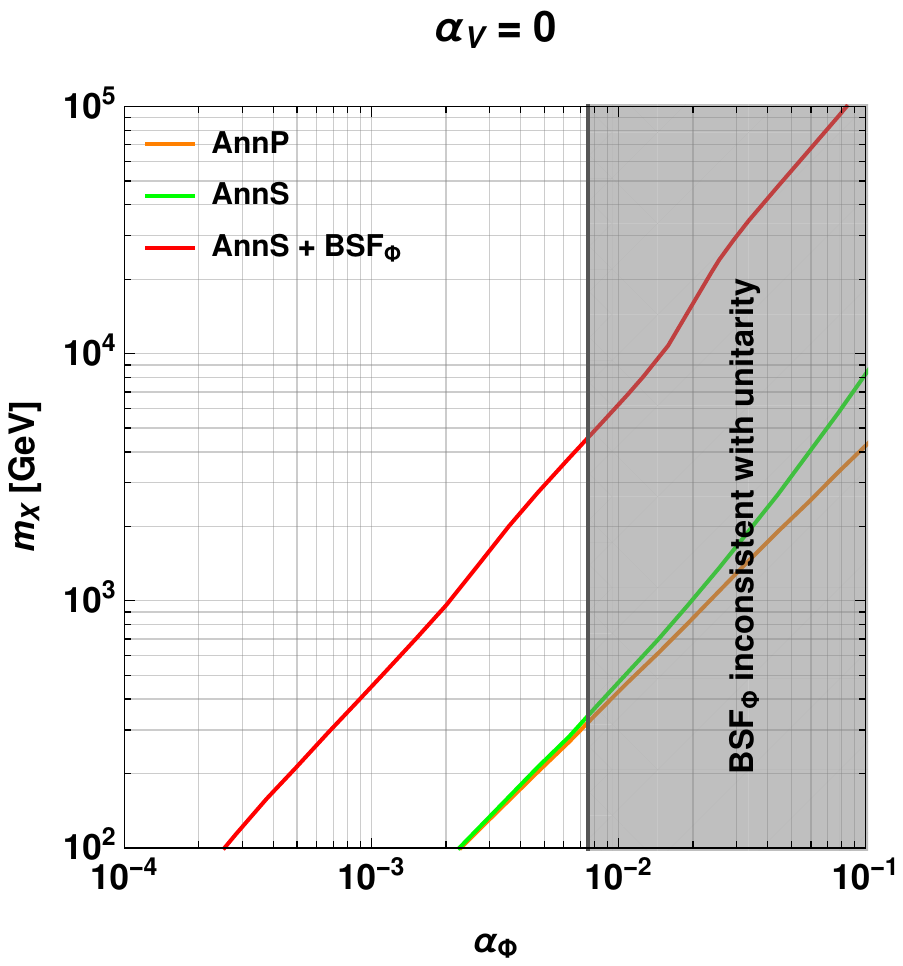}
\hfill
\includegraphics[width=0.45\textwidth]{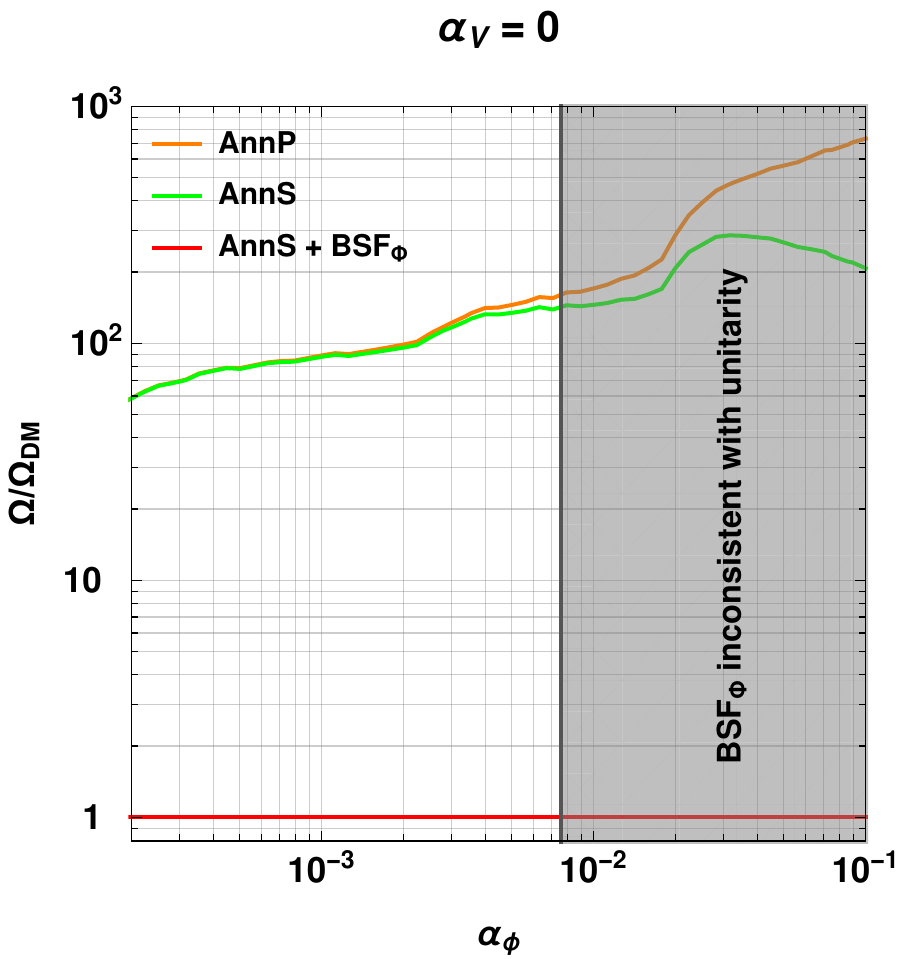}
\hfill{}
	
\caption{\label{fig:mX_vs_aF_aVzero} 
In the limit of the local symmetry becoming global, $\aV = 0$, we show the relation between the DM mass $\mX$ and the coupling $\aF$ to the scalar mediator (\emph{left}), and the effect of \BSFF \ on the DM density (\emph{right}).}
\end{figure}

\begin{figure}[t!]
\centering
{}\hfill
\includegraphics[width=0.45\textwidth]{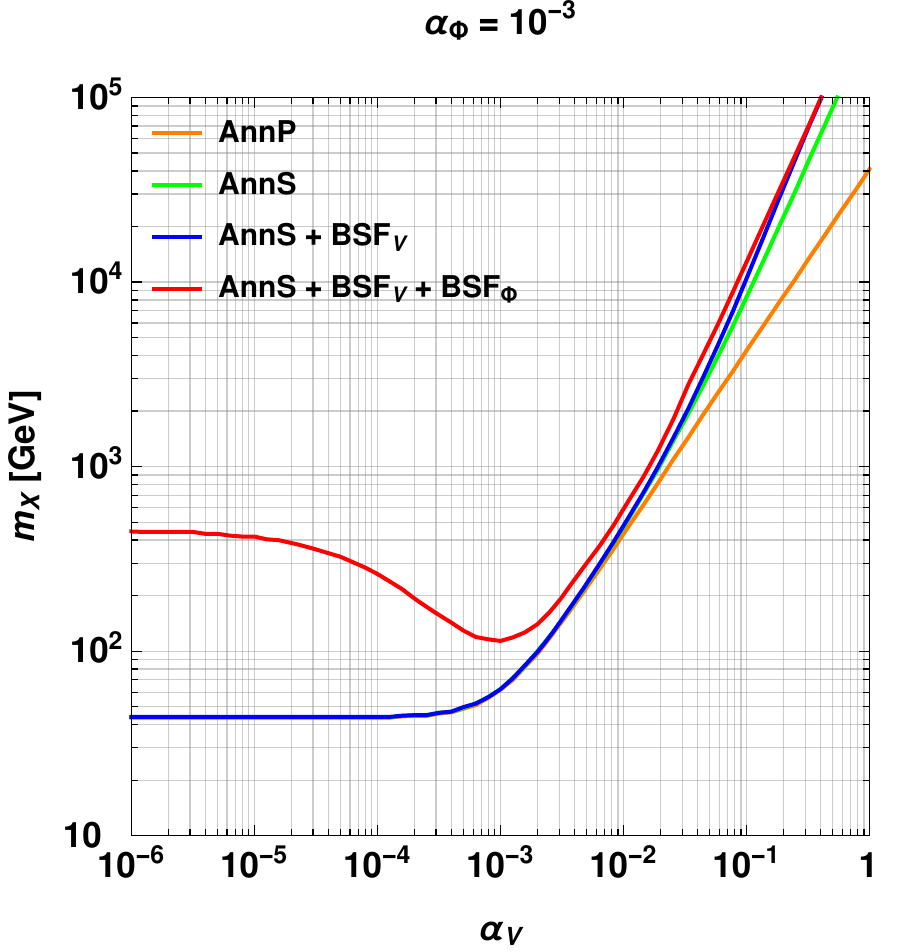}
\hfill
\includegraphics[width=0.45\textwidth]{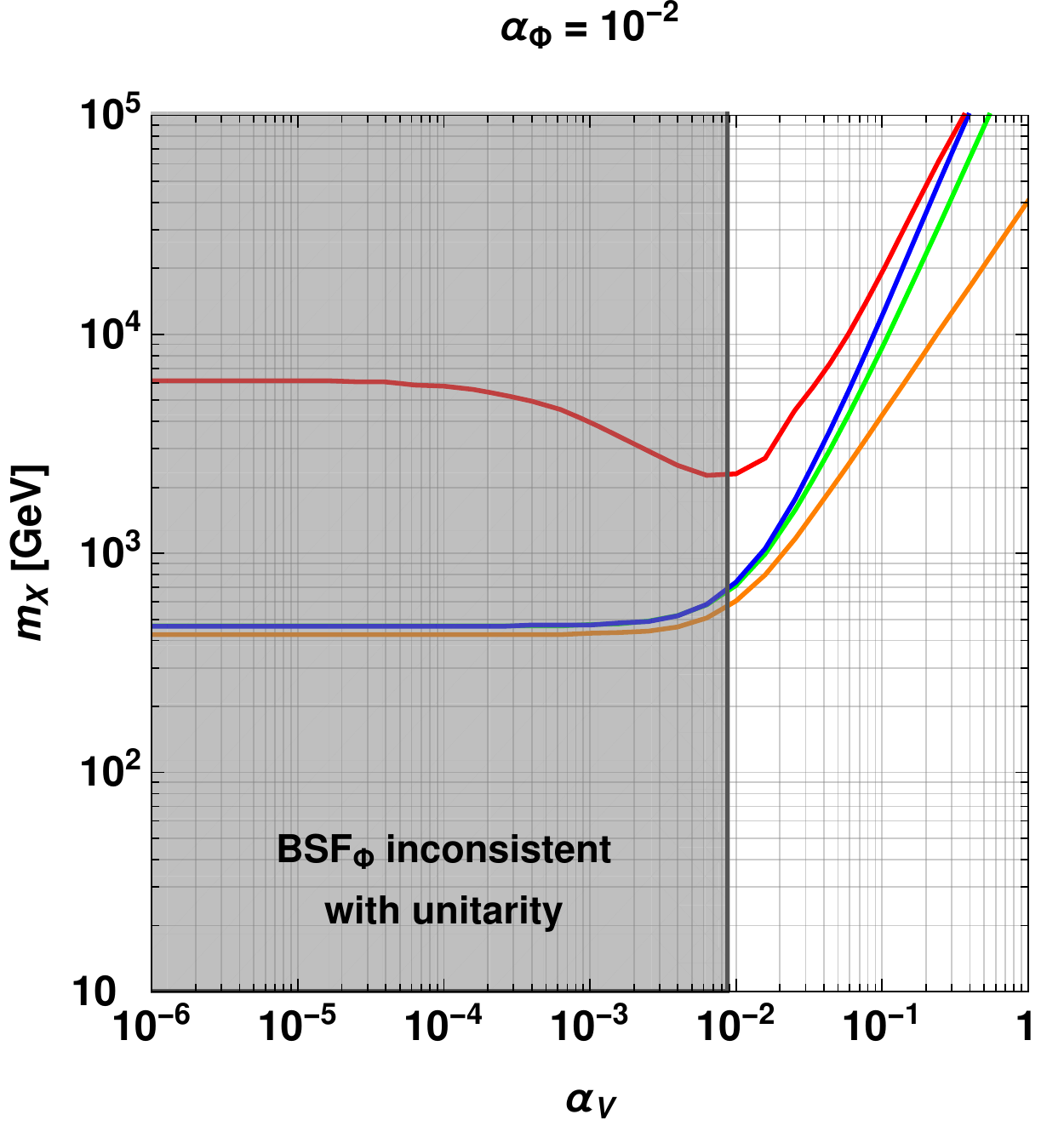}
\hfill{}

\bigskip

{}\hfill
\includegraphics[width=0.45\textwidth]{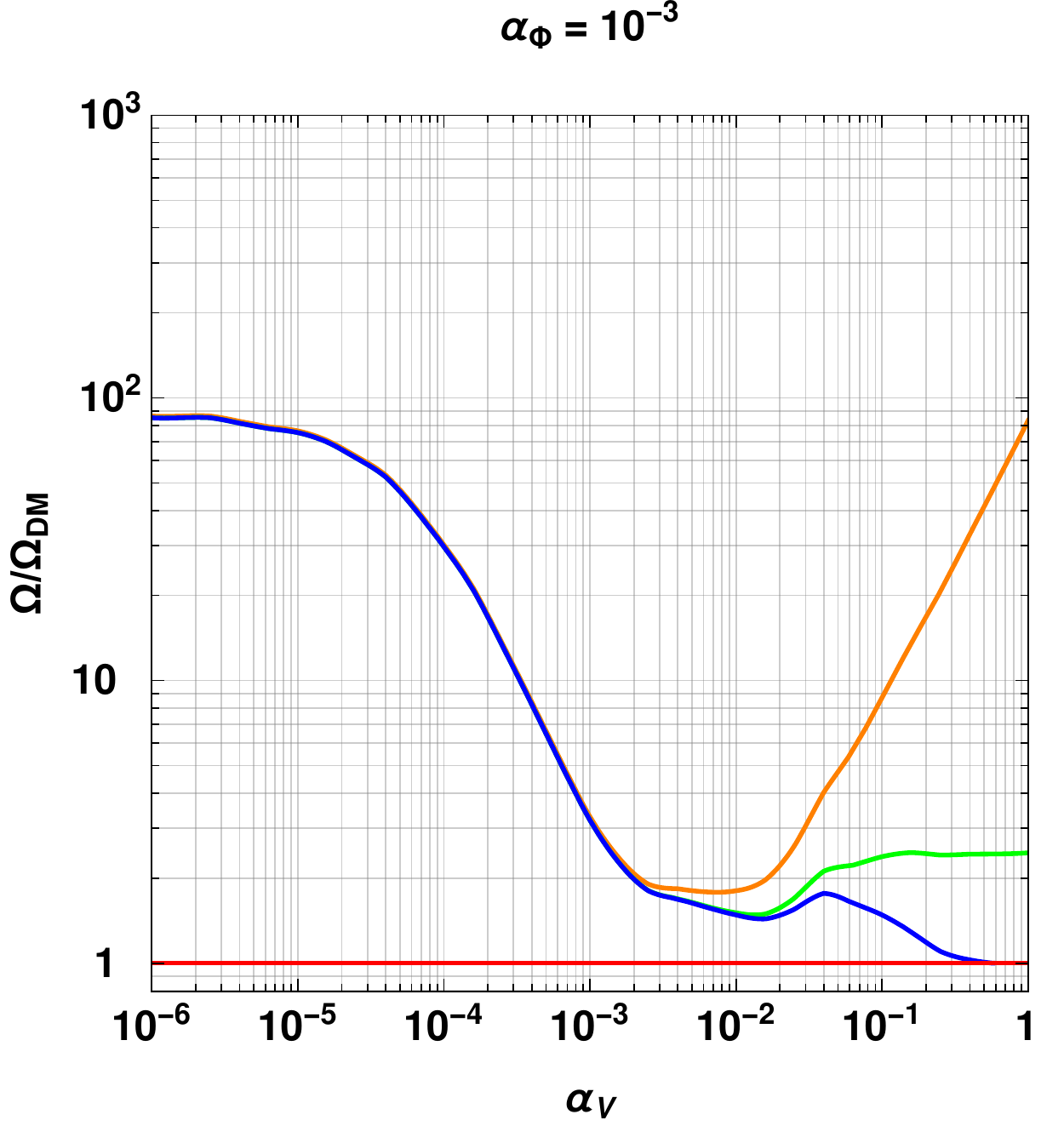}
\hfill
\includegraphics[width=0.45\textwidth]{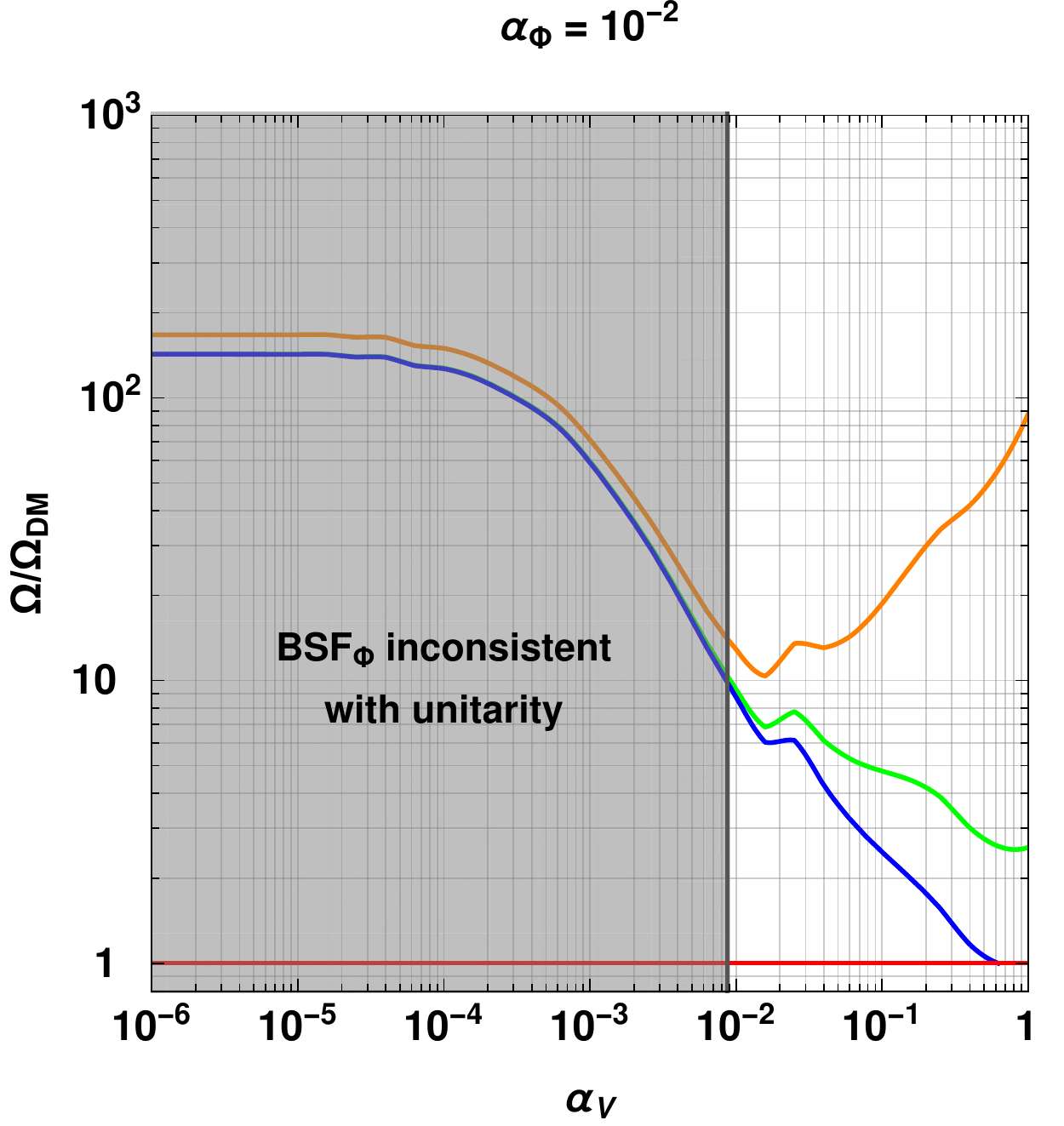}
\hfill{}

\caption{\label{fig:mx_vs_aV_AND_omega}
\emph{Top:} The relation between the DM mass $\mX$ and the coupling to the gauge boson $\aV$ that reproduces the observed DM density, when considering different contributions to the DM depletion. In the gray-shaded region, our computation of the \BSFF \ cross-section violates unitarity within a range of velocities.  
\emph{Bottom:} The $X,X^\dagger$ relic density when taking into account only some of the DM depletion processes, normalised to the observed DM density. For each value of $\aV$ we have chosen $\mX$ such that the observed DM density is reproduced by the \AnnS+\BSFV+\BSFF \ calculation (red line in plots above).}
\end{figure}

\begin{figure}[t!]
\centering
{}\hfill
\includegraphics[width=0.45\textwidth]{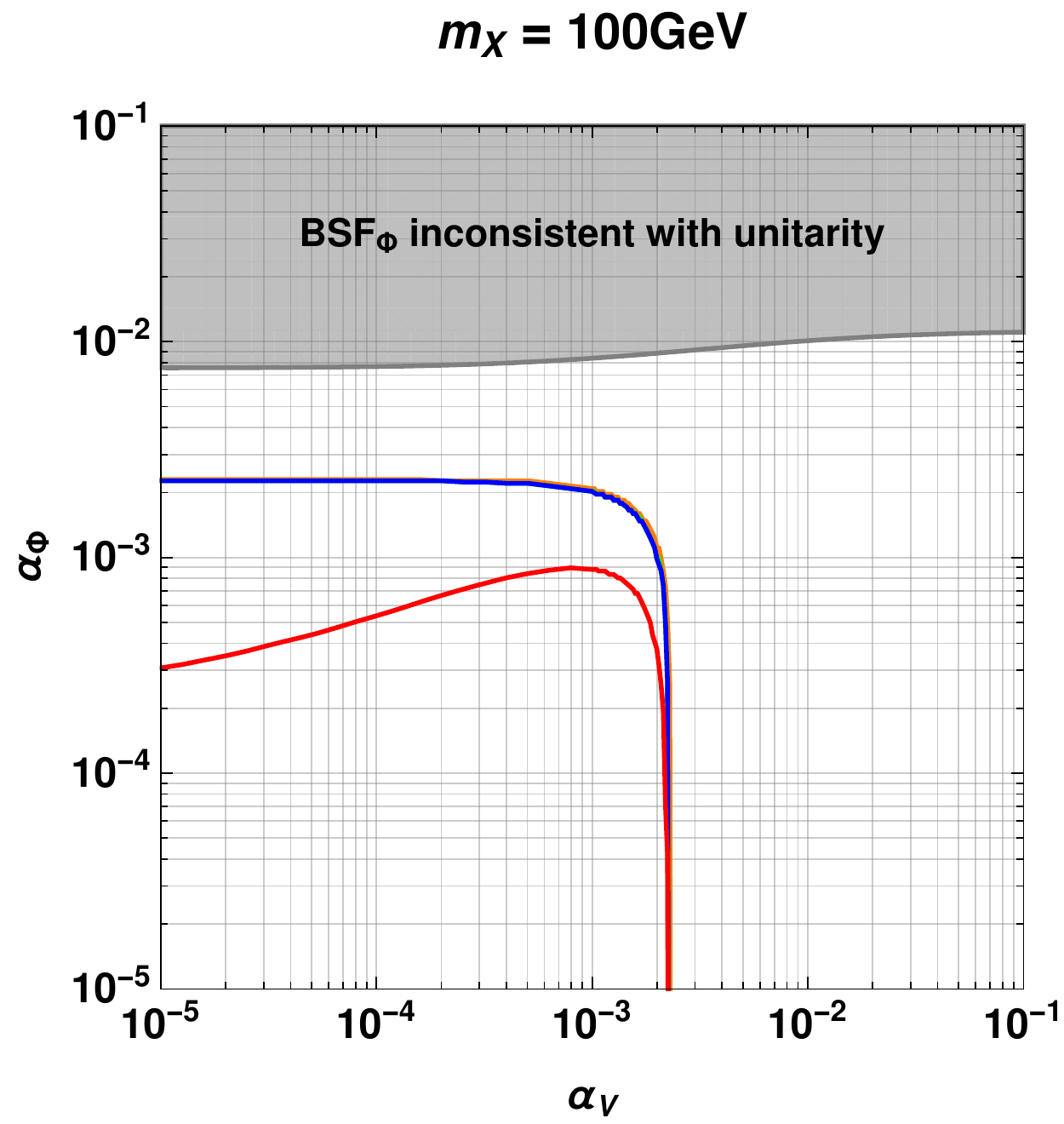}
\hfill
\includegraphics[width=0.45\textwidth]{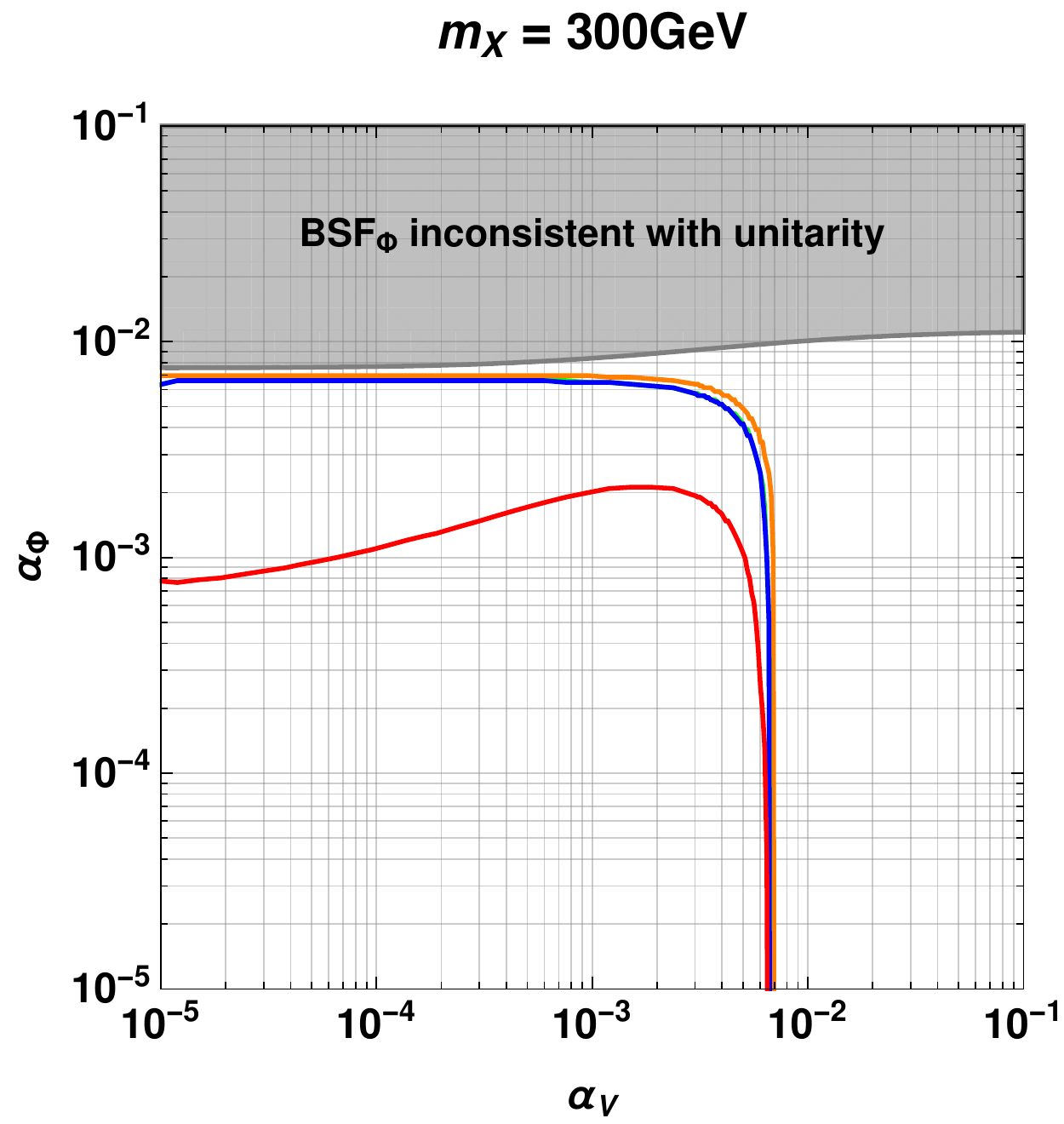}
\hfill{}
	
\bigskip
	
{}\hfill
\includegraphics[width=0.45\textwidth]{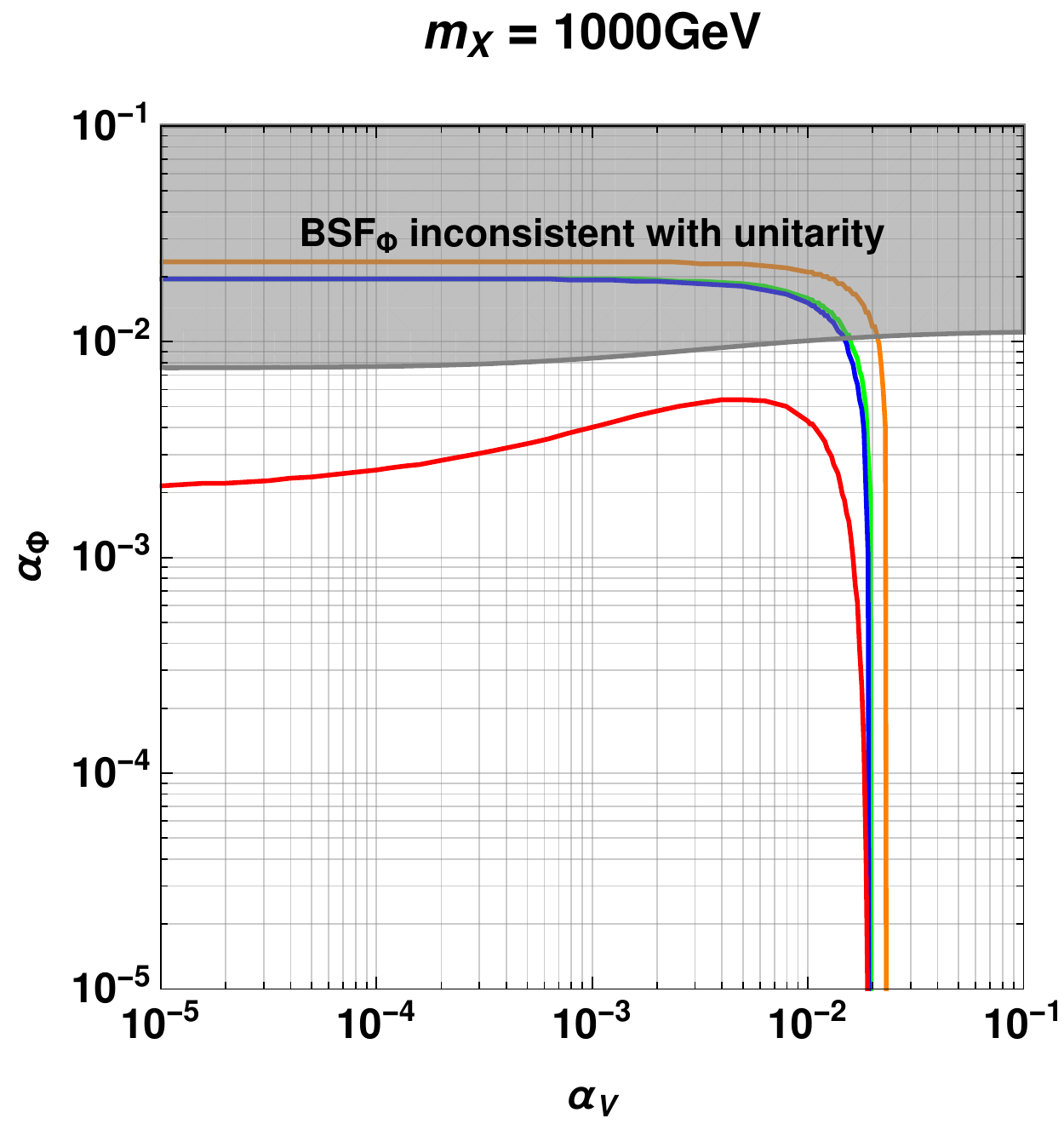}
\hfill
\includegraphics[width=0.45\textwidth]{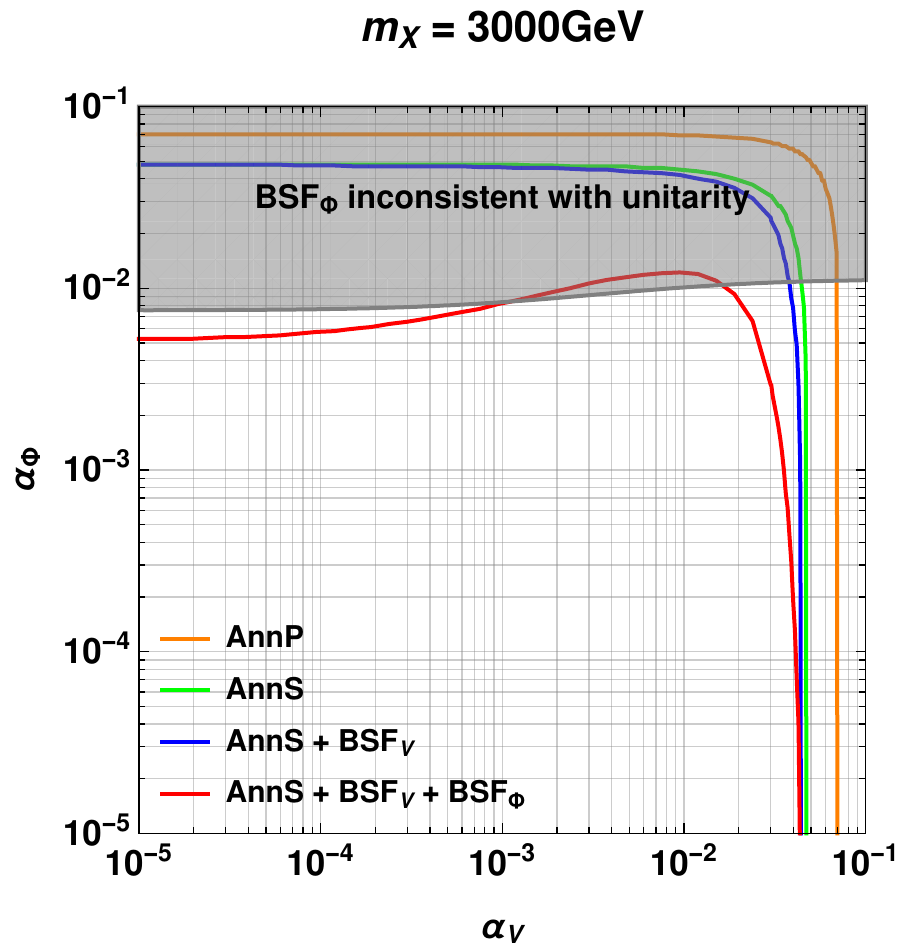}
\hfill{}
	
\caption{\label{fig:aF_vs_aV}
The combination of the couplings $\aV,\aF$ that reproduces the observed DM abundance, for fixed values of the DM mass, when considering different contributions to the DM depletion.}
\end{figure}

\subsubsection{Mass-coupling relation \label{sec:FreezeOut_CouplingMass}}

We solve the Boltzmann \cref{eq:BoltzmannEqs_Eff} numerically and determine the relation between $\aV,\aF$ and $\mX$ that reproduces the observed DM density. In \cref{fig:mx_vs_aV_AND_omega,fig:aF_vs_aV,fig:mX_vs_aF_aVzero}, we present our results. 

From the previous discussion, we expect that the effect of \BSFF \ is more pronounced at small $\aV$, and in particular for $\aV \lesssim \aF$. For this reason, in \cref{fig:mX_vs_aF_aVzero}, we focus on the limit of global symmetry, $\aV =0$, and determine the relation between $\mX$ and $\aF$, while in \cref{fig:mx_vs_aV_AND_omega,fig:aF_vs_aV} we consider also the dependence on $\aV$. 

In all cases, we see that taking \BSFF \ into account changes the predicted couplings or mass very significantly, even by an order of magnitude. The effect on the DM density is illustrated in the right plot of \cref{fig:mX_vs_aF_aVzero} and the bottom plots of \cref{fig:mx_vs_aV_AND_omega}. We pick the combination of parameters that reproduce the observed DM abundance when considering \AnnS+\BSFV+\BSFF, and then calculate the final density attained if only \AnnP, \AnnS, or \AnnS+\BSFV \ are taken into account. We observe the \BSFF \ can deplete the DM density by more than two orders of magnitude.

Because of the interplay of the couplings $\aV$ and $\aF$ in $\<\sigma \vrel\>_{\eff}$, the DM relic density does not always vary monotonically with the parameters. In particular, for fixed $\mX$ and $\aF$ within some range, we observe in \cref{fig:mx_vs_aV_AND_omega,fig:aF_vs_aV} that there are two values of $\aV$ that reproduce the observed DM density: a value in the range $\aV > \aF$ where \BSFF \ has little effect, and a value in the range $\aV < \aF$ where \BSFF \ has significant impact.

\subsection{Validity of the Coulomb approximation  \label{sec:FreezeOut_CoulombApprox}}

Throughout this paper, we have neglected the mass of the charged scalar $\Phi$ (as well as the possibility of a non-vanishing $V$ mass). The calculation of \cref{Sec:BSFviaScalarEmission} can be generalised to include non-zero masses for $\Phi$ and $V$ by evaluating numerically the wavefunctions and the overlap integral, as in refs.~\cite{Petraki:2016cnz,Harz:2017dlj,Harz:2019rro}. Here we examine the validity of the Coulomb approximation in the computation of the DM relic density.

In the present model, $\mF$ affects \BSFF \ via the bound-state wavefunction and the phase-space suppression due to $\Phi$ emission. Note that $\mF$ does not affect the $XX$ and $X^\dagger X^\dagger$ scattering states that participate in \BSFF. The conditions for the Coulomb approximation to be valid are as follows.
\begin{enumerate}[(i)]
\item
For a pure Yukawa potential, the ground state is Coulombic if the mediator mass is much smaller than the Bohr momentum. In the absence of $V$, this would imply $\mF \ll \mu \aF$~\cite{Petraki:2016cnz}. The presence of the $V$-mediated attractive Coulomb potential relaxes this condition~\cite{Harz:2019rro}.  Indicatively we note that, neglecting the $V$-generated potential, the binding energy is larger than 90\% of its Coulomb value if $\mF < \mu \aF/10$~\cite[fig.~13]{Petraki:2016cnz}. On the other hand, for $\aV=\aF$, this occurs if $\mF < \mu \aF/2$~\cite[fig.~6]{Harz:2019rro}. For simplicity, we shall thus assume the following condition for the Coulomb approximation
\begin{align}
{\rm few} \times \mF \lesssim \mu \aB = \mu (\aV+\aF) .
\label{eq:CoulombApprox_BoundStates}
\end{align}
%

\item
\BSFF \ is kinematically accessible if $\mF < {\cal E}_{\bf k} + |\EB| \simeq (\mu/2) [(\aV+\aF)^2+\vrel^2]$. In the thermal bath, $\<\mu \vrel^2/2 \> = 3T/2$. During the first freeze-out the temperature is large, $T \gg |\EB|$, and ${\cal E}_{\bf k}$ dominates the energy available to be dissipated. However, the recoupling of the DM depletion processes occurs at $T \sim |\EB|$ (if at all), when $\<{\cal E}_{\bf k}\> \sim |\EB|$. Therefore, we require
\begin{align}
\mF \lesssim \mu \aB^2/2 = (\mu/2) (\aV+\aF)^2 .
\label{eq:CoulombApprox_PhaseSpace}
\end{align}
\end{enumerate}
The condition \eqref{eq:CoulombApprox_PhaseSpace} is stronger than \eqref{eq:CoulombApprox_BoundStates}. Particularly for the small values of $\aF$ and $\aV$ we have considered here, it ensures that the bound states are very nearly Coulombic.

For \BSFV, there is no kinematic blocking, provided that $V$ is massless. However, a non-vanishing $\mF$ affects the $XX^\dagger$ scattering state, as well as the $XX^\dagger$ bound state. For the latter, the condition for the Coulomb approximation is \eqref{eq:CoulombApprox_BoundStates}. 
We briefly discuss the scattering state. In the case of a pure attractive Yukawa potential, the Coulomb limit is obtained if the mediator mass is lower than the average momentum transfer, $\mF \lesssim \mu \vrel$~\cite{Petraki:2016cnz}. For a pure repulsive Yukawa potential the condition is somewhat stronger. On the other hand, this condition is relaxed by the superposition of the $V$-mediated attractive Coulomb force~\cite[fig.~2]{Harz:2017dlj}. Since \BSFV \ can be important only at early times, when the average kinetic energy is still fairly large, and provided that $\aV+\aF$ is sufficiently large, the Coulomb approximation is typically justified~(see e.g.~discussion in ref.~\cite{Cirelli:2016rnw}). Regardless of the validity of the approximation, \BSFV \ is not the focus of this paper, thus we do not elaborate on this issue further.

Finally, we note that if the charged scalar obtains a VEV, $v_{\mathsmaller{\Phi}}$, the symmetry-breaking phase transition is expected to occur at temperature $T_{\mathsmaller{\rm PT}} \lesssim v_{\mathsmaller{\Phi}}$. Then, if $v_{\mathsmaller{\Phi}} < |\EB|$, the DM chemical decoupling -- including both the first freeze-out and the recoupling epoch -- takes place essentially in the unbroken phase, and the computation of this section is applicable. Assuming that \eqref{eq:CoulombApprox_PhaseSpace} holds, the condition $v_{\mathsmaller{\Phi}} < |\EB|$ is satisfied in models where $v_{\mathsmaller{\Phi}} \lesssim \mF$. However, the DM chemical decoupling may occur in the unbroken phase even for $v_{\mathsmaller{\Phi}} \gg \mF$, if both $v_{\mathsmaller{\Phi}}$ and $\mF$ are much lower than $|\EB|$.

\section{Conclusion \label{Sec:Conclusion}}

The existence of bound states is a generic feature of theories with light force mediators. The formation of stable or metastable bound states has severe implications for the phenomenology of DM today. Scalar force mediators have been invoked in a variety of theories, including self-interacting DM and Higgs portal models residing in the multi-TeV regime.

Here, we computed the cross-sections for the radiative capture of non-relativistic particles into bound states via emission of a scalar that is charged under either a local or global symmetry. The emission of a charged scalar alters the Hamiltonian between the interacting particles, and precipitates extremely rapid transitions. We have provided analytical formulae in the Coulomb approximation for the capture into any bound level [cf.~\cref{eq:sigmav_BSF_ScalarEmission}]. While we carried out our calculations in the context of a minimal $U(1)$ model, our results are readily generalisable to more complex models, including perturbative non-Abelian theories, and can thus be relevant to the phenomenology of various scenarios, e.g.~\cite{Harz:2017dlj,Harz:2019rro,Lonsdale:2014wwa,Lonsdale:2017mzg,Lopez-Honorez:2017ora}. Importantly, our results can be recast to compute BSF via scattering on a bath of relativistic particles, through exchange of a charged scalar, according to ref.~\cite{Binder:2019erp}. This can be particularly important for the chemical decoupling of multi-TeV WIMP DM coupled to the 125~GeV Higgs~\cite{Lopez-Honorez:2017ora}; in this regime the Higgs can indeed act as a light mediator~\cite{Harz:2017dlj,Harz:2019rro}, even if its on-shell emission in capture processes is not kinematically allowed.

The phenomenological implications of the processes we computed can be striking. Here, we demonstrated that the formation of particle-antiparticle bound states via emission of a charged scalar and their subsequent decay can deplete the DM density by as much as two orders of magnitude. While for simplicity we considered only capture into the ground state, the computed cross-sections strongly suggest that the capture into excited states during the DM chemical decoupling should also be significant, thereby producing an even more important effect. The depletion of DM via these processes in the early universe alters the predicted relation between the DM mass and couplings rather dramatically. This in turn implies very different predictions for the DM signals in collider, direct and indirect detection experiments.

For indirect detection, the modification in the predicted relation between the DM mass and couplings implies that the signals arising from the direct annihilation processes and BSF via vector emission are very suppressed with respect to what expected when neglecting BSF via charged scalar emission during freeze-out. This essentially invalidates any existing constraints. On the other hand, BSF via charged scalar emission occurring during CMB or inside halos today may itself produce very significant radiative signals that result in strong constraints. 
For direct detection experiments, the implications are again varied. The larger predicted DM mass can bring a model previously thought to reside in the sub-GeV regime, within the threshold of current detectors. On the other hand, it can relax existing constraints for models already within the experimental sensitivity. 
Finally, the large \BSFF \ cross-sections may imply late kinetic decoupling of DM from radiation in the early universe, as well as strong DM self-interactions inside halos today; both features can potentially affect the galactic structure very significantly.

However, as discussed earlier, the magnitude of the computed cross-sections eventuates in the apparent violation of unitarity already at rather low values of the relevant couplings. Reliable phenomenological studies therefore necessitate first the appropriate treatment of this issue.

\paragraph{Note added:} 

While we were finalising this manuscript, ref.~\cite{Ko:2019wxq} appeared on the arXiv, which considers a similar setup: fermionic DM coupled to a dark $U(1)$ gauge force and to a doubly charged complex scalar, which is assumed to obtain a VEV. Reference~\cite{Ko:2019wxq} points out the non-cancellation of the leading order term in the overlap integral governing the transitions between eigenstates of different potentials that occur via emission of the Goldstone mode, which, in that setup, has been absorbed by the gauge boson.  In the present work, we do not assume a VEV for the scalar mediator, and our computation demonstrates the effect also in the limit of the underlying symmetry being global. Moreover, we provide analytical formulas in the Coulomb limit for the cross-sections of BSF via charged scalar emission, which are readily generalisable to non-Abelian theories. We note that the fermionic and scalar models exhibit different symmetry properties that determine whether the contributions under consideration survive (cf.~\cref{sec:BSFviaScalarEmission_Amplitude}). Both ref.~\cite{Ko:2019wxq} and the present work find that transitions proportional to the overlap of the incoming and outgoing wavefunctions without any momentum suppression, result in the recoupling of the DM depletion processes at late times.

\section*{Acknowledgements}
We thank Iason Baldes, Andreas Goudelis, Alex Kusenko, Eric Laenen and Marieke Postma for useful discussions. This work was supported by the ANR ACHN 2015 grant (``TheIntricateDark" project), and by the NWO Vidi grant ``Self-interacting asymmetric dark matter".

\appendix
\section*{Appendices}

\section{Non-relativistic potential: $t$ and $u$ channels \label{App:Potential}}

\begin{figure}[h!]
\centering
\begin{tikzpicture}[line width=1pt, scale=1.2]
\begin{scope}[shift={(0,0)}]
\node at (-2.3,1){$X$};			\node at (2.3,1){$X$};
\node at (-2.3,0){$X^\dagger$};	\node at (2.3,0){$X^\dagger$};
\draw[momentum] (-1.9, 1.2) 	-- (-1.4, 1.2);\draw[momentum] (+1.4, 1.2) 	-- (+1.9, 1.2);
\draw[momentum] (-1.9,-0.2) 	-- (-1.4,-0.2);\draw[momentum] (+1.4,-0.2) 	-- (+1.9,-0.2);
\node at (-1.6, 1.5){$P/2 +p$};	\node at ( 1.6, 1.5){$P/2 +p'$};
\node at (-1.6,-0.5){$P/2 -p$};	\node at ( 1.6,-0.5){$P/2 -p'$};
\draw (-2,1) -- (2,1);\draw[fermion]    (-2,1) -- (-0.4,1);\draw[fermion]    (0.7,1) -- (2,1);
\draw (-2,0) -- (2,0);\draw[fermionbar] (-2,0) -- (-0.4,0);\draw[fermionbar] (0.7,0) -- (2,0);
\draw[fill=lightgray,shift={(0,0.5)}] (-0.5,-0.5) rectangle (0.5,0.5);
\node at (0,0.5){${\cal A}^{\twoPI}$};
\end{scope}
\end{tikzpicture}
\caption[]{\label{fig:2PI_general} 
Momentum decomposition of the 2PI diagrams of the $XX^\dagger$ interaction.}
\end{figure}
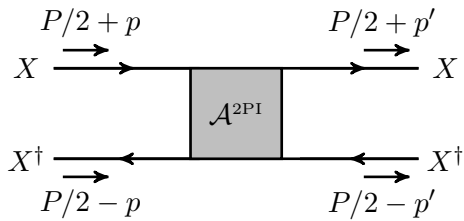

We consider a particle-antiparticle pair $XX^\dagger$ and derive the general formula for the non-relativistic potential arising from $t$-channel and $u$-channel diagrams. The momentum decomposition for the $XX^\dagger$ interaction is shown in \cref{fig:2PI_general}. 

We begin with the Schr\"odinger equation in momentum space (see e.g.~\cite[eq.~(2.78)]{Petraki:2015hla}),
\begin{align}
\(-\frac{{\bf p}^2}{2\mu} + {\cal E}_{n\ell m} \) \tilde{\psi}_{n\ell m} ({\bf p}) = 
\frac{\im}{4M\mu} \int \frac{d^3p'}{(2\pi)^3}
\, \im {\cal A}^\twoPI ({\bf p}, {\bf p}') \ \tilde{\psi}_{n\ell m} ({\bf p}') , 
\label{eq:SchrodingerEq_MomentumSpace}
\end{align}
where ${\cal A}^\twoPI ({\bf p}, {\bf p}')$ is the sum of the 2PI diagrams. We will be interested in $t$-channel and $u$-channel contributions that have the form
\begin{align}
\im {\cal A}^\twoPI ({\bf p}, {\bf p}') 
= \im {\cal A}_{t}^\twoPI ({\bf p} - {\bf p}')
+ \im {\cal A}_{u}^\twoPI ({\bf p} + {\bf p}') .
\end{align}
As in refs.~\cite{Petraki:2015hla,Petraki:2016cnz,Harz:2018csl}, we shall use the Fourier transforms
\begin{align}
\tilde{\psi}_{n\ell m}^{}({\bf p}) 
= \int d^3 r \ \psi_{n\ell m}^{}({\bf r}) \ e^{-\im {\bf p} \cdot {\bf r}} ,
\qquad
\psi_{n\ell m}^{}({\bf r}) 
= \int \frac{d^3 p}{(2\pi)^3} \ \tilde{\psi}_{n\ell m}^{}({\bf p}) \ e^{\im {\bf p} \cdot {\bf r}} .
\end{align}
Applying the operator $\int d^3p / (2\pi)^3 \, \exp(\im {\bf p \cdot r})$  on \cref{eq:SchrodingerEq_MomentumSpace}, we obtain
\begin{align}
&\(\frac{\nabla^2}{2\mu} + {\cal E}_{n\ell m} \) \psi_{n\ell m} ({\bf r}) = 
\frac{\im}{4M\mu} \int \frac{d^3p}{(2\pi)^3} \frac{d^3p'}{(2\pi)^3} 
\, e^{\im {\bf p \cdot r}}
\[ \im {\cal A}_{t}^\twoPI ({\bf p} - {\bf p}') 
+  \im {\cal A}_{u}^\twoPI ({\bf p} + {\bf p}') \] 
\tilde{\psi}_{n\ell m} ({\bf p}') 
\nn \\
&= 
\frac{\im}{4M\mu} \left\{
\[ \int \frac{d^3q}{(2\pi)^3}  \, e^{ \im {\bf q \cdot r}}  \ \im {\cal A}_{t}^\twoPI ({\bf q}) \]
\psi_{n\ell m} ({\bf r}) 
+
\[ \int \frac{d^3q}{(2\pi)^3}  \, e^{ \im {\bf q \cdot r}}  \, \im {\cal A}_{u}^\twoPI ({\bf q}) \]
\psi_{n\ell m} (-{\bf r}) 
\right\}
\nn \\
&= \[ V_t ({\bf r}) + V_u ({\bf r}) \] \psi_{n\ell m} ({\bf r}) ,
\label{eq:SchrodingerEq_CoordinateSpace_prelim}
\end{align}
where we used that $\psi_{n\ell m} (-{\bf r}) = (-1)^\ell \psi_{n\ell m} ({\bf r})$, and
\begin{subequations}
\label{eq:Potentials_tANDuChannels}
\label[pluralequation]{eqs:Potentials_tANDuChannels}
\begin{align}
V_t ({\bf r}) &\equiv \frac{\im}{4M\mu} 
\int \frac{d^3q}{(2\pi)^3} \, e^{\im {\bf q \cdot r}} 
\ \im {\cal A}_{t}^\twoPI ({\bf q}) ,
\label{eq:Potentials_tChannel}
\\
V_u ({\bf r}) &\equiv \frac{\im}{4M\mu} \ (-1)^\ell
\int \frac{d^3q}{(2\pi)^3} \, e^{\im {\bf q \cdot r}} 
\ \im {\cal A}_{u}^\twoPI ({\bf q}) .
\label{eq:Potentials_uChannel}
\end{align}
\end{subequations}
We observe that the $u$-channel contribution depends on the angular momentum mode of the eigenstate. 
\Cref{eq:SchrodingerEq_CoordinateSpace_prelim} can now be rewritten in the familiar order
\begin{align}
\[-\frac{\nabla^2}{2\mu} +  V_t ({\bf r}) + V_u ({\bf r})\]  \psi_{n\ell m} ({\bf r}) 
= {\cal E}_{n\ell m} \psi_{n\ell m} ({\bf r}) .
\label{eq:SchrodingerEq_CoordinateSpace}
\end{align}

\section{Overlap intergral for charged scalar emission \label{App:OverlapIntegral}}

\subsection{Wavefunctions \label{app:OverlapIntegral_WF}}

We consider scattering and bound states in two different Coulomb potentials
\begin{subequations}
\label{eq:SchrodingerEqs}
\label[pluralequation]{eqs:SchrodingerEqs}
\begin{align}
\(-\frac{\nabla^2}{2\mu} - \frac{\aS}{r} \) \phi_{\bf k}({\bf r}) 
&= {\cal E}_{\bf k} \, \phi_{\bf k}({\bf r}) ,
\label{eq:SchrEq_Scatt}
\\
\(-\frac{\nabla^2}{2\mu} - \frac{\aB}{r} \) \psi_{n\ell m}({\bf r}) 
&= {\cal E}_{n} \, \psi_{n\ell m}({\bf r}) .
\label{eq:SchrEq_Bound}
\end{align} 
\end{subequations}
The expectation value of the momentum of each particle in the CM frame in the scattering state, and the Bohr momenta for the scattering and bound states are
\begin{subequations}
\label{eq:Momenta}
\label[pluralequation]{eqs:Momenta}
\begin{align}
{\bf k} &\equiv \mu \vrel , \label{eq:k}    \\
\kappaS &\equiv \mu \aS , \label{eq:kappaB} \\
\kappaB &\equiv \mu \aB . \label{eq:kappaS}
\end{align} 
\end{subequations}
For convenience, we define the parameters
\begin{subequations}
\label{eq:zetas}
\label[pluralequation]{eqs:zetas}
\begin{align}
\zetaS &\equiv \kappaS/k = \aS / \vrel , \label{eq:zetaS} \\
\zetaB &\equiv \kappaB/k = \aB / \vrel , \label{eq:zetaB}
\end{align}
\end{subequations}
as well as the space variables
\begin{subequations}
\label{eq:SpaceVariables}
\label[pluralequation]{eqs:SpaceVariables}
\begin{align}
\xS &\equiv k r , \label{eq:xS} \\
\xB &\equiv \kappaB r . \label{eq:xB}
\end{align}
\end{subequations}
The energy eigenvalues of the scattering and bound states are
\begin{subequations}
\label{eq:EnergyEigenvalues}
\label[pluralequation]{eqs:EnergyEigenvalues}
\begin{align}
{\cal E}_{\bf k}
&= \frac{\bf k^2}{2\mu} = \frac{\mu \vrel^2}{2} ,
\label{eq:EnergyEigenvalue_Scatt}
\\
{\cal E}_{n}
&=-\frac{\kappaB^2}{2\mu\, n^2}
 =-\frac{\mu \aB^2}{2n^2} ,
\label{eq:EnergyEigenvalue_Bound}
\end{align} 
\end{subequations}
and the corresponding wavefunctions are\footnote{
For the spherical harmonics, we assume the normalisation 
$\int d\Omega \, Y_{\ell m} (\Omega) Y_{\ell' m'}^* (\Omega) = \delta_{\ell \ell'} \delta_{m m'}$.}
\begin{subequations}
\label{eq:WaveFunctions}
\label[pluralequation]{eqs:WaveFunctions}
\begin{align}
\phi_{\bf k} ({\bf r}) 
&= 4\pi \sqrt{S_0(\zetaS)} \, \sum_{\ellS=0}^{\infty} \ \sum_{\mS=-\ellS}^{\ellS}
Y_{\ellS \mS}^* (\Omega_{\bf k}) \ Y_{\ellS \mS} (\Omega_{\bf r})
\nn \\
&\times
\ \frac{(-\im^{\ellS})}{(2\ellS+1)! }
\ \frac{\Gamma(1+\ellS-\im \zetaS)}{\Gamma(1-\im \zetaS)}
\ e^{-i \xS}
\ (2\xS)^{\ellS} 
\ {}_1F_1 (1+\ellS + \im \zetaS; ~ 2\ellS + 2; ~ 2 \im \xS) ,
\label{eq:phi}
\\
\psi_{n\ell m} ({\bf r}) &= 
\kappaB^{3/2}  \, Y_{\ell m} ({\Omega_{\bf r}}) 
\ \frac{2}{n^2 (2\ell+1)!} \[\frac{(n+\ell)!}{(n-\ell-1)!}\]^{1/2} \times
\nn \\
&\times 
e^{-\xB/n} \ \(\frac{2\xB}{n}\)^{\ell} 
{}_1F_1 \(-n+\ell+1; ~ 2\ell+2; ~ \frac{2\xB}{n} \) ,
\label{eq:psi}
\end{align}
\end{subequations}
where $S_0(\zetaS) \equiv 2\pi \zetaS / (1-e^{-2\pi\zetaS})$, and in \cref{eq:psi}, we have expressed the bound state wavefunction in terms of the confluent hypergeometric function ${}_1F_1$ rather than the Laguerre polynomials. 
Note that the wavefunctions \eqref{eq:WaveFunctions} assume \emph{distinguishable} interacting particles. We include the necessary (anti)symmetrization factors for identical particles in \cref{Sec:BSFviaScalarEmission,Sec:FreezeOut}, where we discuss the processes of interest.\footnote{
For clarity, in \cref{Sec:BSFviaScalarEmission,Sec:FreezeOut} we denote the wavefunctions of distinguishable particles with the superscript $\DP$. In this appendix, we have omitted this superscript since there is no risk of confusion.}

\subsection{Integral \label{app:OverlapIntegral_Integral}}

We are interested in computing the overlap integral
\begin{align}
{\cal R}_{{\bf k}, n\ell m} \equiv 
\kappaB^{3/2} \, \int d^3 r \ \psi_{n\ell m}^* ({\bf r}) \, \phi_{\bf k} ({\bf r})
\label{eq:Rcal_def_app}
\end{align}
for the wavefunctions of \cref{app:OverlapIntegral_WF}. Note that the prefactor in \cref{eq:Rcal_def_app} has been chosen such that ${\cal R}_{{\bf k},n\ell m}$ is dimensionless. 

Substituting \cref{eqs:WaveFunctions} into \eqref{eq:Rcal_def_app} and performing the angular integration, picks out the $\ellS=\ell$ mode of the scattering state. Then, setting $t \equiv 2\xB /n = 2 \zetaB \xS /n$, we obtain
\begin{align}
{\cal R}_{{\bf k}, n\ell m} &=
\frac{\sqrt{S_0(\zetaS)}}{(\zetaB/n)^\ell}  
\, Y_{\ell m}^* (\Omega_{\bf k}) 
\times
\frac{(-\im^\ell)\pi n}{[(2\ell+1)!]^2} \[\frac{(n+\ell)!}{(n-\ell-1)!}\]^{1/2} 
\ \frac{\Gamma(1+\ell-\im\zetaS)}{\Gamma(1-\im\zetaS)}
\nn \\
&\times \int_0^\infty dt
\ t^{2\ell+2} 
\ e^{-\(1+\frac{\im \, n}{\zetaB}\) \frac{t}{2}}
{}_1F_1 \(1+\ell- n; ~ 2\ell+2; ~ t \)
{}_1F_1 \(1+\ell+ \im \zetaS; ~ 2\ell+2; ~ \frac{\im \, n}{\zetaB}t \) .
\label{eq:Rcal_intermediate}
\end{align}
The confluent hypergeometric functions ${}_1F_1$ obey the identity~\cite[section~7.622]{Integrals_GradshteynRyzhik}
\begin{align}
&\int_0^\infty dt \ t^{c-1} \ e^{-\rho t}
{}_1F_1 \(a; ~ c; ~ t \)
{}_1F_1 \(b; ~ c; ~ \lambda t \) = 
\nn \\ 
&=
\Gamma (c) (\rho-1)^{-a} (\rho - \lambda)^{-b} \rho^{a+b-c}
{}_2F_1 \[ a;b;c;~\lambda (\rho-1)^{-1} (\rho-\lambda)^{-1} \] 
\equiv h(\rho; a,b,c,\lambda) , 
\label{eq:Identity} 
\end{align}
for ${\rm Re}(c) > 0$ and ${\rm Re}(\rho) > {\rm Re}(\lambda) + 1$,  where ${}_2F_1$ is the ordinary hypergeometric function. For $a$ a non-positive integer, ${}_1F_1 (a;~c;~t)$ is a finite polynomial in $t$, and we have checked numerically that \cref{eq:Identity} remains valid for ${\rm Re}(c) > 0$ and ${\rm Re}(\rho) > {\rm Re}(\lambda)$, which encompasses the parameter range of interest. Differentiating \eqref{eq:Identity} over $\rho$, and setting
\begin{subequations}
\label{eq:HypergeometricParameters}
\label[pluralequation]{eqs:HypergeometricParameters}
\begin{align}
a&=1+\ell-n ,\\
b&=1+\ell+\im \zetaS ,\\  
c&=2\ell+2 , \\
\rho &= 1/2+\im\,n/(2\zetaB) , \\   
\lambda &= \im\,n/\zetaB ,
\end{align}
\end{subequations}
we obtain the integral needed to compute the second line of \cref{eq:Rcal_intermediate},
\begin{align}
-\frac{dh}{d\rho} 
&= 
2^{2\ell+4} n (2\ell+1)! \,
\(1-\frac{\zetaS}{\zetaB} \) 
\( \frac{\zetaB^2/n^2}{1+\zetaB^2/n^2} \)^{\ell+2}
\ e^{-2\zetaS \, {\rm arccot} (\zetaB/n)}
\nn \\ 
&\times 
{}_2F_1 \( 
1+\ell-n; ~ 
1+\ell +\im \zetaS; ~
2\ell+2; ~
\frac{4 \im \zetaB/n}{(1+\im \zetaB/n)^2}
\)
e^{\im 2(n-\ell-1) \arctan (\zetaB/n)} .
\label{eq:dgdrho} 
\end{align}
Note that the hypergeometric function in \cref{eq:dgdrho} is a finite polynomial in its last argument because its first argument of is a non-positive integer, $1+\ell-n \leqslant 0$. The last factor in \cref{eq:dgdrho} is an unimportant overall phase. Combining \cref{eq:Rcal_intermediate,eq:dgdrho}, we find
\begin{align}
\sum_{m=-\ell}^{\ell}
\int d\Omega_{\bf k} \ 
&|{\cal R}_{{\bf k}, n\ell m}|^2 
= \frac{2^{4(\ell+2)} \, \pi^2 n^4}{2\ell+1}  \
\frac{(n+\ell)!}{(n-\ell-1)!}
\[ \frac{\ell!}{(2\ell)!} \]^2
\nn \\
&\times
\(1-\frac{\zetaS}{\zetaB} \)^2 
\[ S_0(\zetaS) \prod_{j=1}^{\ell} \(1+\frac{\zetaS^2}{j^2}\) \]
\frac{(\zetaB^2/n^2)^{\ell+4}}{(1+\zetaB^2/n^2)^{2\ell+4}} 
\ e^{-4\zetaS \, {\rm arccot} (\zetaB/n)}
\nn \\ 
&\times 
\left|
{}_2F_1 \( 
1+\ell-n; ~ 
1+\ell +\im \zetaS; ~
2\ell+2; ~
\frac{4 \im \zetaB/n}{(1+\im \zetaB/n)^2}
\)
\right|^2 .
\label{eq:RcalSquared}
\end{align}
${\cal R}_{{\bf k}, n\ell m}$ vanishes if $\zetaB=\zetaS$, as expected from the orthogonality of the wavefunctions.

\section{Bound-state formation via vector emission \label{App:BSFV}}

\begin{figure}[t!]
\centering
\begin{tikzpicture}[line width=1pt, scale=0.86]
\begin{scope}[shift={(-6,0)}]
\node at (0.55,3.2){$V$};
\draw[vector]	(0,2) -- (0.4,3);
\draw[momentum]	(-0.1,2.5) -- (0.2,3);
\node at (-0.3,2.8){$P_{\mathsmaller{V}}$};
\node at (-2.3,2){$X$};
\node at (-2.3,0){$X^{\dagger}$};
\node at ( 2.3,2){$X$};
\node at ( 2.3,0){$X^{\dagger}$};
\node at (-1.7,2.6){$K/2+k^{(\prime)}$};		\node at (1.5,2.6){$P/2+p$};
\draw[momentum] 	(-1.8,2.3) 	-- (-1.1, 2.3);	\draw[momentum] 	(1.1, 2.3) 	-- (1.8, 2.3);
\draw[fermion] 		(-2,  2) 	-- ( 0,   2);	\draw[fermion] 		(0,   2) 	-- (2,   2);
\draw[fermionbar]	(-2,  0) 	-- ( 2,   0);
\draw[momentum] 	(-1.8,-0.3) -- (-1.1, -0.3);\draw[momentum] 	(1.1,-0.3) 	-- (1.8,-0.3);
\node at (-1.7,-0.6){$K/2-k^{(\prime)}$};		\node at (1.5,-0.6){$P/2-p$};
\end{scope}
\node at (-3,1){$+$};
\begin{scope}[shift={(0,0)}]
\node at (-2.3,2){$X$};
\node at (-2.3,0){$X^\dagger$};
\node at ( 2.3,2){$X$};
\node at ( 2.3,0){$X^{\dagger}$};
\node at (-1.7,2.6){$K/2+k^{(\prime)}$};		\node at (1.5,2.6){$P/2+p$};
\draw[momentum] 	(-1.8,2.3) 	-- (-1.1, 2.3);	\draw[momentum] 	(1.1, 2.3) 	-- (1.8, 2.3);
\draw[fermion] 		(-2,  2) 	-- ( 2,   2);
\draw[fermionbar]	(-2,  0) 	-- ( 0,   0);	\draw[fermionbar] 	( 0,  0) 	-- ( 2,  0);
\draw[momentum] 	(-1.8,-0.3) -- (-1.1, -0.3);\draw[momentum] 	(1.1,-0.3) 	-- (1.8,-0.3);
\node at (-1.7,-0.6){$K/2-k^{(\prime)}$};		\node at (1.5,-0.6){$P/2-p$};
\node at (0.55,-1.2){$V$};
\draw[vector]	(0,0) -- (0.4,-1);
\draw[momentum]	(-0.1,-0.5) -- (0.1,-1);
\node at (-0.3,-0.8){$P_{\mathsmaller{V}}$};
\end{scope}
\node at (+3,1){$+$};
\begin{scope}[shift={(6,0)}]
\node at (-2.3,2){$X$};
\node at (-2.3,0){$X^\dagger$};
\node at ( 2.3,2){$X^{\dagger}$};
\node at ( 2.3,0){$X$};
\node at (-1.7,2.6){$K/2+k^{(\prime)}$};		\node at (1.5,2.6){$P/2-p$};
\draw[momentum] 	(-1.8,2.3) 	-- (-1.1, 2.3);	\draw[momentum] 	(1.1, 2.3) 	-- (1.8, 2.3);
\draw[fermion] 		(-2,  2) 	-- ( 0,   2);	\draw[fermionbar]	( 0,  2) 	-- ( 2,  2);
\draw[fermionbar]	(-2,  0) 	-- ( 0,   0);	\draw[fermion]	 	( 0,  0) 	-- ( 2,  0);
\draw[momentum] 	(-1.8,-0.3) -- (-1.1, -0.3);\draw[momentum] 	(1.1,-0.3) 	-- (1.8,-0.3);
\node at (-1.7,-0.6){$K/2-k^{(\prime)}$};		\node at (1.5,-0.6){$P/2+p$};
\draw[scalar]	(0,2) -- (0,1);					\draw[scalar]	(0,1) -- (0,0);
\node at (1.3,1){$V$};
\draw[vector]	(0,1) -- (1,1);
\draw[momentum]	(0.2,1.4) -- (0.6,1.4);
\node at (0.85,1.4){$P_{\mathsmaller{V}}$};
\end{scope}
\end{tikzpicture}
\caption[]{\label{fig:BSF_Vector_RadiativePart} 
The leading order contributions to the radiative part of transitions via vector emission $XX^\dagger \to XX^\dagger +V$, in the model of \cref{eq:L}. The arrows on the field lines denote the flow of the $U(1)_D$ charge.}
\end{figure}
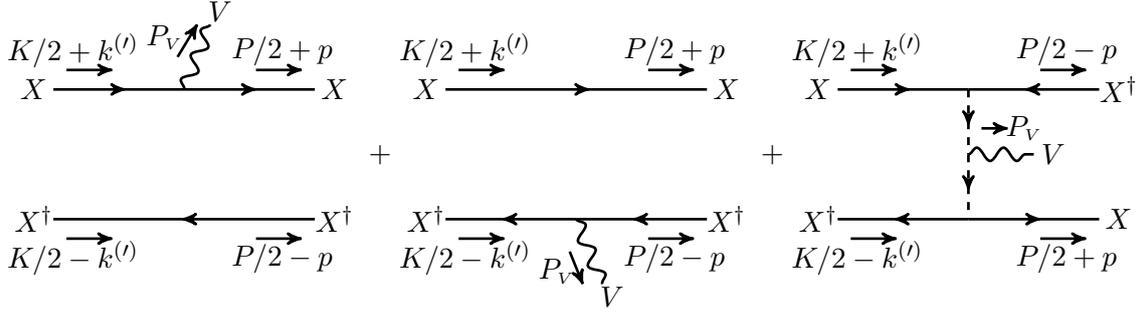

The \BSFV amplitude is~\cite{Petraki:2015hla}
\begin{align}
\im \boldsymbol{{\cal M}}_{{\bf k} \to n\ell m}^{\mathsmaller{V}} 
\simeq 
\int 
\frac{d^3{\bf k'}}{(2\pi)^3}
\frac{d^3{\bf p}}{(2\pi)^3} 
\  \tilde{\phi}_{\bf k}^\XXdagger ({\bf k'})
\ \im \AV ({\bf k',p}) 
\ \frac{[\tilde{\psi}_{n\ell m}^\XXdagger ({\bf p})]^*}{\sqrt{2\mu}} ,
\label{eq:BSF_VectorEmission_M_def}
\end{align}
where the leading order contributions to the perturbative transition amplitude $\AV ({\bf k',p})$ are shown in \cref{fig:BSF_Vector_RadiativePart}. Note that the third diagram does not appear in more minimal $U(1)$ theories where the interacting particles do not couple to a doubly charged scalar. This diagram is akin to the one that appears in non-Abelian theories, where the final-state gluon is radiated from a gluon exchanged between the interacting particles, via the trilinear gluon vertex. Such diagrams seem naively to be of higher order than those with emission directly from one of the interacting particles. However, the momenta exchanged along the propagators scale with the couplings and render these diagrams of the same order as those with emission from the legs~\cite{Asadi:2016ybp}. See also ref.~\cite{Oncala:2018bvl} for analogous contributions arising from couplings in the scalar potential.

Following \cite{Petraki:2015hla,Harz:2018csl}, we find $\AV ({\bf k',p})$ to be
\begin{align}
\im \AV ({\bf k',p}) \simeq -\im g 
&\[
 2m_X \, 2{\bf p} \, (2\pi)^3 \delta^3 ({\bf k'} - {\bf p} - \PVvec/2)
+2m_X \, 2{\bf p} \, (2\pi)^3 \delta^3 ({\bf k'} - {\bf p} + \PVvec/2)
\right. \nn \\ &\left.
+2y^2 M \mu \, \frac{2(\bf k' + p)}{[({\bf k'} + {\bf p})^2 +\mF^2]^2}
\] ,
\label{eq:BSF_VectorEmission_AT}
\end{align}
where we took into account that $\qX=1$ and $\qF=2$. Following \cite{Harz:2018csl}, we define
\begin{subequations}
\label{eq:OverlapIntegrals_JY_def}
\label[pluralequation]{eqs:OverlapIntegrals_JY_def}
\begin{align}
\boldsymbol{{\cal J}}_{{\bf k},n \ell m} ({\bf b}) 
&\equiv \int \frac{d^3p}{(2\pi)^3} 
\, {\bf p} 
\, [\tilde{\psi}_{n\ell m}^{\XXdagger} ({\bf p})]^* 
\, \tilde{\phi}_{\bf k}^{\XXdagger} ({\bf p+b})
= \im \int d^3 r 
\, [\nabla \psi_{n\ell m}^{\XXdagger} ({\bf r})]^* 
\, \phi_{\bf k}^{\XXdagger} ({\bf r})
\, e^{-\im {\bf b} \cdot {\bf r}} ,
\label{eq:OverlapIntegral_J_def}
\\
\boldsymbol{{\cal Y}}_{{\bf k},n \ell m} 
&\equiv 8\pi \mu \aF
\int \frac{d^3k'}{(2\pi)^3}  \frac{d^3p}{(2\pi)^3} 
\, \frac{\bf k'-p}{({\bf k'-p})^4}
\, [\tilde{\psi}_{n\ell m}^{\XXdagger} ({\bf p})]^* 
\, \tilde{\phi}_{\bf k}^{\XXdagger} ({\bf k'})
\nn \\
&
= -\im \mu \aF \int d^3 r 
\, [\psi_{n\ell m}^{\XXdagger} ({\bf r})]^* 
\, \phi_{\bf k}^{\XXdagger} ({\bf r})
\, \hat{{\bf r}} .
\label{eq:OverlapIntegral_Y_def}
\end{align}
\end{subequations}
Then, the amplitude \eqref{eq:BSF_VectorEmission_M_def} becomes
\begin{align}
\im \boldsymbol{{\cal M}}_{{\bf k} \to n\ell m}^{\mathsmaller{V}} 
\simeq 
-\im g \, \frac{4M}{\sqrt{2\mu}} 
\[
  \frac12 \boldsymbol{{\cal J}}_{{\bf k},n \ell m} ( \PVvec/2)
+ \frac12 \boldsymbol{{\cal J}}_{{\bf k},n \ell m} (-\PVvec/2) 
+ \frac{(-1)^\ell y^2}{8\pi \aF} \, \boldsymbol{{\cal Y}}_{{\bf k},n\ell m} 
\] , 
\label{eq:BSF_VectorEmission_M}
\end{align}
where energy-momentum conservation sets $|\PVvec| = \mu (\aB^2+\vrel^2)/ 2$ [cf.~\cref{eq:omega}]. The leading order terms in the $\boldsymbol{{\cal J}}$ integrals are the zeroth order terms in the $\PVvec$ expansion~\cite{Petraki:2016cnz}. For capture into the ground state, both the $\boldsymbol{{\cal J}}$ and the $\boldsymbol{{\cal Y}}$ contributions arise from the $\ell_{\mathsmaller{S}}=1$ mode of the scattering state wavefunction~\cite{Petraki:2016cnz,Harz:2018csl}. In the Coulomb limit~\cite{Harz:2018csl}
\begin{subequations}
\label{eq:OverlapIntegrals_JY_100}
\label[pluralequation]{eqs:OverlapIntegrals_JY_100}
\begin{align}
\boldsymbol{{\cal J}}_{{\bf k},100} ({\bf 0}) 
&= \hat{\bf k} \(\frac{2^6\pi}{k} 
\, S_0 (\zetaS) \, (1+\zetaS^2)
\, \frac{\zetaB^5 \, e^{-4\zetaS {\rm arccot} (\zetaB)}}{(1+\zetaB^2)^4} \)^{1/2},
\label{eq:J100}
\\
\boldsymbol{{\cal Y}}_{{\bf k},100} &= 
(\aF / \aB) \, \boldsymbol{{\cal J}}_{{\bf k},100} ({\bf 0}) ,
\label{eq:Y100}
\end{align}
\end{subequations}
where $\zetaS\equiv \aS/\vrel$ for the scattering state, $\zetaB \equiv \aB/\vrel$ for the bound state, and  $S_0(\zetaS) \equiv 2\pi\zetaS / (1-e^{-2\pi\zetaS})$. Considering the potential \eqref{eq:V_XXdagger}, for the process of interest we find
\begin{align}
\zetaS = \zetaV-\zetaF 
\qquad \text{and} \qquad 
\zetaB = \zetaV+\zetaF .
\end{align}
Collecting the above, we find the amplitude for capture into the ground state to be
\begin{align}
|\boldsymbol{{\cal M}}_{{\bf k} \to 100}^{\mathsmaller{V}} |^2
\simeq  \frac{M^2}{\mu} \aV \(1+ \frac{2\aF}{\aB} \)^2
\ \frac{2^{11}\pi^2}{k} 
\, S_0 (\zetaS) \, (1+\zetaS^2)
\, \frac{\zetaB^5 \, e^{-4\zetaS \, {\rm arccot} (\zetaB)}}{(1+\zetaB^2)^4} .
\label{eq:BSF_VectorEmission_M_100}
\end{align}
Then, the cross-section~\cite{Petraki:2015hla,Harz:2018csl}
\begin{align}
\sigma_{100}^{\mathsmaller{V}} \vrel 
&= 
\frac{|\PVvec|}{2^6 \pi^2 M^2 \mu}
\int d\Omega \(
|\boldsymbol{{\cal M}}_{{\bf k} \to 100}^{\mathsmaller{V}}|^2
- |\hat{\bf P}_{\mathsmaller{V}} \cdot \boldsymbol{{\cal M}}_{{\bf k} \to 100}^{\mathsmaller{V}}|^2
\) 
\label{eq:BSF_VectorEmission_sigma_100_def}
\end{align}
is found to be
\begin{align}
\sigma_{100}^{\mathsmaller{V}} \vrel 
&= 
\frac{2^7 \pi}{3\mu^2} \aV \aB \(1+ \frac{2\aF}{\aB} \)^2
\, S_0 (\zetaS) \, (1+\zetaS^2)
\, \frac{\zetaB^4}{(1+\zetaB^2)^3}
\, e^{-4\zetaS {\rm arccot} (\zetaB)} .
\label{eq:BSF_VectorEmission_sigma_100}
\end{align}
Note that \cref{eq:BSF_VectorEmission_sigma_100} holds also for fermionic DM. This has been already established for the two diagrams on the left of \cref{fig:BSF_Vector_RadiativePart}. In the third diagram, the contraction of two spinor pairs due to the two fermion-antifermion-scalar vertices gives rise to an extra factor $2^2$. Combined with the fact that $y^2 = 4\pi\aF$ for fermions~\cite[appendix~A]{Oncala:2018bvl}, reproduces \cref{eq:BSF_VectorEmission_sigma_100}.

Despite the different potentials in the initial and final states -- which in fact occur commonly in \BSFV \ in non-Abelian theories where the emitted gauge vector boson carries away non-Abelian charge~\cite{Harz:2018csl} -- \BSFV \ is suppressed by $\aB^2$ with respect to the \BSFF \ processes computed in \cref{Sec:BSFviaScalarEmission}, due to the momentum dependence of the vector emission vertices. (For the third diagram of \cref{fig:BSF_Vector_RadiativePart}, and its interference with the other two diagrams, the suppression is actually of order $\aF^2$ and $\aF \aB$ respectively, at the level of the cross-section.).

\bibliography{Bibliography.bib}

\providecommand{\href}[2]{#2}\begingroup\raggedright\begin{thebibliography}{10}

\bibitem{Harz:2017dlj}
J.~Harz and K.~Petraki, \emph{{Higgs Enhancement for the Dark Matter Relic
  Density}}, \href{https://doi.org/10.1103/PhysRevD.97.075041}{\emph{Phys.
  Rev.} {\bfseries D97} (2018) 075041}
  [\href{https://arxiv.org/abs/1711.03552}{{\ttfamily 1711.03552}}].

\bibitem{Harz:2019rro}
J.~Harz and K.~Petraki, \emph{{Higgs-mediated bound states in dark-matter
  models}}, \href{https://doi.org/10.1007/JHEP04(2019)130}{\emph{JHEP}
  {\bfseries 04} (2019) 130}
  [\href{https://arxiv.org/abs/1901.10030}{{\ttfamily 1901.10030}}].

\bibitem{Wise:2014jva}
M.~B. Wise and Y.~Zhang, \emph{{Stable Bound States of Asymmetric Dark
  Matter}}, \href{https://doi.org/10.1103/PhysRevD.90.055030}{\emph{Phys.Rev.}
  {\bfseries D90} (2014) 055030}
  [\href{https://arxiv.org/abs/1407.4121}{{\ttfamily 1407.4121}}].

\bibitem{Petraki:2016cnz}
K.~Petraki, M.~Postma and J.~de~Vries, \emph{{Radiative bound-state-formation
  cross-sections for dark matter interacting via a Yukawa potential}},
  \href{https://doi.org/10.1007/JHEP04(2017)077}{\emph{JHEP} {\bfseries 04}
  (2017) 077} [\href{https://arxiv.org/abs/1611.01394}{{\ttfamily
  1611.01394}}].

\bibitem{Oncala:2018bvl}
R.~Oncala and K.~Petraki, \emph{{Dark matter bound states via emission of
  scalar mediators}},
  \href{https://doi.org/10.1007/JHEP01(2019)070}{\emph{JHEP} {\bfseries 01}
  (2019) 070} [\href{https://arxiv.org/abs/1808.04854}{{\ttfamily
  1808.04854}}].

\bibitem{Migdal:QualitativeQM}
A.~Migdal, \emph{Qualitative Methods In Quantum Theory}, Advanced Books
  Classics. Avalon Publishing, 2000.

\bibitem{vonHarling:2014kha}
B.~von Harling and K.~Petraki, \emph{{Bound-state formation for thermal relic
  dark matter and unitarity}},
  \href{https://doi.org/10.1088/1475-7516/2014/12/033}{\emph{JCAP} {\bfseries
  12} (2014) 033} [\href{https://arxiv.org/abs/1407.7874}{{\ttfamily
  1407.7874}}].

\bibitem{Harz:2018csl}
J.~Harz and K.~Petraki, \emph{{Radiative bound-state formation in unbroken
  perturbative non-Abelian theories and implications for dark matter}},
  \href{https://doi.org/10.1007/JHEP07(2018)096}{\emph{JHEP} {\bfseries 07}
  (2018) 096} [\href{https://arxiv.org/abs/1805.01200}{{\ttfamily
  1805.01200}}].

\bibitem{Kats:2009bv}
Y.~Kats and M.~D. Schwartz, \emph{{Annihilation decays of bound states at the
  LHC}}, \href{https://doi.org/10.1007/JHEP04(2010)016}{\emph{JHEP} {\bfseries
  04} (2010) 016} [\href{https://arxiv.org/abs/0912.0526}{{\ttfamily
  0912.0526}}].

\bibitem{Petraki:2015hla}
K.~Petraki, M.~Postma and M.~Wiechers, \emph{{Dark-matter bound states from
  Feynman diagrams}},
  \href{https://doi.org/10.1007/JHEP06(2015)128}{\emph{JHEP} {\bfseries 1506}
  (2015) 128} [\href{https://arxiv.org/abs/1505.00109}{{\ttfamily
  1505.00109}}].

\bibitem{Sommerfeld:1931}
A.~Sommerfeld, \emph{{{\"U}ber die Beugung und Bremsung der Elektronen}},
  {\emph{Ann. Phys.} {\bfseries 403} (1931) 257}.

\bibitem{Sakharov:1948yq}
A.~D. Sakharov, \emph{{Interaction of an Electron and Positron in Pair
  Production}},
  \href{https://doi.org/10.1070/PU1991v034n05ABEH002492}{\emph{Zh. Eksp. Teor.
  Fiz.} {\bfseries 18} (1948) 631}.

\bibitem{Baldes:2017gzw}
I.~Baldes and K.~Petraki, \emph{{Asymmetric thermal-relic dark matter:
  Sommerfeld-enhanced freeze-out, annihilation signals and unitarity bounds}},
  \href{https://doi.org/10.1088/1475-7516/2017/09/028}{\emph{JCAP} {\bfseries
  1709} (2017) 028} [\href{https://arxiv.org/abs/1703.00478}{{\ttfamily
  1703.00478}}].

\bibitem{Baldes:2017gzu}
I.~Baldes, M.~Cirelli, P.~Panci, K.~Petraki, F.~Sala and M.~Taoso,
  \emph{{Asymmetric dark matter: residual annihilations and
  self-interactions}},
  \href{https://doi.org/10.21468/SciPostPhys.4.6.041}{\emph{SciPost Phys.}
  {\bfseries 4} (2018) 041} [\href{https://arxiv.org/abs/1712.07489}{{\ttfamily
  1712.07489}}].

\bibitem{Cirelli:2018iax}
M.~Cirelli, Y.~Gouttenoire, K.~Petraki and F.~Sala, \emph{{Homeopathic Dark
  Matter, or how diluted heavy substances produce high energy cosmic rays}},
  \href{https://doi.org/10.1088/1475-7516/2019/02/014}{\emph{JCAP} {\bfseries
  1902} (2019) 014} [\href{https://arxiv.org/abs/1811.03608}{{\ttfamily
  1811.03608}}].

\bibitem{Fukuda:2018ufg}
H.~Fukuda, F.~Luo and S.~Shirai, \emph{{How Heavy can Neutralino Dark Matter
  be?}}, \href{https://doi.org/10.1007/JHEP04(2019)107}{\emph{JHEP} {\bfseries
  04} (2019) 107} [\href{https://arxiv.org/abs/1812.02066}{{\ttfamily
  1812.02066}}].

\bibitem{Pospelov:2008jd}
M.~Pospelov and A.~Ritz, \emph{{Astrophysical Signatures of Secluded Dark
  Matter}},
  \href{https://doi.org/10.1016/j.physletb.2008.12.012}{\emph{Phys.Lett.}
  {\bfseries B671} (2009) 391}
  [\href{https://arxiv.org/abs/0810.1502}{{\ttfamily 0810.1502}}].

\bibitem{Pearce:2015zca}
L.~Pearce, K.~Petraki and A.~Kusenko, \emph{{Signals from dark atom formation
  in halos}},
  \href{https://doi.org/10.1103/PhysRevD.91.083532}{\emph{Phys.Rev.} {\bfseries
  D91} (2015) 083532} [\href{https://arxiv.org/abs/1502.01755}{{\ttfamily
  1502.01755}}].

\bibitem{An:2016gad}
H.~An, M.~B. Wise and Y.~Zhang, \emph{{Effects of Bound States on Dark Matter
  Annihilation}}, \href{https://doi.org/10.1103/PhysRevD.93.115020}{\emph{Phys.
  Rev.} {\bfseries D93} (2016) 115020}
  [\href{https://arxiv.org/abs/1604.01776}{{\ttfamily 1604.01776}}].

\bibitem{Cirelli:2016rnw}
M.~Cirelli, P.~Panci, K.~Petraki, F.~Sala and M.~Taoso, \emph{{Dark Matter's
  secret liaisons: phenomenology of a dark U(1) sector with bound states}},
  \href{https://doi.org/10.1088/1475-7516/2017/05/036}{\emph{JCAP} {\bfseries
  1705} (2017) 036} [\href{https://arxiv.org/abs/1612.07295}{{\ttfamily
  1612.07295}}].

\bibitem{Messiah:1962}
A.~{Messiah}, \emph{{Quantum mechanics}}. North-Holland Pub. Co., 1962.

\bibitem{Asadi:2016ybp}
P.~Asadi, M.~Baumgart, P.~J. Fitzpatrick, E.~Krupczak and T.~R. Slatyer,
  \emph{{Capture and Decay of Electroweak WIMPonium}},
  \href{https://doi.org/10.1088/1475-7516/2017/02/005}{\emph{JCAP} {\bfseries
  1702} (2017) 005} [\href{https://arxiv.org/abs/1610.07617}{{\ttfamily
  1610.07617}}].

\bibitem{Cassel:2009wt}
S.~Cassel, \emph{{Sommerfeld factor for arbitrary partial wave processes}},
  \href{https://doi.org/10.1088/0954-3899}{\emph{J.Phys.} {\bfseries G37}
  (2010) 105009} [\href{https://arxiv.org/abs/0903.5307}{{\ttfamily
  0903.5307}}].

\bibitem{Binder:2019erp}
T.~Binder, K.~Mukaida and K.~Petraki, \emph{{Rapid bound-state formation of
  Dark Matter in the Early Universe}},
  \href{https://arxiv.org/abs/1910.11288}{{\ttfamily 1910.11288}}.

\bibitem{Kim:2016zyy}
S.~Kim and M.~Laine, \emph{{Rapid thermal co-annihilation through bound states
  in QCD}}, \href{https://doi.org/10.1007/JHEP07(2016)143}{\emph{JHEP}
  {\bfseries 07} (2016) 143}
  [\href{https://arxiv.org/abs/1602.08105}{{\ttfamily 1602.08105}}].

\bibitem{Biondini:2017ufr}
S.~Biondini and M.~Laine, \emph{{Re-derived overclosure bound for the inert
  doublet model}}, \href{https://doi.org/10.1007/JHEP08(2017)047}{\emph{JHEP}
  {\bfseries 08} (2017) 047}
  [\href{https://arxiv.org/abs/1706.01894}{{\ttfamily 1706.01894}}].

\bibitem{Biondini:2018xor}
S.~Biondini, \emph{{Bound-state effects for dark matter with Higgs-like
  mediators}}, \href{https://doi.org/10.1007/JHEP06(2018)104}{\emph{JHEP}
  {\bfseries 06} (2018) 104}
  [\href{https://arxiv.org/abs/1805.00353}{{\ttfamily 1805.00353}}].

\bibitem{Griest:1989wd}
K.~Griest and M.~Kamionkowski, \emph{{Unitarity Limits on the Mass and Radius
  of Dark Matter Particles}},
  \href{https://doi.org/10.1103/PhysRevLett.64.615}{\emph{Phys.Rev.Lett.}
  {\bfseries 64} (1990) 615}.

\bibitem{Aghanim:2018eyx}
{\scshape Planck} collaboration, N.~Aghanim et~al., \emph{{Planck 2018 results.
  VI. Cosmological parameters}},
  \href{https://arxiv.org/abs/1807.06209}{{\ttfamily 1807.06209}}.

\bibitem{Liew:2016hqo}
S.~P. Liew and F.~Luo, \emph{{Effects of QCD bound states on dark matter relic
  abundance}}, \href{https://doi.org/10.1007/JHEP02(2017)091}{\emph{JHEP}
  {\bfseries 02} (2017) 091}
  [\href{https://arxiv.org/abs/1611.08133}{{\ttfamily 1611.08133}}].

\bibitem{Binder:2018znk}
T.~Binder, L.~Covi and K.~Mukaida, \emph{{Dark Matter Sommerfeld-enhanced
  annihilation and Bound-state decay at finite temperature}},
  \href{https://doi.org/10.1103/PhysRevD.98.115023}{\emph{Phys. Rev.}
  {\bfseries D98} (2018) 115023}
  [\href{https://arxiv.org/abs/1808.06472}{{\ttfamily 1808.06472}}].

\bibitem{vandenAarssen:2012ag}
L.~G. van~den Aarssen, T.~Bringmann and Y.~C. Goedecke, \emph{{Thermal
  decoupling and the smallest subhalo mass in dark matter models with
  Sommerfeld-enhanced annihilation rates}},
  \href{https://doi.org/10.1103/PhysRevD.85.123512}{\emph{Phys.Rev.} {\bfseries
  D85} (2012) 123512} [\href{https://arxiv.org/abs/1202.5456}{{\ttfamily
  1202.5456}}].

\bibitem{Binder:2017lkj}
T.~Binder, M.~Gustafsson, A.~Kamada, S.~M.~R. Sandner and M.~Wiesner,
  \emph{{Reannihilation of self-interacting dark matter}},
  \href{https://doi.org/10.1103/PhysRevD.97.123004}{\emph{Phys. Rev.}
  {\bfseries D97} (2018) 123004}
  [\href{https://arxiv.org/abs/1712.01246}{{\ttfamily 1712.01246}}].

\bibitem{Gondolo:1990dk}
P.~Gondolo and G.~Gelmini, \emph{{Cosmic abundances of stable particles:
  Improved analysis}},
  \href{https://doi.org/10.1016/0550-3213(91)90438-4}{\emph{Nucl.Phys.}
  {\bfseries B360} (1991) 145}.

\bibitem{Lonsdale:2014wwa}
S.~J. Lonsdale and R.~R. Volkas, \emph{{Grand unified hidden-sector dark
  matter}}, \href{https://doi.org/10.1103/PhysRevD.90.083501,
  10.1103/PhysRevD.91.129906}{\emph{Phys. Rev.} {\bfseries D90} (2014) 083501}
  [\href{https://arxiv.org/abs/1407.4192}{{\ttfamily 1407.4192}}].

\bibitem{Lonsdale:2017mzg}
S.~J. Lonsdale, M.~Schroor and R.~R. Volkas, \emph{{Asymmetric Dark Matter and
  the hadronic spectra of hidden QCD}},
  \href{https://doi.org/10.1103/PhysRevD.96.055027}{\emph{Phys. Rev.}
  {\bfseries D96} (2017) 055027}
  [\href{https://arxiv.org/abs/1704.05213}{{\ttfamily 1704.05213}}].

\bibitem{Lopez-Honorez:2017ora}
L.~Lopez~Honorez, M.~H.~G. Tytgat, P.~Tziveloglou and B.~Zaldivar, \emph{{On
  Minimal Dark Matter coupled to the Higgs}},
  \href{https://doi.org/10.1007/JHEP04(2018)011}{\emph{JHEP} {\bfseries 04}
  (2018) 011} [\href{https://arxiv.org/abs/1711.08619}{{\ttfamily
  1711.08619}}].

\bibitem{Ko:2019wxq}
P.~Ko, T.~Matsui and Y.-L. Tang, \emph{{Dark Matter Bound State Formation in
  Fermionic $Z_2$ DM model with Light Dark Photon and Dark Higgs Boson}},
  \href{https://arxiv.org/abs/1910.04311}{{\ttfamily 1910.04311}}.

\bibitem{Integrals_GradshteynRyzhik}
I.~S. {Gradshteyn}, I.~M. {Ryzhik}, A.~{Jeffrey} and D.~{Zwillinger},
  \emph{{Table of Integrals, Series, and Products}}. 2007.

\end{thebibliography}\endgroup

\end{document}